\documentclass[a4paper,11pt]{article}

\usepackage{geometry}
\usepackage{cite}
\usepackage{amsmath,amssymb,amsfonts, amsthm}
\usepackage{graphicx}
\usepackage{textcomp}
\usepackage{xcolor}

\usepackage{stmaryrd}
\usepackage{hyperref}
\usepackage{MnSymbol}

\usepackage{adjustbox}
\usepackage{bussproofs}\EnableBpAbbreviations
\usepackage{subcaption}
\usepackage{listings}
\usepackage{algorithm2e}
\usepackage{algpseudocode}

\usepackage{tikz}
\usetikzlibrary{cd, angles, shapes}
\tikzcdset{scale cd/.style={every label/.append style={scale=#1},
    cells={nodes={scale=#1}}}}

\usetikzlibrary{calc}
\tikzset{curve/.style={settings={#1},to path={(\tikztostart)
    .. controls ($(\tikztostart)!\pv{pos}!(\tikztotarget)!\pv{height}!270:(\tikztotarget)$)
    and ($(\tikztostart)!1-\pv{pos}!(\tikztotarget)!\pv{height}!270:(\tikztotarget)$)
    .. (\tikztotarget)\tikztonodes}},
    settings/.code={\tikzset{quiver/.cd,#1}
        \def\pv##1{\pgfkeysvalueof{/tikz/quiver/##1}}},
    quiver/.cd,pos/.initial=0.35,height/.initial=0}

\providecommand{\keywords}[1]
{
  \small	
  \textbf{\textit{Keywords---}} #1
}

\newtheorem{remark}{Remark}[section]
\newenvironment{varitemize}
{
	\begin{list}{\labelitemi}
		{\setlength{\itemsep}{0pt}
			\setlength{\topsep}{0pt}
			\setlength{\parsep}{0pt}
			\setlength{\partopsep}{0pt}
			\setlength{\leftmargin}{15pt}
			\setlength{\rightmargin}{0pt}
			\setlength{\itemindent}{0pt}
			\setlength{\labelsep}{5pt}
			\setlength{\labelwidth}{10pt}
	}}
	{
	\end{list} 
}

\newcounter{numberone}

\newenvironment{varenumerate}
{
	\begin{list}{\arabic{numberone}.}
		{
			\usecounter{numberone}
			\setlength{\itemsep}{0pt}
			\setlength{\topsep}{0pt}
			\setlength{\parsep}{0pt}
			\setlength{\partopsep}{0pt}
			\setlength{\leftmargin}{15pt}
			\setlength{\rightmargin}{0pt}
			\setlength{\itemindent}{0pt}
			\setlength{\labelsep}{5pt}
			\setlength{\labelwidth}{15pt}
	}}
	{
	\end{list} 
}

\newcommand{\NINF}{\BB N^\infty}

\newcommand{\RS}{\BB R_{\geq 0}^\infty}
\newcommand{\TS}{\BB{T}}
\newcommand{\BS}{\{0,1\}}

\newcommand{\LEFT}{\Big\langle}
\newcommand{\RIGHT}{\Big\rangle}

\newcommand{\VN}{\mathbf{VN}}
\newcommand{\trop}{\mathsf t^!}

\newcommand{\TIT}{\B{P}_{\mathrm{trop}}}
\newcommand{\IT}{\B{P}}

\newcommand{\pPCF}{$\mathrm{PCF}\langle \vec X\rangle$}
\newcommand{\pPCFn}{$\mathrm{PCF}\langle X_1,\dots, X_n\rangle$}
\newcommand{\pPCFX}[1]{$\mathrm{PCF}\langle #1 \rangle$}

\newcommand{\choice}[1]{\oplus_{#1}}
\newcommand{\Bool}{\mathrm{Bool}}

\newcommand{\YY}{\mathrm{Y}}

\newcommand{\NAT}{\mathbf{N}
}
\newcommand{\SUCC}[1]{\mathsf{succ}\ #1}
\newcommand{\PRED}[1]{\mathsf{pred}\ #1}
\newcommand{\ZZERO}{{\mathsf 0}}
\newcommand{\ZERO}{\mathsf 0}
\newcommand{\ONE}{\mathsf 1}

\newcommand{\ITE}[3]{\mathsf{ifz}(#1,#2,#3)}
\newcommand{\RED}[1]{\stackrel{#1}{\twoheadrightarrow}}
\newcommand{\onestep}[1]{\stackrel{#1}{\to}}

\newcommand{\modd}[1]{\llbracket#1\rrbracket }

\newcommand{\set}[1]{\{ #1 \}}

\newcommand{\fmset}[1]{!{#1}}
\newcommand{\fps}[2]{{#1}\{\!\!\{#2\}\!\!\}}
\newcommand{\fpp}[2]{{#1}\{#2\}}

\newcommand{\B}[1]{\mathbf{#1}}

\newcommand{\SELECT}{\mathrm{merge}\ }

\newcommand{\PROB}{\mathbb{P}}

\newcommand{\model}[1]{\modd{#1}}

\newcommand{\an}[1]{{#1}\textbf{An}}
\newcommand{\rel}[1]{{#1}\textbf{Rel}}

\newcommand{\TT}[1]{\mathtt{#1}}

\newcommand{\C}[1]{\mathcal{#1}}
\newcommand{\BB}[1]{\mathbb{#1}}

\newcommand{\OV}[1]{\overline{#1}}
\newcommand{\F}[1]{\mathfrak{#1}}

\newcommand{\To}{\Rightarrow}

\newcommand{\N}{\BB N}
\newcommand{\R}{\BB R}

\providecommand{\dotdiv}{% Don't redefine it if available
  \mathbin{% We want a binary operation
    \vphantom{+}% The same height as a plus or minus
    \text{% Change size in sub/superscripts
      \mathsurround=0pt % To be on the safe side
      \ooalign{% Superimpose the two symbols
        \noalign{\kern-.35ex}% but the dot is raised a bit
        \hidewidth$\smash{\cdot}$\hidewidth\cr % Dot
        \noalign{\kern.35ex}% Backup for vertical alignment
        $-$\cr % Minus
      }%
    }%
  }%
}

\newcommand{\DEL}[1]
{
{
\color{violet}%\bfseries
}
}

\newcommand{\deriv}[2]{#1 \ \ \vdots \ \ #2}

\newcommand{\appr}[2]{#1 \triangleleft #2}

\def\BibTeX{{\rm B\kern-.05em{\sc i\kern-.025em b}\kern-.08em
    T\kern-.1667em\lower.7ex\hbox{E}\kern-.125emX}}

\newtheorem{example}{Example}[section]
\newtheorem{definition}{Definition}[section]

\newtheorem{theorem}{Theorem}[section]

\newtheorem{corollary}{Corollary}[section]
\newtheorem{lemma}[theorem]{Lemma}

\newtheorem{proposition}[theorem]{Proposition}

\begin{document}

\title{Tropical Mathematics and the Lambda-Calculus II:

Tropical Geometry of Probabilistic Programming Languages}

%\author{\IEEEauthorblockN{Anonymous Authors}}
%
%
\author{{Davide Barbarossa}
{\textit{University of Bath} }\\
Bath, UK \\
db2437@bath.ac.uk
\and
{Paolo Pistone}
{\textit{Universit\'e Claude Bernard Lyon 1} }\\
Lyon, France \\
paolo.pistone@ens-lyon.fr}

\maketitle

\begin{abstract}
In the last few years there has been a growing interest towards methods for statistical inference and learning based on %ideas from 
computational %algebraic
geometry %, \NEW{linear optimisation} 
and, notably, %from
tropical geometry, that is, the study of algebraic varieties over the min-plus semiring.
At the same time, recent work has demonstrated the possibility of interpreting higher-order probabilistic programming languages in the framework of tropical mathematics, by exploiting %the weighted relational semantics from linear logic.
algebraic and categorical tools coming from the semantics of linear logic.
In this work we combine %\DEL{By combining}
these two worlds, %\DEL{in this work we show}
showing that tools and ideas from tropical geometry can be used 
%{allow} 
to perform statistical inference over
%study %infer the most likely behaviours of 
higher-order probabilistic programs.
%
%In particular, we show that each such program can be associated with an $n$-dimensional polyhedron that encodes its most likely reduction paths, and we show  that this polyhedron can be reconstructed in a compositional and efficient way through a suitable intersection type system. We finally show that this approach can be applied to estimate the differential privacy of probabilistic protocols.
%
Notably, we first show that each such program can be associated with a \emph{degree} and a \emph{$n$-dimensional polyhedron} that encode its \emph{most likely} %reduction paths
runs. %We then put in action our approach: 
Then, we use these tools in order to design an intersection type system that estimates most likely runs in a compositional and efficient way.
\end{abstract}


%expressed in our language.
%can be used to estimate its differential privacy.
% that 
%
%we propose an interaction between programming language principles and linear optimisation ones, in order to study probabilistic processes:
%In this paper we
%put % in contact:
%, by showing thatactual methods from tropical geometry can indeed be exploited to perform statistical inference on higher-order programs. 
%For example, we show that the problem of describing the most-likely behavior of a probabilistic PCF program reduces to %computing a
%studying a tropical polynomial function %
%associated with the program. 
%%thanks to the second toolbox, we introduce the notion of tropical degree of a program, an estimation of the length of its most-likely behaviour and, thanks to the first toolbox, we associate a program with a $n$-dimensional polyhedron encoding its most-likely behaviours.
%%We then put our approach in action:
%%we use the polyhedron in order to design
%%%Then we
%%%We also  design
%%an intersection type system %capable of capturing 
%%that %captures such polynomials
%%%estimates the most-likely behaviours.
%%allows to perform statistical inference over programs and, also,
%%%As an application of our approach
%%%Finally, 
%%we use the tropical degree %polynomial associated with a 
%%%of a program 
%%to estimate the differential privacy of probabilistic protocols. %expressed in our language.
%%%can be used to estimate its differential privacy.

%
%\begin{IEEEkeywords}
%Probabilistic lambda-calculus, Tropical geometry, Relational Semantics, Differential Privacy
%\end{IEEEkeywords}

\section{Introduction}\label{sec:intro}

% !TEX root = Tropical_II.tex

\paragraph{From Probabilistic Models to Probabilistic Programming Languages}
Probabilistic models play a fundamental role in many areas of computer science, such as, just to name a few, machine learning, bioinformatics, speech recognition, robotics and computer vision. For many common problems (like, for example, identifying the regions of DNA that code for some specific protein or tracking the location of a vehicle from the data produced by possibly faulty sensors)
finding an exact solution requires to enumerate an impossibly large list of possibilities; by contrast, a probabilistic model may allow one to focus 
only on those (usually, much less) possibilities which are \emph{more likely} to occur, under normal circumstances.
In this respect, models like Bayesian Networks (BN) or Hidden Markov Models (HMM) provide an extremely well-studied and modular approach making  
the representation of (our current knowledge of) the system under study independent from the inference algorithms that can be applied in order to answer specific questions about it.

\DEL{
At the same time, the pervasiveness of probabilistic methods to extract information from raw data may raise concerns about the exposure of sensible or critical information. Approaches  like \emph{differential privacy} (DP) have been developed as means to ensure that statistical queries, while producing relevant global information, may not not leak sensible data.
}

While probabilistic models provide a description of a system under conditions of uncertain knowledge, probabilistic programming languages (PPL) provide ways to \emph{specify such models via programs}: the execution of the program produces the model, in the sense that the probabilistic reductions of the program describe the trajectories of the model. 
The inference tasks associated with a model are thus naturally related to the study of the probabilistic execution of the corresponding program (e.g., what is the probability that the program will return $\mathrm{True}$? Or that it terminates?). 

The goal of PPL like e.g.~Church or Anglican (to name two languages that rely on the LISP architecture),    
%the description of the model (the program) is conceptually separated from the inference tasks and methods that can be associated with it. Here the goal
 is to streamline the activity of probabilistic modeling, by exploiting features of programming languages like {compositionality} and higher-order functions.
% (a complex program can be analyzed as the composition of several, simpler, ones), \emph{higher-order} (programs are allowed to operate on other programs as functions) or even \emph{abstract} (e.g.~involving forms of \emph{polymorphism}, so that the same piece of code can be re-used in different situations). 
 This becomes particularly relevant when considering possibly \emph{infinitary} models that take {temporality} into account, like e.g.~template-based Bayesian Networks, which can be conveniently described in higher-order functional languages, cf.~\cite{Vanoni2024}.  
 The study of PPLs has recently seen a flourishing of research directions, going from more foundational/category-theoretic approaches \cite{Jacobs_Zanasi_2020, Dahlqvist2018, Heunen2017, Staton2017}, to others more oriented towards inference algorithms and their efficiency like \cite{Vanoni2024}. 
\DEL{The PPL perspective has also been successfully applied to the problem of differential privacy:
higher-order languages like e.g.~System FUZZ \cite{Reed2010}, ensure, by construction, that well-typed programs will respect the required privacy conditions.}

\paragraph{The Tropical Geometry of Probabilistic Models }

The application of methods from computational algebraic geometry in areas like machine learning and statistical inference is well investigated. Among such methods a growing literature has explored the application of ideas from tropical geometry to the study of deep neural networks and graphical probabilistic models  \cite{Maragos2021, Maragos2017, Zhang2018, Pachter2004, Pachter2004b}.

Tropical geometry is the study of polynomials and algebraic varieties defined over the min-plus (or the max-plus) semiring: a tropical polynomial is obtained from a standard polynomial by replacing $+$ with min and $\times$ with $+$. 
Several computationally difficult problems expressible in the language of algebraic geometry admit a tropical counterpart which is purely combinatorial and, in some cases, tractable in an effective way.
For example, while finding the roots of a  polynomial is a paradigmatic undecidable problem, tropical roots can be computed in linear time and used to approximate the actual roots of the polynomial \cite{Noferini2015, Porzio2021}. 

Concerning probabilistic models, it has been observed that several inference algorithms based on convex optimization, like the \emph{Viterbi algorithm}, have a ``tropical flavor'' \cite{Maragos2018}. Usually, graphical probabilistic models express the probability of an event as a polynomial $p_E$, which intuitively adds up the probabilities $p_i$ of the (so many) mutually independent situations $i$ that might produce $E$. A typical problem, for instance when computing Bayesian posteriors, is to know, given the knowledge that the event $E$ occurred, which situations $i$ are the \emph{most likely} to have produced $E$. While comparing \emph{all} the situations $i$ is certainly not feasible, 
works like \cite{Pachter2004, Pachter2004b} have shown that the study of the \emph{Newton polytope} of the tropical polynomial associated to $p_E$ provides an efficient method to select the potential solutions $i$.

\paragraph{The Tropical Geometry of PPLs}

While ideas from tropical mathematics have been applied successfully to probabilistic models like HMM or BN, in this paper we are concerned with the following, broader, question: would it be possible to exploit the computational toolkit of tropical geometry as an inference engine for the programs of some higher-order programming language, and, as a consequence, for the large class of probabilistic models that this language may represent?

%
%\TODO{The goal of this work is to demonstrate that the statistical methods coming from tropical geometry can be  statistical inference on a ...here! }
Our approach relies on a recent line of work \cite{BarbaPist2024} that has demonstrated the possibility of interpreting higher-order probabilistic languages within the setting of tropical mathematics. This interpretation relies on the \emph{weighted relational semantics} (WRS) \cite{Manzo2013}, a well-studied class of models of PCF and related languages that is parametric on the choice of a continuous semiring $Q$. The WRS arises from the literature on linear logic and has been at the heart of numerous investigations and results about programming languages with non-determinism, probabilities or even quantum primitives \cite{Pagani2018, EhrPagTas2018, Laird2016, Laird2020, Pagani2014, Forest2024}.

When $Q$ is the min-plus semiring {(on $\R_{\geq0}\cup\set{\infty}$ or $\N\cup\set{\infty}$)}, one obtains a semantics of probabilistic PCF (pPCF) that has been shown to capture the \emph{most likely} behavior of a program {\cite{Manzo2013,BarbaPist2024}}. For example, of the many ways in which a program $M$ of type $\mathrm{Bool}$ may reduce to $\mathrm{True}$, only those which have the \emph{highest} probability to occur are represented in the semantics. 
Therefore, the tropical WRS intuitively ``solves'' the inference task of selecting the most likely execution paths of a pPCF- program, and, along with them, the most likely trajectories in the corresponding probabilistic model. However, as this semantics is neither finitary nor computable in general, the obvious question is: to which extent could such ``solutions'' be produced in an effective way?
%
%Another way to phrase this is as follows: would it be possible to exploit the tropical relational semantics, together with the computational toolkit of tropical geometry, to efficiently solve statistical tasks over the programs of some higher-order programming language, and, as a consequence, over the large class of probabilistic models that this language may represent?

This paper provides an answer to this question, obtained in two steps: first, we exploit the tropical setting to show that the problem of describing a \emph{most likely} trajectory for a pPCF program can be reduced to a search problem within some \emph{finite} (albeit possibly very large) space. This fact is non-trivial, since, unlike standard BN, the probabilistic models corresponding to pPCF programs may have an \emph{infinite} trajectory space, and is established via a non-constructive argument. 
{Nonetheless, by combining ideas from tropical geometry and programming language theory, 
we demonstrate that it is possible to design a recursive procedure that, 
% enumerates the most likely trajectories of a pPCF program,G
given a higher-order program of ground type, explores the trajectories of the corresponding infinitary model in a compositional way, and
converges to a solution to the following two standard inference tasks (cf.~Theorem \ref{thm:stats}): 
\begin{varitemize}
\item[(I1)] select the most likely trajectories of the model,
\item[(I2)] for a fixed trajectory $\theta$, identify the values of the probabilistic parameters making $\theta$ most likely.
\end{varitemize}
The convergence of this procedure is non-recursive, as the underlying problem is undecidable, 
%Notice that one cannot hope to be able to recognize in any situation \emph{when} the method has actually reached an \emph{exact} solution, since this would require to know the size of the underlying finite trajectory space, un undecidable problem. Still, 
but the partial solutions provided at each stage of the computation are correct for a restricted set of trajectories approximating the behaviour of the program. Furthermore, we show that in several interesting cases (e.g.~when considering an \emph{affine} higher-order program) our method does indeed produce a correct answer, and it does so \emph{efficiently}: while the number of possible trajectories is exponential in the size of the program, the most likely ones are found in polynomial time.

}
%
%
%Second, by combining ideas from tropical geometry and programming language theory, we design a procedure to enumerate the most-likely trajectories of a program in a way that is \emph{both} compositional and efficient. 
%
%
%The outcome of this strategy is a method that, given a higher-order program of ground type, describing an infinitary model with a finite set of parameters, solves the following two standard inference tasks (cf.~Theorem \ref{thm:stats}): 
%\begin{varitemize}
%\item[(I1)] select the most likely trajectories of the model,
%\item[(I2)] for a fixed trajectory $\theta$, identify the values of the underlying probabilistic parameters that make $\theta$ a most likely one.
%
%
%\end{varitemize}
%

\paragraph{The Tropical Degree}
As we said, our first result shows that the trajectory space for a pPCF program can be reduced to a finite one. 
This is obtained by exploiting a representation of such programs via \emph{tropical polynomials}.
%
%This is obtained via the notion of \emph{tropical degree}, introduced in Section \ref{sec:tropdeg}.
Indeed, graphical probabilistic models like BN and HMM are often presented algebraically via families of polynomials in a given set of parameters. The WRS extends this presentation to pPCF programs, except that, due to their higher-order nature,  programs correspond to \emph{power series}, not just polynomials, in the parameters. Intuitively, if a finite sum of monomials is enough to add up finitely many independent trajectories that may lead to the same result, an infinite sum is required when the number of trajectories is infinite.
%This setting is indeed especially convenient for exploring infinitary graphical models like \emph{Dynamical Bayesian Networks} (DBN) as well as  HMM. 
When considering the interpretation of pPCF over the tropical semiring, 
these power series are turned into \emph{tropical analytic functions} (\emph{taf}, for short), 
that is, continuous functions that can be written as $f(x)=\inf_{i\in I}\phi_i(x)$, i.e.~an $\inf$ of possibly infinitely many linear maps $\phi_i(x)$. While tropical polynomials and their geometric properties are very well-studied, the literature on taf is still scarce \cite{Porzio2021, BarbaPist2024}. 

Our central result here, Corollary \ref{thm:collapse} from Section \ref{sec:tropdeg}, is that any program $M$ of ground type, say $\Bool$, is represented by a taf that is in fact a tropical polynomial (in other words, that $f$ can be written as $f(x)=\min_{i\in J\subseteq I}\phi_i$ for some \emph{finite} subset $J\subseteq I$). This follows from a general result (cf.~Proposition \ref{prop:collapse}) about tafs with discrete coefficients.
The meaning of this result is that, among the many trajectories that may lead to the same event, only a \emph{finite} portion has a chance of occurring ``most likely''. Intuitively, if we think of $M$ as describing a probabilistic model that iterates a given procedure until it produces a given result $o$ (a typical \emph{Las Vegas} algorithm), then the probability that $o$ was obtained after \emph{no less} than $n$ iterations will reach its maximum after a finite number $D$ of steps: the greater the number of iterations in a reduction of $M$, the lower the probability that this reduction may actually have occurred. 
This number $D$ is what we call \emph{tropical degree of $M$}, and coincides with the (finite) degree of the tropical polynomial that captures the behavior of $M$.

\paragraph{Statistical Inference via Intersection Types and the Newton Polytope}

Even once we have reduced the trajectories of our program to a finite set, this set may still be \emph{too large} to be enumerated in practice. The second step is thus to show that the trajectory space can be further reduced to one of a more \emph{tractable} size, in order to have a hope to access it in a feasible computational way.
Our first result here is to associate a program $M$ with a $n$-dimensional polyhedron $NP_{\min}(M)$, a variant of the standard Newton polytope called the \emph{minimal Newton polytope}, that encodes the most likely runs of $M$ in a more optimized way (cf.~Theorem \ref{thm:np}). Then, we design a type system $\TIT$ that captures the most likely executions of the program compositionally in $M$ (cf.~Theorem \ref{thm:correct}),
{ and we show that this system can be used to design a method that reconstructs the polytope $NP_{\min}(M)$ in a recursive way. As anticipated before, while the termination condition for this method is non-recursive, we discuss the situations in which it can be used to provide correct solutions efficiently.}
The system $\TIT$ uses 
 \emph{non-idempotent intersection types} \cite{decarvalho2018, Kesner2017, Pagani2018, Kesner2023},
 a well-studied technique to capture the quantitative behavior of higher-order programs. 
In $\TIT$ a \emph{single} type derivation may explore a \emph{plurality} of possible executions of the program, at the same time \emph{selecting} a {small} enough set of most likely trajectories; to do that it makes use of an algorithm to \emph{compose} graphical models inspired by the Viterbi algorithm for HMM and relying on the computation of the minimal {Newton polytopes} of the underlying polynomials (cf.~Theorem \ref{thm:vn}).

\DEL{
\paragraph{Differential Privacy via Tropical Geometry}

As an application of our results, we explore a connection with differential privacy.
A key factor to show that a program may not extract private information is that the program need not be too \emph{sensitive} to 
small changes in the input. When a program $M$ has a low sensitivity, well-established probabilistic methods (e.g.~the Laplace mechanism \cite{Dwork2014}) can be applied to turn $M$ into a differentially-private program.

By exploiting the fact that tropical polynomials are always Lipschitz continuous, we show that the tropical degree of a probabilistic program can be used to produce an estimation of its differential privacy. Since the tropical degree of a complex probabilistic program may be quite high in general, it is not obvious that the produced estimations are of practical use, beyond simple explanatory cases. 
At the same time, our results highlight a surprising conceptual connection between these two areas that we think it might be worth to explore further.
}

\paragraph{Outline of the Paper}
In Section \ref{sec:PCF} we introduce the language \pPCF, a parameterized version of probabilistic PCF inspired from \cite{Pagani2018, Manzo2013} that will serve as our base language, and we illustrate how its programs describe potentially infinite discrete probabilistic models. 
Section \ref{sec:inference} contains an informal overview on the problem of finding most likely explanations for \pPCF\ programs and of our main ideas to overcome them.
In Section \ref{sec:QRel} we introduce a parameterized version of the weighted relational semantics, yieding a model of \pPCF\ in terms of formal power series. Section \ref{sec:tropdeg} contains our first result, that is, that first-order type programs of \pPCF\ have a finite tropical degree. Section \ref{sec:geo} contains our results on the minimal Newton polytope and our variant of the Viterbi algorithm to compute the tropical product of polynomials. In Section \ref{sec:inters_types} we introduce the intersection type system $\TIT$ {and the method, relying on it, to enumerate the most likely runs}. Finally, in Section \ref{sec:conc} we discuss related work and future directions.
\DEL{In Section \ref{sec:dp} we discuss potential applications of our results to differential privacy.}

\section{Parametric PCF: Specifying Models via Higher-Order Programs}\label{sec:PCF}
% !TEX root = Tropical_II.tex

%\TODO{The general goal must be crystal clear: in PPLs the execution of the program computes the model; at the same time PPLs are more general than usual finite discrete models like BNs; our goal is to do statistical inference on the execution of PPLs.}

In this section 
{we introduce \pPCF, a variant of probabilistic PCF \cite{Pagani2018, Manzo2013}, that will serve as our base language in the rest of the paper. We then illustrate in which sense the programs of \pPCF\ describe a class of discrete (infinitary) probabilistic models that we will explore in the next sections.
%, and we illustrate the general goal of our work through a few examples. 
}

\subsection{The language \pPCF}

The language \pPCF\ differs from standard probabilistic PCF \cite{Pagani2018, Manzo2013} in that real probabilities are replaced by a finite number of \emph{parameters} $X_1,\dots, X_n$. 
For instance, a probabilistic term $M\choice{p} N$, corresponding to a choice yielding $M$ with probability $p$ and $N$ with probability $1-p$, is replaced in \pPCF \ by a parametric term  $M\choice{X} N$, intuitively corresponding to a choice between $M$ and $N$ depending on some \emph{unknown} parameter $X$. {A similar language with probabilistic parameters was used already in \cite{Pagani2018} to establish the full abstraction of the semantics of \emph{probabilistic coherent spaces}.}

{The reason for considering, in this work, a language where explicit probabilities are replaced by parameters is twofold}. 
Firstly, in probabilistic models like BN or HMM it is standard to take the underlying basic probabilities as parameters of the model, for instance when considering problems
like that of computing the \emph{maximum likelihood} of an event. In the next section we will make more precise the role of parameters in the inference tasks we consider in this work.

%
% {as we illustrate in the next section,} we are interested in doing statistical inference on programs. Notably, we want to consider questions like: given a certain probabilistic event (e.g.~the program $M$ reduced to $\True$), what is the reduction of $M$ that has the most chances to have occurred (the \emph{most likely explanation} of the event?) And how do explanation vary depending on the parameters?
%
%
%we want to consider questions like: what are the probability assignments $\vec X\mapsto \vec q$ that \emph{maximize} a certain probabilistic event (e.g.~"the program reduces to $\True$"? Or, how does the \emph{most likely} behavior of $M$ change according to the choice of assignments $\vec X\mapsto \vec q$?
The second reason is that we will explore, \emph{in parallel}, an interpretation of \pPCF\ that associates parameters with actual probabilities $ q\in [0,1]$ as well as a second interpretation that associates the same parameters with \emph{negative log-probabilities} $z=-\ln q\in \RS$. Indeed, the methods based on tropical geometry that we develop exploit the latter, more combinatorial, viewpoint as a means to \emph{gain knowledge} about the former.

\begin{definition}
Let $X_1,\dots, X_n$ be $n$ distinct formal variables. The terms of \pPCF\ are defined by the following grammar {(and quotiented by usual $\alpha$-equivalence)}:
\begin{align*}
M::=\  \ZERO\mid \SUCC M\mid \PRED M\mid \ITE{M}{M}{M} \mid x\mid \lambda x.M\mid MM \mid \YY M\mid M\choice{X_i}M
%\tag{Booleans and integers}\\
%& \mid x\mid \lambda x.M\mid MM \mid \YY M \tag{$\lambda$-calculus}\\
%& \mid M\choice{X_i}M %\ (i\in \{1,\dots n\})
%%\ { \mid \ITE{M}{M}{M}} 
%\tag{parametric choice}
 \end{align*}
We let $\mathsf n:=\SUCC^{\!\!n}(\ZERO)$.
The types of \pPCF\ are defined by 
$
A::= \Bool \mid \NAT \mid  A\to A
$. 
The typing rules are presented in Fig.~\ref{fig:typerules} {(where contexts are finitely many variable declarations.
Also, observe that we overload $\ZERO$ and $\ONE$ as being both Booleans and integers)}.

%Consider $n$ more formal variables $\OV{X_1},\dots, \OV{X_n}$. 
%We indicate a finite multiset $\mu$ over the $2n$ formal variables $X_1,\OV{X_1},\dots, X_n, \OV{X_n}$ as a monomial $\Pi_{i=1}^n X_i^{\mu(2i)}\overline{X_i}^{\mu(i+1)}$.
For any set $\Sigma$, let $!\Sigma$ indicate the set of finite multisets over $\Sigma$.
We indicate a multiset $\mu\in\ !\Sigma$ as a formal monomial $\prod_{a\in \Sigma} a^{\mu(a)}$.
{The operational semantics is given by a}
reduction relation $M\RED{\mu}~N$, where $\mu\in !\{X_1,\OV{X_1},\dots, X_n, \OV{X_n}\}$, 
defined by {the relexive and transitive closure of} the rules in Fig.~\ref{fig:reductions}, which include standard PCF weak head {CbN} reductions, as well as parametric reductions for the choice operator.
{For the reflexive closure, we set $M\RED{[]}M$; for the transitive closure, we set $M\RED{\mu\cdot \nu}P$ whenever $M\RED{\mu}N$ and $N\RED{\nu}P$.}
\end{definition}

\begin{figure}[t]
\begin{subfigure}{0.95\textwidth}
\fbox{
\begin{minipage}{0.95\textwidth}
{\centering\footnotesize

\AXC{\phantom{$i$}}
\UIC{$\Gamma\vdash \ZZERO:\NAT$}
\DP
\qquad
\AXC{$\Gamma\vdash M:\NAT$}
\UIC{$\Gamma\vdash\SUCC M, \PRED M:\NAT$}
\DP
\qquad
\AXC{\phantom{$i$}}
\UIC{$\Gamma\vdash \ZERO, \ONE:\Bool$}
\DP
\qquad
{\AXC{$\Gamma\vdash M:\Bool$}
\UIC{$\Gamma\vdash M:\NAT$}
\DP}

\vspace{3mm}

\AXC{\phantom{$i$}}
\UIC{$\Gamma, x:A\vdash x:A$}
\DP
\qquad
\AXC{$\Gamma, x:A\vdash M:B$}
\UIC{$\Gamma\vdash \lambda x.M: A\to B$}
\DP
\qquad
\AXC{$\Gamma\vdash M:A\to B$}
\AXC{$\Gamma\vdash N:A$}
\BIC{$\Gamma\vdash MN:B$}
\DP

\vspace{3.mm}

\AXC{$\Gamma\vdash M:\NAT$}
\AXC{$\Gamma\vdash N,P:A$}
\BIC{$\Gamma\vdash \ITE{M}{N}{P}:A$}
\DP
\qquad
\AXC{$\Gamma\vdash M:A$}
\AXC{$\Gamma\vdash N:A$}
\BIC{$\Gamma\vdash M\choice{X} N:A$}
\DP
\qquad
\AXC{$\Gamma\vdash M:A\to A$}
\UIC{$\Gamma\vdash \YY M:A$}
\DP

%\vskip0.1mm
%$$
%\AXC{$\Gamma\vdash M:\Bool$}
%\AXC{$\Gamma\vdash N:A$}
%\AXC{$\Gamma\vdash P:A$}
%\TIC{$\Gamma\vdash \ITE{M}{N}{P}:A$}
%\DP
%$$
%\vskip0.1mm
%$$
%\AXC{$\Gamma\vdash M:A\to A$}
%\UIC{$\Gamma\vdash \YY M:A$}
%\DP
%$$
}
\end{minipage}
}
\caption{Typing rules.}
\label{fig:typerules}
\end{subfigure}

\medskip

\begin{subfigure}{0.95\textwidth}
\fbox{
\begin{minipage}{0.95\textwidth}
{\footnotesize
%\begin{minipage}{0.7\textwidth}
\[\begin{array}{cccc}
\ITE{\ZERO}{M}{N} \onestep{{[]}} M &
(\lambda x.M)N \onestep{{[]}}M[N/x] & \PRED{\ZZERO} \onestep{{[]}}{\ZZERO} & M\choice{X_i}N \onestep{X_i} M \\[0.3ex]
\ITE{\mathsf{\SUCC\mathsf n}}{M}{N} \onestep{{[]}} N & \YY M \onestep{{[]}} M(\YY M) & \PRED{(\SUCC M)} \onestep{{[]}}{M} & M\choice{X_i}N \onestep{\OV{X_i}} N \\[0.9ex]
{\ITE{M}{P}{Q}} {\onestep{\mu}\ITE{N}{P}{Q}} & MP \onestep{\mu}NP & {\PRED M} {\onestep{\mu}\PRED N} & {\SUCC M} {\onestep{\mu}\SUCC N}
\end{array}\]
%\end{minipage}

%\begin{minipage}{0.26\textwidth}
%%\vskip-3mm
%\hfill
%{\AXC{$M\RED{\mu}N$}
%\AXC{$N\RED{\nu}P$}
%\BIC{$M\RED{\mu\cdot \nu}P$}
%\DP}
%\end{minipage}

}
\end{minipage}
}
\caption{Parametric reduction rules. {In the last line, we suppose $M\onestep{\mu}N$.}}
\label{fig:reductions}
\end{subfigure}
\caption{\small Rules of \pPCF.}
\label{fig:reductionss}
\end{figure}

\begin{example}\label{example1}

Consider the term 
$
M_1=(\ONE\choice{X}\ZERO)\choice{X}((\ONE \choice{X} \ZERO)\choice{X}(\ZERO\choice{X}\ONE)).
$
There are three reductions $M\RED{\mu}\ZERO$, that give %$\mu_1=[X,\OV X], \mu_2=\mu_3=[X,\OV X,\OV X]%=\mu_3=\{\OV X,\OV X, X\}
%$ 
$\mu_1=X\OV X, \mu_2=\mu_3=X\OV X^2$
and three reductions $M\RED{\nu}\ONE$, with $\nu_1=X^2, \nu_2=X^2\OV X, \nu_3=\OV X^3$.

\end{example}

\begin{remark}[relation with probabilistic PCF]
By reading the parameters $X_1,\dots, X_n$ as reals $q_1,\dots, q_n\in [0,1]$ the typing and reduction rules of \pPCF\ are just rules for a standard PCF with biased choice operators $M\choice{q_i}N$ (where instead of adding to a multiset, we take the product in $[0,1]$). %. Notice that a multiset $\mu$ becomes the number $\prod_{i=1}^n q_i^{\mu(2i)}(1-q_i)^{\mu(2i)+1}\in [0,1]$.
In this way, standard properties like e.g.~subject reduction are easily deduced from those of pPCF.
\end{remark}

\begin{remark}\label{rem:monomial}
We could have chosen to label reductions with finite \emph{words} over $X_i,\overline{X_i}$ instead of multisets, so that each label $\mu$ in $M\RED{\mu}N$ univocally determines one reduction of $M$. We chose multisets because this is more natural in view of the formal manipulations discussed in the next sections. We will quickly go back at the possibility of using words instead at the end of Section~\ref{sec:inters_types}.
\end{remark}

\subsection{\pPCF\  and Graphical Probabilistic Models}

%\TODO{Add here discussion.
%What is the relation between PCF and graphical models? Why do we care? Why statistical inference on PCF programs is relevant? What can we do with it?
%}

%
%
%In this subsection we illustrate, through a few examples, how the programs of \pPCF\ can be seen as specifying some statistical models, and we introduce the statistical inference tasks that we are interested in in this work.  

Consider again the term $M_1$ from Example \ref{example1}.
What kind of probabilistic model does $M_1$ specify? A {possible} illustration is provided in Fig.~\ref{fig:dag1}: a simple Hidden Markov Model model with: one \emph{observable} variable $o\in \{\ZERO,\ONE\}$, corresponding to the result of the execution of $M_1$; three \emph{hidden} {independent random} variables $h_1,h_2,h_3\in \{\ZERO,\ONE\}$ (read: left, right), {each} corresponding to the {choice made at the $i$-th step of} a reduction of $M_1$; 
%\davide{
%The value of the observed value $o$ depends on the values of the hidden ones $h_1,h_2,h_3$, that is, on the choice of some reduction path. At the same time, all these dependencies can be expressed via polynomials in the only two \emph{parameters} of the model, that is, $X,\OV X$. For instance, the probability that $o=\ONE$ is expressed as the sum of the probabilities $p_{\theta}$ of those trajectories $\theta=\theta_1,\theta_2,\theta_3\in \{\ZERO,\ONE\}^3$ that lead to $\ONE$. As we can take $p_{\theta_i}(X,\overline X)=\nu_i$ (recall that $M\RED{\nu_i}\ONE$), we have then
%\begin{equation}\label{eq:mp1}
%\PROB(o=\ONE)=p_{\theta_1}(X,\OV X)+p_{\theta_2}(X,\OV X)+ p_{\theta_3}(X,\OV X)= X^2+X^2\OV X+\OV X^3.
%\end{equation}
%%corresponding to the sum of the monomials $\nu_1,\nu_2,\nu_3$ labeling the three reductions $M\RED{\nu}\ONE$. 
%}{
{a conditional probability $\PROB(o=\ONE\mid h_1,h_2,h_3)$ (and similarly for $o=\ZERO$), modeling the probability of $M_1$ reducing to $\ONE$ following the reduction specified by the $h$'s.
Notice that we choose three variables $h$'s, while sometimes $M_1$ only performs two choices, for example in $M\RED{X^2}\ONE$.
To solve this mismatch, we can declare that the previous reduction of $M_1$ corresponds to both $(h_1,h_2,h_3)=(0,0,0)$ and $(h_1,h_2,h_3)=(0,0,1)$ in the HMM.
In this way, we have, parametrically in $X,\OV X$, that $\PROB(o=\ONE)= (X^3+X^2\OV{X})+X^2\OV X+\OV X^3$.
For each probability assignment $X:=p,\OV X:=1-p$ of the parameters, the above probability now becomes the correct
\begin{equation}\label{eq:mp1}
\PROB(o=\ONE)= p^2+p^2(1-p)+(1-p)^3.
\end{equation}
%While 
%Anyway, these considerations only regard the encoding of (the behaviour of) a term as a HMM, while the content of the paper regards terms.
%The purpose of this section is simply to justify the fact that terms can be thought of as complex HMM, enhancing the motivating to focus on the former.

}

\begin{figure}
\begin{subfigure}{0.25\textwidth}
%\fbox{
%\begin{minipage}{.99\textwidth}
\adjustbox{scale=0.7,center}{
$
\begin{tikzpicture}

\node[draw, circle, fill=gray!20, minimum width=0.8cm](s0) at (-0.2,0) {$h_1$};
\node[draw, circle, fill=gray!20, minimum width=0.8cm](s1) at (1,0) {$h_2$};
\node[draw, circle, fill=gray!20, minimum width=0.8cm](s2) at (2.2,0) {$h_3$};
\node[draw, circle, fill=gray!20, minimum width=0.8cm](o2) at (1,-1.5) {$o$};

\draw[->] (s0) to (o2);
\draw[->] (s1) to (o2);
\draw[->] (s2) to (o2);

\end{tikzpicture}
$
}
%\end{minipage}
%}
\caption{Bayesian Network for Example \ref{example1}.}
\label{fig:dag1}
\end{subfigure}
\hskip6mm 
\begin{subfigure}{0.28\textwidth}
%\fbox{
%\begin{minipage}{.99\textwidth}
\adjustbox{scale=0.7,center}{
\begin{tikzpicture}

\node[rectangle, draw, minimum width=1cm, minimum height=3cm, rounded corners, gray, thick](rect1) at (-0.5,-0.75) {};
\node[rectangle, draw, minimum width=1cm, minimum height=3cm, rounded corners, gray, thick](rect1) at (1,-0.75) {};

\node(w) at (-0.5,1) {\footnotesize time $0$};
\node(w) at (1,1) {\footnotesize time $t+1$};
\node[draw, circle, fill=gray!20, minimum width=0.8cm](s0) at (-0.5,0) {$D$};
\node[draw, circle, fill=gray!20, minimum width=0.8cm](s1) at (1,0) {\footnotesize $ND$};
\node[draw, circle, fill=gray!20, minimum width=0.8cm](o) at (1,-1.5) {$O$};

\draw[->] (s0) to (s1);
\draw[->] (s1) to (o);
\end{tikzpicture}
}
%\end{minipage}
%}
\caption{Dynamical Bayesian Network for Example \ref{example2}}
\label{fig:dag2}
\end{subfigure}
\hskip6mm
\begin{subfigure}{0.29\textwidth}
%\fbox{
%\begin{minipage}{.99\textwidth}
\adjustbox{scale=0.7,center}{
\begin{tikzpicture}

\node[draw, circle, fill=gray!20, minimum width=0.8cm](s0) at (-0.2,0) {$D$};
\node[draw, circle, fill=gray!20, minimum width=0.8cm](s1) at (1,0) {$D_1$};
\node[draw, circle, fill=gray!20, minimum width=0.8cm](o1) at (1,-1.5) {$O_1$};
\node[draw, circle, fill=gray!20, minimum width=0.8cm](s2) at (2.2,0) {$D_2$};
\node[draw, circle, fill=gray!20, minimum width=0.8cm](o2) at (2.2,-1.5) {$O_2$};
\node[draw, circle, fill=gray!20, minimum width=0.8cm](s3) at (3.4,0) {$D_3$};
\node[draw, circle, fill=gray!20, minimum width=0.8cm](o3) at (3.4,-1.5) {$O_3$};
\node(o5) at (4.2,-0.75) {$\dots$};

\node(o5) at (4.2,-2.2) {\ };

\draw[->] (s0) to (s1);
\draw[->] (s1) to (o1);
\draw[->] (s1) to (s2);
\draw[->] (s2) to (o2);
\draw[->] (s2) to (s3);
\draw[->] (s3) to (o3);

\end{tikzpicture}
}
%\end{minipage}
%}

\caption{Unrolled Bayesian Network \\ \ }
\label{fig:dag3}
\end{subfigure}
%
%
%\bigskip
%
%
%\begin{subfigure}{0.75\textwidth}
%\adjustbox{scale=0.8,center}{
%$
%\begin{tikzpicture}
%
%
%\node[draw, circle, fill=gray!20, minimum width=0.8cm](s0) at (-0.2,0) {$h_1$};
%\node[draw, circle, fill=gray!20, minimum width=0.8cm](s1) at (1,0) {$h_2$};
%\node[draw, circle, fill=gray!20, minimum width=0.8cm](s2) at (2.2,0) {$h_3$};
%\node[](s22) at (3.4,0) {$\dots$};
%\node[draw, circle, fill=gray!20, minimum width=0.8cm](sn) at (4.6,0) {$h_n$};
%\node[](snn) at (5.8,0) {$\dots$};
%
%
%
%\node[draw, circle, fill=gray!20, minimum width=0.8cm](o2) at (2.2,-1.5) {$o$};
%
%
%\draw[->] (s0) to (o2);
%\draw[->] (s1) to (o2);
%\draw[->] (s2) to (o2);
%\draw[->] (sn) to (o2);
%
%
%\end{tikzpicture}
%$
%}
%\caption{Flat model with infinitely hidden many variables.}
%\label{fig:dag4}
%\end{subfigure}

\caption{\small Examples of Bayesian Networks.}
\end{figure}

{In fact, it} is well-known that graphical models like HMM or BN can be encoded as terms of some PCF-like language \cite{Gordon2014, Vanoni2024}, {and the overall goal of this section is to suggest that PCF programs can be thought as specifying complex HMM of some kind. However, due to their higher-order nature, 
}
%. Conversely, due to its higher-order nature, 
as well as the possibility of defining functions recursively via the fixpoint $\YY$, a general program of \pPCF\ need \emph{not} describe a finite probabilistic models like those illustrated so far.
%
%
%
%Still, as the example below shows, in some cases it is still possible to describe these models via some form of \emph{iterated} Bayesian Network.

\begin{example}\label{example2}
{Let $M_2= \YY
%\davide{
%\Big(
%\lambda fx.\ITE{Ox}{\ONE}{f(Nx)}
%\Big)}
{B}(ND) : \Bool$, where}
\[
%\davide{M_2= \YY\davide{
%\Big(
%\lambda fx.\ITE{Ox}{\ONE}{f(Nx)}
%\Big)}{B}(ND) : \Bool}
{
B:=\lambda fx.\ITE{x}{\ITE{O\ZERO}{f(N\ZERO)}{\ONE}}{\ITE{O\ONE}{f(N\ONE)}{\ONE}}:(\Bool\to\Bool)\to\Bool\to\Bool
},
\]
$D=\ZERO\choice{X_0}\ONE:\Bool$  %The type $\Bool^n$ stands for a $n$-fold product of 
 represents an initial Distribution of Booleans, $N=\lambda x.\ITE{x}{\ZERO\choice{X_1}\ONE}{\ZERO\choice{X_2}\ONE}:\Bool\to \Bool$ a probabilistic protocol to turn a distribution into a New one, and $O=\lambda x.\ITE{x}{\ZERO\choice{X_3}\ONE}{\ZERO\choice{X_4}\ONE}:\Bool\to \Bool$ another probabilistic protocol to Observe a Boolean value. The behaviour of $M_2$ 
% {(see also end of Section~\ref{subsec:explanations_tropdeg})} 
 can be recast in {pseudocode as follows: %can be unfolded as follows:
\[d=\mathsf{sample}(D)\,;\,\mathsf{while}(\mathsf{true})\,\set{%\mathsf{do}
\,d=\mathsf{sample}(Nd)\,;\,o=\mathsf{sample}(Od)\,;\,\mathsf{if}(o==\ONE)\,\set{\mathsf{return}\,\ONE}\,}%\mathsf{od}
.\]}
%first apply $N$ to $D$, then apply $N$ until the observed value is $\ONE$. Observe that the computation of $M_2$ may be arbitrarily long, and might even diverge.
{We see that $M_2$ describes thus a} \emph{dynamic} Bayesian Network (cf.~\cite{Daphne2009}, ch.~6) as the one illustrated in Figg.~\ref{fig:dag2}
and \ref{fig:dag3}: a potentially infinite DAG constructed following an iterative pattern. 
Notice that the number of hidden and observed variables is potentially infinite: each iteration produces a new hidden variable $D_i$ (corresponding to the value produced by applying $N$ $i$ times to $D$) and a new observation $O_i$. 
By contrast, the number of parameters of the model is finite, as it consists of the parameters $X_0-X_4$ in the terms $D,N,O$. 

\end{example}

In the example above, in order to reconstruct the model described by the program, an essential ingredient is to be able to identify a certain \emph{pattern} that is repeated over and over, so as to provide a compact graphical representation of how the hidden and observed variables relate to each other. Obviously, extracting such patterns may be very hard, or even undecidable, to do for an arbitrary \pPCF\ program. Another solution would be to consider a somehow \emph{flat} model with one observed variable $o$, corresponding to the final result of the execution (if any), conditioned on \emph{infinitely many} hidden variables $h_1,h_2,h_3,\dots$, corresponding to the unbounded number of choices made during a terminating reduction (which may be arbitrarily long).
%, 
% as illustrated in Fig.~\ref{fig:dag4}. 
 
 However, in a such a model the dependency of $o$ on any variable $h_i$ may be rather difficult to track explicitly, since \emph{infinitely many} and \emph{arbitrarily long} different reductions may lead to the same result. As we'll see, this dependency is in general not expressed by a polynomial as in standard HMM or BN, but by a \emph{power series} in the parameters. 
 At the same time, the flat model provides no insight on how such complex dependencies might be decomposed following the structure of the program. In other words, it is not \emph{compositional}.
 
Reconstructing the probabilistic model underlying a program of \pPCF\ is indeed tantamount to reconstructing the semantics of the program. Still, as we'll see, linear logic and programming language theory provide us with precisely the good methods to, first, design a probabilistic semantics capturing the relevant power series and, second, design a syntactic method (i.e.~a type system) to fully approximate the semantics of the program. This is the approach we describe through Sections \ref{sec:QRel}-\ref{sec:inters_types}.  
But before delving into that, let us look more closely at the inference tasks that we consider.
%  
%By contrast, in the next section we introduce a semantics of power series for \pPCF\ programs, corresponding to a parameterized version of the weighted relational semantics of linear logic, which captures these dependencies in a compositional and flexible way (since, as we'll see, the parameters of a program may be interpreted either as probabilities $q\in[0,1]$ or as \emph{negative log-probabilities} $-\ln q\in [0,+\infty]$). Moreover, in Section \ref{sec:inters_types} we introduce an intersection type system for \pPCF\ that captures the interpretation of a program in this model by enumerating the corresponding probabilistic trajectories in a compositional and efficient way.
%

%
%The following is proved similarly to standard probabilistic PCF.
%\begin{proposition}[subject reduction]
%If $\Gamma\vdash M:A$ and $M\RED{\mu}N$, then $\Gamma\vdash N:A$.
%\end{proposition}
\section{Can We Do Statistical Inference over PCF Programs?}\label{sec:inference}

The overall goal of this work is to demonstrate the possibility of inferring the most likely trajectories in the models specified by higher-order probabilistic programs. 
 In this section we provide an informal overview on the difficulties lying ahead of this goal, and a first intuitive illustration of our two main ideas to overcome them: the notion of \emph{tropical degree} (cf.~Section \ref{sec:tropdeg}), and an adaptation of the \emph{Viterbi algorithm} from HMM to higher-order programs (cf.~Sections \ref{sec:geo} and \ref{sec:inters_types}).

\subsection{Most Likely Explanations and the Tropical Degree}\label{subsec:explanations_tropdeg}

 The typical inference task for a Bayesian Network is to compute the \emph{marginal probabilities}
associated with an assignment $\vec \sigma$ to the {observed} variables. This corresponds, intuitively, to summing up the probabilities of all trajectories producing the outcomes $\vec\sigma$, that is, of all possible assignments $\vec \theta$ to the hidden variables, as in \eqref{eq:mp1}. 
In a \emph{finite} BN the marginal probabilities can be expressed as polynomials in the parameters and can be computed via algorithms like e.g.~the \emph{sum-product} algorithm \cite{Wymeersch_2007}.

In this work we are not interested in the problem of computing marginal probabilities. Indeed, when considering probabilistic models, like HMM, with a marked distinction between hidden and observed variables, a natural question is to predict the \emph{most likely explanation} for a given outcome: supposing that we observed that our program returned $\mathrm{True}$, what are the trajectories (i.e.~the values of the hidden variables) that have the most chances of having produced this result? 

%A second natural inference task is the following: supposing  that we focus on some given trajectory $\gamma$, what are the values of the parameters that make $\gamma$ the most likely explanation for the outcome?

More precisely, we are interested in the following two inference problems:
\begin{enumerate}
\item[(I1)] given both the observation $\vec\sigma$ and the values $\vec q$ assigned to the parameters, compute the \emph{maximum a posteriori probabilities} 
\begin{equation}\label{eq:pp1}
\max\Big\{ \PROB(\vec o=\vec \sigma,\vec h=\vec \theta, \vec X=\vec q)\ \Big \vert \ {\vec \theta\text{ assignment to the hidden variables}}\Big\}
\end{equation}
or, equivalently, the \emph{minimum} a posteriori \emph{negative log-}probabilities:
\begin{equation}\label{eq:pp2}
\min\Big\{-\ln \PROB(\vec o=\vec \sigma,\vec h=\vec \theta, \vec X=-\ln\vec q)\ \Big \vert \ {\vec \theta\text{ assignment to the hidden variables}}\Big\}
\end{equation}
\noindent and produce one such assignment $\vec \theta$ to the hidden variables that provides a \emph{most likely} explanation for the observation $\vec \sigma$;
\item[(I2)] given both the observation $\vec \sigma$ and the hidden data $\vec\theta$, identify the values of the parameters $\vec X$ that make the assignment $\vec\theta$ the most likely explanation of $\vec \sigma$, i.e.~compute the set
\begin{equation}
\Big \{\vec q\in [0,1]^n\ \Big \vert \ \vec \theta=\mathrm{argmax}_{\vec \rho}\{\PROB(\vec o=\vec\sigma,\vec h=\vec \rho,\vec X= \vec q)\}\Big \}
\end{equation}

\end{enumerate} 

Coming back again to our running example $M_1$, let us show how to compute solutions to both problems.
For (I1), considering the observation $\sigma=\ONE$, and fixing values e.g.~$X=\OV X=\frac{1}{2}$ for the parameters, we are led from \eqref{eq:mp1} and \eqref{eq:pp2} to compute
\begin{equation}\label{eq:prob2}
\min\{ -\ln X^2, -\ln X^2\OV X, -\ln \OV X^3\}=
\min\{ 2z,2z+\OV z, 3\OV z\}= 2z = 2\ln 2,
\end{equation}
where $z=-\ln X, \OV z=-\ln \OV X$, showing that the leftmost reduction is the most-likely to produce the outcome $\ONE$ under the parameter assignment $X,\OV X\mapsto\frac{1}{2}$. 
For (I2), for instance, we may wish to know for which values of $X,\OV X$ the rightmost trajectory  becomes the most likely to produce the result $\ONE$. Using \eqref{eq:prob2}, we are led to find $z,\OV z$ such that  
$
\min\{ 2z,2z+\OV z, 3\OV z\}= 3\OV z,
$
yielding the condition $\OV z\leq \frac{2}{3}z$, that is, $X\leq \OV X^{\frac{2}{3}}$ (e.g.~$X=\frac{1}{4},\OV X=\frac{3}{4}$).

At this point the connection with tropical geometry strikes the eye: the expression $\min\{ 2z,2z+\OV z, 3\OV z\}$, obtained by 
replacing, in \eqref{eq:mp1}, the outer sum by a min, is an example of a \emph{tropical polynomial}, i.e.~a polynomial with $\min$ in place of $+$ and $+$ in place of $\times$. In fact, solving problems like (I1) essentially amounts to computing marginal probabilities in a tropical setting (i.e.~in which sums are replaced by $\min$s and multiplications by sums).

For \emph{finite} HMMs, well-known algorithms like the \emph{Viterbi algorithm} can be used to compute, in an efficient way, solutions to problems (I1) and (I2). This algorithm can indeed be seen as a ``tropical'' variant of the sum-product algorithm \cite{Maragos2018}.
Still, as we discussed above, we are here considering models with infinitely many hidden variables, and thus, with infinitely many trajectories.
% In cases like the program $M_2$ above, the marginal probabilities are no more computed as polynomials, since the number of possible trajectories to consider may be infinite, but rather as {power series}. Correspondingly, the minimum log-probabilities are not expressed by tropical polynomials, but by \emph{tropical power series}, that is, $\inf$s of infinitely many log probabilities: once the program $M_2$ has produced $\ONE$, the most likely explanation is to be searched for within an infinite space. 
%
How could one solve such an infinitary optimization problem? Here is where our fundamental idea comes into play: while a \pPCF\ program may well produce infinitely many, arbitrary long, different trajectories, one might well guess that, since the probability assigned with a trajectory is obtained by multiplying the same finite number of parameters at each iteration, such probabilities should start to \emph{decrease} after a sufficiently long number of reduction steps.

For example, consider the experiment of repeatedly tossing a coin with bias $X$ until a head is produced. This is represented in \pPCF\ by the program below
$$
M_3= Y ( \lambda x.x\choice{X}\ONE ) :\Bool
$$
The probability of getting the first head at iteration $n+1$ is thus $X\overline X^{n}$ and the total probability is expressed by the \emph{power series} $\PROB(M\RED{}\ONE)=\sum_{n=1}^\infty X\overline X^n$, that sums over infinitely many trajectories. At the same time, it is clear that, across all these trajectories, the most likely explanation for a head is that we obtained it at the \emph{first} iteration, since $q>q(1-q)^n$ for all possible choice $q$ for $X$. Indeed, all this can be restated as the observation that, for $z=-\ln X, \OV z=-\ln \OV X\in[0,+\infty]$, the $\inf$ of the sequence below is reached by its first element:
$$
\inf_n\big\{ -\ln (X\OV{X}^n)\big\}=
\inf_n\big\{ z+n\OV{z}\} = z.
$$
Consider now the term $M_2$ from Example \ref{example2}. {We will} see (cf.~Example \ref{example78}) that, in a reduction $M_2\RED{\mu}\ONE$ with $n$ {calls to $\YY$}, the monomial $\mu$ has degree {$2n+3$}, corresponding to {$2n+3$} independent probabilistic choices, {and} the probability of getting $\ONE$ starts to decrease after the {second} iteration. 
This {implies} that a reduction  $M_2\RED{\mu}\ONE$ of \emph{maximum} probability can always be found among those with {$\deg\mu\leq 5$}. In more algebraic terms, in the power series describing the probability $\PROB(M_2\RED{}\ONE)$, all monomials of degree greater than $5$ are in fact \emph{dominated} by some monomial of degree $\leq 5$. Moving to the negative $\ln$s, the $\inf$ over \emph{all} trajectories must reduce to a \emph{finite} $\min$, in fact a tropical polynomial, containing only the monomials of degree $\leq 5$:
\[
\inf_n\big\{\text{monomials of degree }n\big\}= \min \big\{ \text{monomials of degree }n\leq 5\big\}.
\]

This value $5$ is what we call the \emph{tropical degree} of $M_2$, noted $\F d(M_2)$: it is the smallest number $n$ for which we can find a \emph{finite} set $S$ of trajectories such that all trajectories in $S$ correspond to reductions with at most $n$ choices, and  any trajectory of the program is dominated by some trajectory in $S$. 
Our first result is that \emph{any} program $M:\Bool$ of \pPCF\ has a finite tropical degree $\F d(M)$ (Corollary \ref{thm:collapse}), that is, that the most likely explanations for the results produced by $M$ can always be found within a \emph{finite} trajectory space. 

\subsection{Most Likely Explanations \emph{Efficiently}, via the Newton Polytope}\label{sec:unfeasable}

Even once the space of trajectories for an arbitrary program $M:\Bool$ has been reduced to a finite one, this space may still be \emph{too large} to explore in practice.  
{
For instance, consider  the following higher-order %probabilistic 
 pPCF program
$$
M_4=(\lambda x.x \choice{p_1}x)(\lambda x.x \choice{p_2}x)\dots
(\lambda x.x \choice{p_n}x)\ONE,
$$
where $p_1,\dots, p_n\in [0,1]$ are fixed probabilities.
Observe that $M_4$ always terminates on the normal form $1$, and there are exactly $2^n$ reductions $M_4\RED{}{1}$, each of probability $q_{\theta_1,1}\cdots q_{\theta_n,n}$, where for $\theta\in\set{0,1}^n$ we set $q_{0,i}:=p_i$ and $q_{1,i}:=1-p_i$.

Suppose now we want to find the {probability of a} most likely reduction path of $M_4$ (to $\ONE$).
}
Writing $z_{0,i}$ for $-\ln p_i$ and $z_{1,i}$ for $-\ln (1-p_i)$, the maximum probability above %corresponds to computing 
is the minimum of the corresponding negative log-probabilities:
\[
\min_{\theta\in \{0,1\}^n } \big\{z_{\theta_1,1}+\dots + z_{\theta_n,n}\big\}.
\]
{Remark also that computing the $\arg\min$, instead of the $\min$ above, gives the most likely trajectories. In either cases,}
a na\"ive approach to this computation would inspect \emph{all} possible trajectories.
However, this leads to computing and comparing $2^n$ different sums of positive reals, which is hardly feasible in practice. 
By contrast, a more efficient strategy for computing the same minimum is to compare (negative-log-)probabilities piece after piece, that is, to compute:
\[
\min\big\{z_{0,1}, z_{1,1}\big\} +\dots+ \min\big\{z_{0,n}, z_{1,n}\big\}.
\]
In this case we are computing $n$ $\min$s and summing $n$ reals.  Moreover, if we keep track, each time we compute a $\min$, of the value $\theta_i\in\{0,1\}$ producing the minimum, at the end of the computation we even obtain the most likely trajectories $\theta\in \{0,1\}^n$.

This simple example illustrates the idea behind the already mentioned {Viterbi algorithm}. Both this algorithm and the {sum-product} algorithm for Bayesian networks can be seen as instances of a general "distributive law" algorithm \cite{Aji2000}. 
Very roughly, the algorithm exploits the remark that in occurrences of the distributive law of (semi)rings like e.g.~$(x+y)\cdot (z+w)=xz+xw+yz+yw$ there are, often, \emph{less} operations to perform to evaluate the left-hand term, compared to the right-hand.
So, whenever one is asked to evaluate a possibly too large sum of monomials, it is wise to try to use distributivity \emph{from right to left} as much as possible, so as to express this sum as a product of simpler polynomials. In the case above, we reduced the problem of computing a $\min$ across $2^n$ {(tropical) monomials $\mu_{\theta}:= z_{\theta_1,1}+\dots + z_{\theta_n,n}$} to that of computing the sum (indeed, the tropical product) of $n$ polynomials $m_i:=\min\{z_{0,i},z_{1,i}\}$. 

{Let us write now the term $M_4$ in \pPCF, by replacing probabilities with formal parameters,}
%Suppose now to replace in term $P$ the positive reals $p_i$ with parameters, :
$
M_4=(\lambda x.x \choice{X_1}x)(\lambda x.x \choice{X_2}x)\dots
(\lambda x.x \choice{X_n}x)\ONE.
$
{Then the distributive law argument as above suggests moving from considering \emph{all} $2^n$ trajectories and look instead at the tropical product:}
%and consider the problem of describing the most likely reductions of this program. 
%Again, we cannot simply compute \emph{all} $2^n$ trajectories. 
%At the same time, the distributive law algorithm suggests to look at the tropical product:
\begin{equation}\label{eq:mins}
\min\{ X_1, \OV{X_1}\}+\dots+ \min\{ X_n, \OV{X_n}\}
\end{equation}
But this time, since the $X_i,\OV{X_i}$ are not reals but just formal variables, it is not clear how to obtain a tropical polynomial from it other than by applying distributivity in \emph{wrong sense}, that is, from left to right, thus getting back to an exponentially large $\min$. 

As we discuss through Sections \ref{sec:geo} and \ref{sec:inters_types}, this is the point where tropical geometry comes to rescue us. In Section \ref{sec:geo} we illustrate how {we manipulate} the \emph{Newton polytope}, a $n$-dimension polyhedron that is an invariant of tropical polynomials, %can be used 
{in order} to extract a \emph{not too large} polynomial from a sum like \eqref{eq:mins} and, more generally, to compute the tropical product of polynomials in an efficient way. 
%Notably, in the case of \eqref{eq:mins} the Newton polytope yields the 
%\begin{equation}\label{eq:mins}
%\min\{ X_1, \OV{X_1}\}+\dots+ \min\{ X_n, \OV{X_n}\}=\min\{ nX_1,\dots, nX_n\}
%\end{equation}
Then, in Section \ref{sec:inters_types}, we will exploit this method to design an intersection type system $\TIT$ that enumerates the most likely trajectories of \pPCF\ programs: intuitively, type derivations \emph{select} the most likely trajectories by applying our tropical version of the Viterbi algorithm recursively on the structure of the program. 
{This will enable us to explore, in polynomial time, a space of trajectories of size exponential in the size of the program, providing a significant speed-up to the search for most-likely reductions.}
%
%In this way the number of the selected trajectories never grows too large, remaining polynomial in the size of the program. 
 
%
%
%computes the tropical polynomial associated with a \pPCF\ program in a compositional and efficient way. 

\section{Parametric Weighted Relational Semantics}\label{sec:QRel}
% !TEX root = Tropical_II.tex
%In this section we design a semantics for \pPCF\ by putting the weighted relational semantics from \cite{Manzo2013} in a parametric setting. This provides an interpretation of \pPCF-programs by \emph{formal power series}, to be evaluated on an arbitrary continuous semiring.
{In this section we introduce a semantics for \pPCF-programs given in terms of \emph{formal power series} whose variables include $\vec X$. This semantics is a parameterized version of the  weighted relational semantics from \cite{Manzo2013}.
%As mentioned in the introduction, this comes from well-established work in semantics of programming languages, especially those coming from linear logic.
%The purpose of this section is to introduce the main mathematical objects on which the paper relies. 
%This section can be read as follows: 
%In Subsection \ref{subsec:fps} we recall the notion of formal power series (and polynomial) in multiple commuting variables, %in a standard algebraic way. These constitute 
%the basic algebraic objects that we manipulate in the paper.
%In Subsection~\ref{subsec:pcf_fps} we recall the usual interpretation of pPCF in the relational semantics, expressing it in terms of formal power series (instead of the equivalent matrix formulation).
While the presentation of this semantics requires us to combine algebraic and categorical language, the key points to look at for the following are equations \eqref{eq:Rrel_pcf}, \eqref{eq:Trel_pcf} and \eqref{eq:Xrel_pcfX}, which show how the formal power series in the semantics are related to the probabilities of the execution paths of \pPCF-programs, as well as the fundamental observation, in Subsection \ref{subsec:QAn}, that \emph{distinct} formal power series may induce \emph{the same} (tropical) analytic function.
%In Subsection \ref{subsec:pcfX_pfs}, we introduce the ``parametric interpretation'' of \pPCF, whose main interest is \eqref{eq:Xrel_pcfX}, showing in which sense this semantics tracks the formal (weights of the) execution paths of a term.
%Finally, in Subsection \ref{subsec:QAn} we associate an analytic function with each formal power series coming from a \pPCF-program.}

%\subsection{Continuous semirings}

%Def, their homomorphisms, fundamental examples: $\mathbb N^\infty$, $\mathbb R_{\geq 0}^\infty$, $\mathbb L$.

%Let us recall some basic the definitions.

\subsection{Formal Power Series%, evaluation, examples with $Q=\mathbb N^\infty, \mathbb R_{\geq 0}^\infty, \mathbb L$, change of base
}\label{subsec:fps}

%In the following, by semiring we silently mean an ordered commutative semiring with units $0$ and $1$.
In the following, by semiring we mean commutative and with units $0$ and $1$.
A semiring is \emph{continuous} if it is ordered (compatible with $+$ and $\cdot$) and (among other properties) it admits infinite sums. %it admits all sups of directed subsets. 
We will consider the following \emph{continuous} semirings (cf.~\cite{Manzo2013}): $\BS$ with Boolean addition and multiplication,
$\NINF$ with standard addition and multiplication, $\RS$ with standard addition and multiplication, and $\TS$, the \emph{tropical semiring} (also noted $\BB L$, cf.~\cite{BarbaPist2024}), corresponding to $\RS$ with reversed order, with $\min$ as $+$ and addition as $\cdot$.

%\paragraph{Definitions and notations}
%For any set $\Sigma$, let $!\Sigma$ indicate the set of finite multisets over $\Sigma$.
{We indicate multisets $\mu\in\ !\Sigma=\Sigma\to\N$ as formal monomials, %in their domain,
which allows us to retain standard notations for polynomials/power series.
For instance, the multiset $\mu\in \ !\set{0,1,2}$ with $\mu(0)=2,\mu(1)=1,\mu(2)=0$ will be denoted as $0^21$ (or $0^212^0$% if we want to stress that the multiplicity of $2$ is $0$
).
Often, for clarity, we introduce a set $x_\Sigma$ of $\sharp\Sigma$ fresh formal variables $x_a$, one for each $a\in \Sigma$, and we denote $\mu\in\ !\Sigma$ by the formal monomial $\prod_{a\in \Sigma} x_a^{\mu(a)}$, also denoted $x^\mu$.
For instance, $0^21$ becomes the standard $x_0^2x_1$ (or $x_0^2x_1x_2^0$).}

%For convenience, we indicate a multiset $\mu\in \ !\Sigma$ as a formal monomial $\prod_{a\in \Sigma} x_a^{\mu(a)}$, denoted $x^\mu$, over a set $x_\Sigma$ of $\sharp\Sigma$ formal variables $x_a$, one for each $a\in \Sigma$. For instance, we note the multiset $0^21\in \ !\set{0,1}$ as $x_0^2x_1$.

% over $\sharp\Sigma$ formal variables $X_a$ (one for each element of $\Sigma$).
% (where we mean that it is $1$ if $\Sigma=\emptyset$ and it is the empty monomial if $\mu=[]$)

Let $\Sigma$ be a set and $Q$ a semiring.  We call $\fps{Q}{\Sigma}$ the set of %$\fmset \Sigma$-indexed sequences in $Q$, i.e.\ 
functions $\fmset\Sigma \to Q$, and its elements are called \emph{formal power series (\emph{fps}, for short) %\footnote{Usually this set is called $Q[[\Sigma]]$, but we slightly change notation because in computer science logic we already vastly use the similar $\model{\,}$. Using the notation $\fps{Q}{\Sigma}$ is also coherent with the habit, in logic, of writing e.g.\ $A\{\Sigma\}$ for a (finite) formal expression $A$ with formal variables in $\Sigma$. Coherently, we will write $Q\{\Sigma\}$ for polynomials, instead of the common $Q[\Sigma]$.%, and fps are the same but potentially infinite.} 
over $Q$ with (commuting) variables the elements of $\Sigma$}.
Given $s\in \fps{Q}{\Sigma}$, the image $s_{\mu}\in Q$ of $\mu\in\fmset\Sigma$ is called the \emph{coefficient of $s$ at $\mu$} and $\mathrm{supp}(s):=\fmset\Sigma-s^{-1}0$ is called the \emph{support $\mathrm{supp}(s)$ of $s$}. A fps $s$ is \emph{all-one} when all coefficients $s_\mu$ are either $0$ or $1$. %{(i.e.\ when $s$ is the characteristic fps of its support)}.  
When $\Sigma$ is finite and the support is finite, $s$ is a \emph{formal polynomial}. 
We let $\fpp{Q}{\Sigma}\subseteq \fps{Q}{\Sigma}$ indicate the set of formal polynomials.
As usual, we visualize a fps $s\in \fps{Q}{\Sigma}$ as the formal sum $s=\sum_{\mu\in\fmset\Sigma} s_\mu x^\mu$,  e.g.~$s={s_{0^01^0}}+s_{0^21}x_0^2x_1+s_{10^2}x_0x_1^2\in \fps{Q}{\set{0,1}}$. 
%
%It is useful to visualize formal power series as formal sums $s=\sum_{\mu\in\fmset\Sigma} s_\mu \mu$.
% Let us explain how to interpret this notation: a multiset $\mu\in\fmset\Sigma$ can be identified with a formal monomial $\prod_{a\in \Sigma} a^{\mu(a)}$ over $\Sigma$ (and we mean that it is $1$ if $\Sigma=\emptyset$ and it is the empty monomial if $\mu=[]$).
%In fact, it is even more convenient (in particular when, as in the remainder of the paper, $\Sigma$ will be some concrete set like $\{0,1\}$, or $\N$), to introduce $\#\Sigma$ many new formal variables $x_a$, for each $a\in\Sigma$, and use $x_a$ instead of $a$ (thus considering $s$ to be actually in $\fps{Q}{x_\Sigma}$, where $x_\Sigma:=\{x_a \mid a\in \Sigma\}$).
%For example, taking $\Sigma=\{a,b\}$, the multiset $\mu=\{a,a,b\}$ indicates the monomial $a^2b=aba=ba^2$.
%In this way we can visualize a formal power series as the more familiar formal sum $\sum_{\mu\in\fmset\Sigma} s_\mu \prod_{a\in \mu}x_a^{\mu(a)}$. 
%An example is the formal polynomial $s=s_{[]}[]+s_{0^21}0^21+s_{10^2}01^2\in \fps{Q}{\set{0,1}}$.
% can be written in the more familiar form  $s_{[]}+s_{0^21}x_0^2x_1+s_{01^2}x_0x_1^2$.
%
If $\Sigma=\Sigma_1+\cdots+\Sigma_n$, then $\fps{Q}{\Sigma}$ is canonically isomorphic to the set of functions $\fmset\Sigma_1\times\cdots\times\fmset\Sigma_n\to Q$, which we call $\fps{Q}{\Sigma_1,\cdots,\Sigma_n}$, whose elements can be visualized as formal power series with multiple sets $x_{\Sigma_1}, \dots, x_{\Sigma_n}$ of variables.
The notations introduced above are implicitly compatible with the fact that $\fps{Q}{\Sigma}$ is a commutative monoid with pointwise addition, with $0$ being the %polynomial
fps $\sum_\mu 0x^\mu$.
In fact, $\fps{Q}{\Sigma}$ is a semiring with multiplication given by the usual Cauchy's formula: $(ss')_\mu := \sum_{\rho+\eta=\mu} s_\rho s'_\eta$ (this is a sum in $Q$ and exists because it is finite, since $\mu$ is), i.e.\ $ss'=\sum_{\rho,\mu} s_\rho s'_\eta \, x^{\rho+\eta}$.
The $1$ for this multiplication is the polynomial $1$ with our notation, i.e.\ $1x^{[]}$.
Polynomials form a sub-semiring for this structure. 
If $Q$ is continuous, $\fps{Q}{\Sigma}$ is also continuous with respect to the pointwise partial order (the bottom element is 0 and supremas are pointwise).
The \emph{evaluation map at $q\in Q^\Sigma$} is the continuous semiring homomorphism $\fps{Q}{\Sigma}\to Q$ sending $\sum_\mu s_\mu x^\mu$ to $\sum_\mu s_\mu q^\mu$, where $q^\mu:=\prod_{a\in \Sigma} q_a^{\mu(a)} \in Q$.

%
%
%With respect to the pointwise order, $\fps{Q}{\Sigma}$ is continuous
%%$\fps{Q}{\Sigma}$ is also a commutative semiring with same addition and pointwise multiplication (called the Hadamard product, with $1$ being the fps $\sum_\mu x^\mu$).
%%Polynomials do not form a sub-semiring for this structure, but polynomials of fixed bound on the degree do.
%%We will not use the Hadamrd semiring structure though, so for us product means the Cauchy one.
%
%%
%%Let us suppose now and for all the paper that $Q$ is complete, i.e.\ it is equipped with a notion of infinite sum (extending the finite one).
%Then the evaluation map at $q\in Q^\Sigma$ is the map $\fps{Q}{\Sigma}\to Q$ sending $\sum_\mu s_\mu x^\mu$ to $\sum_\mu s_\mu q^\mu$, where now the sum and the product $q^\mu:=\prod_a q_a^{\mu(a)}$ exist in $Q$ respectively because $Q$ is complete and $\mu$ is finite.
%With respect to the Cauchy semiring structure, this is a complete semiring homomorphism. %(but it is not with respect to the Hadamard one).
Any continuous semiring homomorphism $Q\to Q'$ lifts to a continuous semiring homomorphism $\fps{Q}{\Sigma}\to \fps{Q'}{\Sigma}$ by acting on the coefficients.
Remark that sum, products, evaluation map and lifts of homomorphisms above, are all compatible with the bijection $\fps{Q}{\Sigma}\simeq\fps{Q}{\Sigma_1,\cdots,\Sigma_n}$ and so they are compatible with the multiple variables notation; for example, the evaluation map at $(q_1,\dots,q_n)\in Q^{\Sigma_1}\times\cdots\times Q^{\Sigma_n}$ would now go from $\fps{Q}{\Sigma_1,\cdots,\Sigma_n}$ to $Q$. %, defined in the obvious way.
Also, remark that for $Q=\fps{Q'}{Z}$, a fps $s\in\fps{Q}{X}=\fps{(\fps{Q'}{Z})}{X}$ is the same data as a fps $s\in\fps{Q'}{Z,X}$. %, namely $s=\sum_{\mu,\rho} s_{\mu,\rho} \mu \rho$.

We have the following folklore result (proven in the appendix), %, that we will use in the next section:
where for any continuous semiring $Q$, $q\in Q$ and $n\in \NINF$, we write $nq:=\sum_{i=1}^n q$ in $Q$.

\begin{proposition}\label{prop:free_ccs}
$\fps{\N^\infty}{\Sigma}$ is
    the free continuous commutative semiring on a finite set $\Sigma$.
    For any continuous {commutative} semiring $Q$ and $q\in Q^\Sigma$, the unique map realizing the universal property is %$\texttt{ev}_q: s\in \fps{\NINF}{\Sigma} \mapsto \sum_{\mu} s_\mu q^\mu  \in Q$. 
    $\texttt{ev}_q: \fps{\NINF}{\Sigma} \to Q$, defined by $\texttt{ev}_q(s):=\sum_{\mu} s_\mu q^\mu$.
\end{proposition}

For a given continuous semiring $Q$, the category $\rel{Q}$ \cite{Manzo2013} has sets as objects and matrices $Q^{X\times Y}$ as arrows $X\to Y$. The category 
$\rel{Q}_!$ is the coKleisli category of $\rel{Q}$ wrt the multiset comonad $!$, so its arrows $X\to Y$ are matrices in $Q^{!X\times Y}$. 
%Composition of $t\in \rel{Q}(X,Y), s\in \rel{Q}{Y,Z}$ given by the "matrix product"
%$(st)_{x,z}=\sum_{y\in Y}s_{y,z}t_{x,y}$
$\rel{Q}_!$ is cartesian closed, with product $X+Y$, terminal object $\B 1=\{\star\}$ and exponential $!X\times Y$.
Observe that sets in $\rel{Q}_!$ play the role of %sets of 
indices, and
%Actually,
%But matrices in $Q^{X\times Y}$ are the same data as $Y$-indexed \emph{formal linear combinations with (commuting) variables in the set $X$}, i.e.\ fps's $s$ such that $\mathrm{supp}(s)\subseteq \{\mu \mid \mu(a)= 1 \textit{ for all }a\in X\}$.
%And 
a matrix $t\in Q^{\fmset X\times Y}$ is the same data as a $Y$-indexed family of formal power series with commuting variables in $X$, namely $t=(\sum_{\mu\in \fmset X} t_{\mu,y}x^\mu )_{y\in Y} \in \fps{Q}{X}^Y$.
{This identification is compatible with units and compositions (and linearity):} %, which we still call $t$.
from now on, for us 
%$Q\mathrm{Rel}$ has sets as objects and arrows $A\to Y$ the $Y$-indexed family of formal linear combinations in $X$, and
%
$\rel{Q}_!$ \emph{has sets as object and $\fps{Q}{X}^Y$ as homsets $X\to Y$}.
%Notice that here sets play the role of sets of formal variables.
%They are related by the same coKleisli construction simply by expressing the usual multiset comonad $!$ on $Q\mathrm{Rel}$, seen as a category of formal linear combinations. %We do not precise it here because it is just about writing the same things differently and it is trivial.

Lastly, notice that for any continuous semiring homomorphism $\theta:Q\to Q'$, the induced one %homomorphism
$\fps{Q}{\Sigma}\to~\fps{Q'}{\Sigma}$ yields a (cartesian closed) identity-on-objects functor $F_\theta:~\rel{Q}_!\to~\rel{Q'}_!$. %: in fact, this functor is immediately induced by the homomorphisms $\fps{Q}{\Sigma}\to \fps{Q'}{\Sigma}$ via the identification $\rel{Q}_!(X,Y)\simeq \fps{Q}{X}^Y$ (in other words, $F_\theta$ simply acts on the coefficients of the matrices of $\rel{Q}_!$).
%\TODO{CCC structure or not?}

\subsection{Interpreting %\pPCF\ 
(probabilistic) PCF-programs as formal power series}\label{subsec:pcf_fps}

{Before introducing the interpretation of \pPCF-{typing derivations}, let us recall the interpretation $\model{-}^Q$ of standard PCF in the category $\rel Q_!$, for any continuous semiring $Q$ \cite{Manzo2013}. 
%
%
%\subsection{The categories $Q\mathrm{Rel}$ and $Q\mathrm{Rel}_!$% as a category of power series
%}
%
%
%For example, let us fix some $s=\sum_{\mu\in \ !\{X\}}s_\mu\mu\in \fps{\NINF}{X}$. Notice that we can identify $!\{X\}$ with $\BB N$. Any assignment $q:\{X\}\to \RS$ induces then a \emph{power series} $F_q(s)=\sum_{n}s_n q^n:\RS\to \RS$ (or a polynomial if $s$ is a formal polynomial).
%
% 
%
%
%
%
%\subsection{Interpreting \pPCF \ in $Q\mathrm{Rel}_!$}
%
%
Actually, \cite{Manzo2013} introduces a language PCF$^Q$ with \emph{weighted terms} $q\cdot M$, for $q$ is an element of $Q$, and a generic choice operator $M\ \mathbf{ or } \ N$, and it is shown that, for any $Q$, PCF$^Q$ can always be interpreted inside $\rel{Q}_!$.
The basic types $\Bool, \NAT$ are interpreted by the sets $\{0,1\}$ and $\BB N$, respectively, and 
arrow types $A\to B$ are interpreted as $!\model{A}\times\model{B}$. A {derivation of} $x_1:A_1,\dots, x_n:A_n\vdash M:B$ is interpreted as %a matrix in $Q^{\fmset{(\model{A_1}+\cdots+\model{A_n})}\times \model B}$, that is,
%%%
%, where $+$ is the disjoint union of sets.
%
%In terms of matrices, we would obtain $\model{\vec{x}:\vec{A}:\vec{A}_n\vdash M:B}\in Q^{\fmset{(\model{A_1}+\cdots+\model{A_n})}\times \model B}$. %, where $+$ is the disjoint union of sets.
%%
%an element of
a fps in $\fps{Q}{\model{A_1},\dots,\model{A_n}}^{{\model B}}$, i.e.\ a ${\model B}$-family of fps with variables in $\model{A_1},\dots, \model{A_n}$.
For instance, a {closed} program $M:\Bool$ is interpreted as %an element of
a fps in $%Q^{!\emptyset\times\set{0,1}}\simeq 
\fps{Q}{\emptyset}^{\set{0,1}}\simeq Q^{\set{0,1}}$, in other words, by two elements\footnote{{As common, we simply write $\model{\Gamma\vdash M:A}$ or even just $\model M$, but we really mean $\model \pi$ for $\pi$ a given typing derivation for $M$.}} $\model{M}_0,\model{M}_1\in Q$.
Weighted and choice terms are interpreted via $\model{q\cdot M}=q\cdot \model{M}$ and $\model{M \ \mathbf{ or } \ N}=\model{M}+\model{N}$.

% where, as mentioned before, we have used the identification $\fps{Q}{\model{A_1}^Q+\dots+\model{A_n}^Q} \simeq \fps{Q}{\model{A_1}^Q,\dots,\model{A_n}^Q}$.
%Importantly, the interpretation of the fixpoint $\YY$ exploits the fact that $Q$ is continuous.
% and the multiple formal variables notation in order to keep track of the declared program variables.

One obtains in this way an interpretation of usual probabilistic PCF \cite{Pagani2018} (\emph{pPCF} for short)
in $\rel{\RS}_!$, translating it into PCF$^{\RS}$ via $M\choice{p}N:= p\cdot M\ \mathbf{or}\ (1-p)\cdot N$. In fact, this interpretation precisely captures the probabilistic execution of closed terms \cite{Manzo2013}: the interpretation $\model{M}^{\RS}\in(\RS)^{\set{0,1}}$ of a program $M:\Bool$ consists in two real numbers $
\model{M}^{\RS}_0$, $
\model{M}^{\RS}_1$, describing the probability that $M$ reduces to $\mathsf i=0,1$ respectively:
\begin{equation}\label{eq:Rrel_pcf}
\model{M}^{\RS}_{\mathsf i}=\PROB( M \RED{}{\mathsf i} )=%\sum\Big\{ p\ \Big \vert \ M\RED{p} \mathsf i\Big \},
\sum_{M\RED{p} \mathsf i} p \quad\textit{(where $M\RED{p}\mathsf i$ indicates a reduction 
%of $M$ to $\mathsf i$ 
of probability $p$).}
\end{equation}

\vspace{-0.2cm}
One also obtains an interpretation of pPCF in $\rel{\TS}_!$ by taking \emph{negative log-probabilities} 
$-\ln p\in \TS$ in place of $p$, that is, $M\choice{p}N:= (-\ln p)\cdot M\ \mathbf{or}\ (-\ln(1-p))\cdot N$.
%
%: the argument works as for $\RS$ except that we use \emph{negative log-probabilities} and the $\min$ replaces the sum, yielding 
%$$\model{M\choice{p}N}^{\TS}:=\min\{ -\ln p\model{M}^{\RS},-\ln(1-p)\model{N}^{\RS}\}.$$
Since $\mathbf{or}$ is now interpreted by the $\min$ operation, this interpretation describes the negative log-probability of a \emph{most likely} reduction: %that is
\begin{equation}\label{eq:Trel_pcf}
\model{M}^{\TS}_{\mathsf i}=\inf
\big\{-\ln p \mid M\RED{p}\mathsf i  \big\} = - \ln \sup
\big\{ p \mid M\RED{p}\mathsf i \big\}.%\quad (i=0,1).
\end{equation}

%
%of terms: the interpretation $\model{M}^{\RS}\in(\RS)^{\set{0,1}}$ of a program $M:\Bool$ consists in two real numbers $
%\model{\tilde M_1}_0$, $
%\model{\tilde M_1}_1$, where 
%$
%\model{\tilde M_1}_i$
%coincides with the probability $\PROB( M \to^*\mathsf i )$
%that $M$ reduces to $\mathsf i$. 

\begin{example}\label{ex:31}
For the closed pPCF term
$
M=\ONE \choice{p}(\ONE\choice{p}\ONE)
$,
%
%$\tilde M_1=(\ONE\choice{p}\ZERO)\choice{p}((\ONE \choice{p} \ZERO)\choice{p}(\ZERO\choice{p}\ONE))$, which corresponds to the $M_1$ of the previous section once we chose an actual probability $X:=p\in[0,1]$ and, thus, the value $1-p$ for $\OV X$.
we have $\model{M}_1^{\RS}=p+p(1-p)+(1-p)^2=1$, i.e.~the sum of the probabilities of all trajectories leading to $\ONE$, and 
$\model{M}_1^{\TS}= \min\{ z,z+w,2w\}=\min\{z,2w\}$, where $z=-\ln p, w=-\ln(1-p)$, yielding e.g.~$-\ln 2$ 
when $p=1-p=\frac{1}{2}$. %, i.e.~the value $-\ln p$ where $p$ is the probability of the most likely reduction. 

\end{example}

\subsection{Interpreting \pPCF-programs as formal power series %in a category of power series
}\label{subsec:pcfX_pfs}

We now show how to %obtain an interpretation of 
interpret \pPCF\ inside \emph{any} category $\rel{Q}_!$.
In fact, we interpret it in a ``free way'', factorizing any {such }interpretation. %in $\rel{Q}_!$.
Let $\mathbb X$ be the set %of $2n$ elements, so that $x_\Sigma$ coincides with the set of parameters 
$\{X_1,\overline{X_1},\dots, X_n,\overline{X_n}\}$. 
We can {translate} \pPCF\ {to} PCF$^{\fps{\NINF}{\mathbb{X}}}$ via $M\choice{X_i}N:= X_i\cdot M\ \mathbf{or} \ \overline{X_i}\cdot N$, and %we obtain 
{$\model{\_}^{\fps{\NINF}{\mathbb{X}}}$ gives}
then an interpretation of \pPCF\ inside $\rel{(\fps{\NINF}{\mathbb{X}})}_!$.
We call it 
the \emph{parametric interpretation% of $M$
} and note it $\model{\_}^{X_1,\dots,X_n}$. That is, %as
$\model{\Gamma\vdash M:A}^{X_1,\dots, X_n} \in
(\fps{(\fps{\NINF}{X_1,\overline{X_1},\dots, X_n,\overline{X_n}})}{\model{\Gamma}})^{\model A}$, which is the same as $(\fps{\NINF}{X_1,\overline{X_1},\dots, X_n,\overline{X_n},\model{\Gamma}})^{\model A}$. %, i.e.\ (f
For e.g.\ $n=1$, it is a {$\model A$-family of} fps ${\sum_\mu(\sum_{i,j} s_{ij\mu} X^i \overline{X}^j) x^\mu=}\sum_{i,j,\mu} s_{ij\mu} X^i \overline{X}^j x^\mu$ ($i,j\in \N, \mu\in \ !\model{\Gamma}$).

\begin{example}\label{ex:32}
%Consider the term 
%$M_1:\Bool$ from the previous section (cf.~Example \ref{ex:31}.
The parametric interpretation of the term $M=\ONE \choice{X}(\ONE\choice{X}\ONE):\Bool$ (the parametrization of the one in Example \ref{ex:31})
consists in two fps
$\model{M}^{X,\overline X}_0,\model{M}^{X,\overline X}_1\in \fps{\NINF}{X,\overline{X}}$, namely 
$\model{M}^{X,\overline X}_0=0$ and 
$\model{M}^{X,\overline X}_1= X+X\overline X+\overline X^2${, which represent the (weights of the) possible reductions}.
%
%
%e.g.~$\model{M_1}^{{X,\overline X}}_1=X_2+X\overline{X}^2+\overline{X}^3$ describes the formal sum of the three reductions $M_1\RED{\mu}\ONE$. 
%The map $\sigma_p $ that sends $X$ onto $p$ recovers then the number 
%$\model{\tilde{M_1}}^{\RS}_1=p_2+p(1-p)^2+p^3$ as the real interpretation of $M_1$.
\end{example}

Directly from \cite[Theorem V.6]{Manzo2013}, we get that for a closed term $M$ of type e.g.\ $\Bool$, and $\mathsf i\in \set{\ZERO,\ONE}$, %the fps $\model{M}_{\mathsf i}^{X_1,\dots, X_n}%%\in \fps{\NINF}{X_1,\overline{X_1},\dots, X_n,\overline{X_n}}
%$ is
\begin{equation}\label{eq:Xrel_pcfX}
\model{M}_{\mathsf i}^{X_1,\dots, X_n}=
\sum\nolimits_{\vec i,\vec j\in \N^n}
\sharp(\vec i,\vec j) X_1^{i_1}\OV X_1^{j_1}\dots X_n^{i_n}\OV X_n^{j_n}
\end{equation}
%\sum\Big\{ \sharp_{\mu,M} %\mu
%\cdot \mu\ \Big \vert \ M\RED{\mu} i \Big \},
%$$
where $\sharp(\vec i,\vec j)
%\in \NINF
$ is the number of reductions to $\mathsf i$ of weight $X_1^{i_1}\OV X_1^{j_1}\dots X_n^{i_n}\OV X_n^{j_n}$. 

\begin{example}\label{ex:33}
    Remember $M_3=\YY(\lambda x. \ONE \choice{X} x):\Bool$ from the previous section. 
    Its parametric interpretation yields two fps $\model{M_1}^{{X,\overline X}}_0,\model{M_1}^{{X,\overline X}}_1$ where $\model{M_2}^{{X,\overline X}}_0=0$, as $M_2$ cannot reduce to $\ZERO$, and $\model{M_2}^{{X,\overline X}}_1= \sum_{n}X\overline{X}^n$ describes the weights $\mu\simeq n$ of the \emph{infinitely many} trajectories %by which
    $M_2\RED{\mu}\ONE$.

%    Let us consider the slightly modified version $\tilde M_2:=Y(\lambda f. \, x \choice{X} f)$, just for the sake of discussing the case of an open term.
%    We have $\model{x:\NAT \vdash Y(\lambda f.\, x \choice{X} f):\NAT}$
\end{example}

{
The parametric interpretation is indeed a parametrisation of the semantics in \cite{Manzo2013}:}
%Observe that, 
by Proposition~\ref{prop:free_ccs}, any %function $\sigma:\Sigma\to Q$, associating the parameters with elements of some $Q$
choice $q\in Q^{\mathbb{X}}$ of actual values of parameters in a $Q$, canonically induces an interpretation of \pPCF\ inside $\rel{Q}_!$ via the functor $F_{\texttt{ev}_q}:\rel{(\fps{\NINF}{\mathbb{X}})}_!\to\rel{Q}_!$. 
One easily checks that, if %$p\in (\RS)^{\mathbb{X}}$ associates $X_i,\overline{X_i}$ with probabilities $p_i,1-p_i$,
{${\tiny\begin{bmatrix}
        X:=p_X\\ \OV X:=p_{\OV X}
        \end{bmatrix}}\in(\RS)^{\mathbb{X}}$}
then 
the produced interpretation of a term $M$ of \pPCF\ coincides with the one of the %corresponding 
PCF$^{\RS}$-term {$M[X:=p_X,\OV X:=p_{\OV X}]$}. Similarly, if $\tau\in\TS^{\mathbb{X}}$ associates
$X_i,\overline{X_i}$ with negative log-probabilities $-\ln p_i,-\ln(1-p_i)$, the produced interpretation of \pPCF\ terms coincides with the one of the corresponding PCF$^{\TS}$-terms.

\begin{example}
For $M$ from Example \ref{ex:32}, choosing the values $p,1-p\in \RS$ for $X,\OV X$ turns the fps $\model{M}^{X,\overline X}_1=X+X\overline X+\overline X^2$ into the real number 
$\model{M}_1^{\RS}=p+p(1-p)+(1-p)^2$. 
Evaluating  $X,\OV X$ as $-\ln p,-\ln(1-p)\in \TS$ turns it into  
$\model{M}_1^{\TS}=\min{\{-\ln p,-2\ln(1-p)\}}$ (cf.~Example \ref{ex:31}).. 
 \end{example}

\begin{example}
Consider $M_3$ from Example \ref{ex:33}; choosing $X,\OV X$ as $p,1-p\in \RS$ turns the fps $\model{M_2}^{X,\overline X}_1=\sum_n X\overline{X}^n$ into 
$\model{M_2}_1^{\RS}=\sum_n p(1-p)^n=\frac{p}{1-p}$ {(if $p\neq0)$}.  
Evaluating  them as $-\ln p,-\ln(1-p)\in \TS$ turns it into  
$\model{M}_1^{\TS}=\inf_n \{-\ln p-n\ln(1-p)\}=-\ln p$.
 \end{example}

\subsection{%The Category $\an{Q}$ of 
Tropical Analytic Functions}\label{subsec:QAn}

%\TODO{QRel and QAn not isomorphic for $Q=[0,+\infty]$ but only for PCOH. Reference?}

By evaluating at points, formal power series define analytic functions via the map $(\_)^!:\fps{Q}{\Sigma} \to [Q^\Sigma \to Q]$, where $s^!(q)$ evaluates $s$ at $q$, i.e.~$s^!(q)=\sum_{\mu\in \ !X}s_{\mu}q^\mu$. The central example we consider is the case of the analytic functions for $Q=\TS$: %sending $s$ to the function which evaluates $s$ at the points of $Q^\Sigma$.

%We call $\an Q(\Sigma)$ its image, the set of analytic functions from $Q^\Sigma$ to $Q$.

%The map $(\_)^!:\fps{Q}{\Sigma}^Y \to [Q^\Sigma \to Q^Y]$, where now $s^!$ is defined by $s^!(q)_y=\sum_{\mu\in \ !X}s_{\mu,y}q^\mu$, \TODO{is still a continuous semiring homomorphism}.

\begin{definition}
Let $\Sigma$ have $n$ elements. 
%A fps $s=\sum_{\mu\in \ !\Sigma}s_\mu\mu\in \fps{\RS}{\Sigma}$ yields an analytic function $s^!:(\RS)^n\to \RS$, where $s^!(x_1,\dots, x_n)=\sum_{\mu}s_{\mu}x_1^{\mu(1)},\dots,x_n^{\mu(n)}$. 
The functions $%s^!:
\TS^n\to \TS$ of shape $s^!$, for a fps $s%=\sum_{\mu}s_\mu x^\mu
\in \fps{\TS}{ \Sigma}$, are called \emph{tropical analytic functions} (\emph{taf} for short, aka tropical power series) \cite{Porzio2021, BarbaPist2024}. % if it is induced by a fps $s \in \fps{\TS}{ \Sigma}$.
Concretely,
\[
s^!(x_1,\dots, x_n)=\inf_{\mu\in \ !\Sigma}\big\{ s_{\mu}+\mu\cdot x\big\}
\]
with $\mu\cdot x:=\sum_{i=1}^n\mu(i)x_i$.
%In particular, 
When $s$ is a formal polynomial, %has finite support, 
the $\inf$ %in the $s^!$ 
above is a $\min$ and $s^!$ is then called a \emph{tropical polynomial function}. These are precisely the piecewise linear functions at the heart of tropical geometry, as we discuss in Section~\ref{sec:geo}.

%$s( \vec x)=\min_\mu\{ \mu\cdot \vec x+s_\mu\}$, where $\mu\cdot x=\sum_{i}\mu(i)x_i$. 
%More generally, a fps $s=\sum_{\mu}s_\mu\mu\in \fps{\TS}{\Sigma}$ is a \emph{tropical power series} (tps) $s(\vec x)=\inf_\mu\{ \mu\cdot \vec x+s_\mu\}$ in the variables $x_1,\dots, x_n$, see \cite{Porzio2021, BarbaPist2024}. 

\end{definition}

%In~\cite[p.~20]{Ehrhard2005} it is proven that when $Q=\RS$ then $(\_)^!$ is injective.
The map  $(\_)^!$ from fps to the corresponding analytic function is \emph{not}, in general, {injective}. This means that different formal power series may well induce \emph{the same} analytic function. Notably, injectivity fails for $Q=\TS$, as the following example shows.
%
%, not even in the most simple cases, as the following counterexamples show:

\begin{example}
%\begin{varitemize}
%    \item
%    Let $p$ be a prime and $Q=\mathbb F_p$ the field with $p$ elements.
%    Let $\Sigma = Y = \set{*}$.
%    Let $t=x^p\in \fps{\mathbb F_p}{x}$ and $s=x\in\fps{\mathbb F_p}{x}$.
%    Then $t\neq s$ but Fermat's little Theorem says that $t^!=s^!$.
%    \item
%    Let $Q:=\RS$, $\Sigma = Y = \set{*}$.
%    For a fixed $p\in\law$, let $t:=p\sum_n x^n\in\fps{\RS}{x}$ and $s:=p\in\fps{\RS}{x}$.
%    Then $t\neq s$ but $t^!=s^!$.
%    In fact $t^!(q)=p+\inf_n nq=p=s^!(q)$ for all $q\in\RS$.
%\end{varitemize}
Let $Q:=\TS$, $\Sigma =  \set{*}$.
    For a fixed $p\in\TS$, let $t:=\sum_n px^n\in\fps{\TS}{x}$ and $s:=p\in\fps{\TS}{x}$.
    Then $t\neq s$ but $t^!=s^!$.
    In fact $t^!(q)=p+\inf_n nq=p=s^!(q)$ for all $q\in\TS$.
\end{example}

%In the following sections we shall discover that
As it will be seen since the next section, it is precisely this mismatch between tropical power series and the corresponding analytic functions that enables a combinatorial and efficient exploration of the most likely behaviour of probabilistic programs.

\begin{remark}
The considerations above could be rephrased by considering a category $\an Q$ whose objects are sets and the homset from $\Sigma$ to $Y$ is $\an Q(\Sigma,Y)$, the set of functions $s^!:Q^\Sigma\to Q^Y$ defined by $s^!(q)_y=\sum_{\mu\in \ !X}s_{\mu,y}q^\mu$, for some $Y$-indexed family $s\in \fps{Q}{\Sigma}^Y$ of fps.
Via the map $(\_)^!$, any program $\Gamma\vdash M:A$ yields then a \emph{function} $\model{M}^!:Q^{\model{\Gamma}}\to Q^{\model{A}}$, and one may ask what is the status of such interpretation.
%However, since $\rel{Q}_!$ is not equivalent to $\an Q$, one must be careful about the categorical structure of the latter and the kind of interpretation $\model{-}^!$ that we get in this way. Notably, the category $\an\TS$ is \emph{not} cartesian closed\footnote{We thank Guy McCusker for discussions on this matter.}.
In the appendix, it is shown that $(\_)^!$ turns the exponential of $\rel{\TS}_!$ into a \emph{weak} exponential in $\an\TS$ (cf.~\cite{Martini1992}), so that the interpretation $\model{-}^!$ produces a \emph{non-extensional} model of \pPCF, that is, one that satisfies the $\beta$-rule but not the $\eta$-rule.
\end{remark}

\section{The Tropical Degree}\label{sec:tropdeg}
% !TEX root = Tropical_II.tex
%This section contains the first main result of this work, stating that the space of the most likely trajectories for a \pPCF-program of ground type is finite.
 
Suppose $M$ is a probabilistic algorithm that iterates a given protocol until a certain condition is satisfied, and suppose that 
the computation of $M$ ends after $n$ iterations producing the value $V$.
As we observed in Section 3, we can expect that the probability of producing $V$ after \emph{no less} than $n$ steps %starts to \emph{decrease} 
does {not} increase when $n$ is large enough.
In this section we show that, in \pPCF, this intuition is correct and reflects a general phenomenon captured by the tropical semantics.

To state our general result, we first show how to associate, canonically, a \emph{discrete} power series (i.e.~with coefficients in $\NINF$) with a taf called the \emph{tropicalization} of the power series. 

The inclusion $\iota\in \fps{Q}{\Sigma}^{\Sigma}$ that sends any element $X\in\Sigma$ onto itself induces the homomorphism $\mathsf{ev}_{\iota}:\fps{\NINF}{\Sigma}\to \fps{Q}{\Sigma}$.
If $Q$ is an idempotent semiring, %(i.e.\ $q+q=q$ for all $q\in Q$), 
one can check that $\mathsf{ev}_{\iota}$ turns all $0$ coefficients into $0\in Q$ and all coefficients $n\neq 0$ onto $1\in Q$.
That is, it gives the characteristic series of the support:
$\mathsf{ev}_{\iota}(s)=\sum_{\mu\in\mathrm{supp}(s)} x^\mu$.
Composed with $(-)^!$, this yields a map 
$\mathsf{ev}_{\iota}^!:\fps{\NINF}{\Sigma}^Y\to \an{Q}(\Sigma, Y)$.
For $Q:=\TS$, which is idempotent since $+=\inf$, the above lines yield the following:

\begin{definition}[tropicalization]\label{def:trop}
Let $\Sigma$ a finite set.
The homomorphism $\mathsf{ev}_{\iota}:\fps{\NINF}{\Sigma}\to \fps{\TS}{\Sigma}$ is called $\mathsf t$.
Remark that, in terms of real numbers, $\mathsf t(s)=\sum_{\mu\in\mathrm{supp}(s)} 0x^\mu$ (the other coefficients, i.e.\ the $\mathsf t(s)_\mu$ for $\mu\notin\mathrm{supp}(s)$, being $+\infty$).
We call $\trop := \mathsf{ev}_{\iota}^!:\fps{\NINF}{\Sigma}^Y\to [\TS^\Sigma\to\TS^Y] $ the \emph{tropicalization map}.
For $s\in \fps{\NINF}{\Sigma}^Y$, the tropicalization $\trop s:\TS^\Sigma\to \TS^Y$ of $s$ is the taf concretely given by: 
\[\trop s(x)_y=\inf_{\mu\in\mathrm{supp}(s_y)} \mu\cdot x=\trop s_y(x).\]
\end{definition}

{
\begin{example}
%\TODO{troppo difficile, e mi sembra poco istruttivo per il resto. Toglierei questo esempio.}
Let $s\in\fps{\NINF}{X,\OV X}$.
Then $\trop s:\TS^2\to\TS$ is $\trop s(X:=x_1, \OV X:=x_2)=\inf_{\mu\in\mathrm{supp}(s)} \mu(X)\cdot x_1+\mu(\OV X)\cdot x_2$.
For instance,
$\trop s(X:=0, \OV X:=+\infty)=\inf_{\mu\in\mathrm{supp}(s)} \set{\mu(\OV X)\cdot\infty}$.
It is immediate to see that this value is $0$ if there is $\mu$ such that $s_\mu\neq 0$ and $\mu(\OV X)=0$, while it is $+\infty$ otherwise.
For instance, for $M=\ONE\choice{X}\ZERO$, we have
$\trop \model{M}_{\ONE}= X\in\fps{\NINF}{X,\OV X}$, and $\trop \model{M}_{\ONE}(X:=0, \OV X:=+\infty)=0$.
Similarly, $\trop \model{M}_{\ZERO}(X:=0, \OV X:=+\infty)=+\infty$.
The first corresponds to the presence of %(at least) one 
the reduction %-path
$M\RED{X%^n
}{\ONE}$, the second to the absence of reduction %path
$M\RED{\OV X%^n
}{\ZERO}$.
In fact, the choice $X:=0, \OV X:=+\infty$ corresponds to choosing %almost surely choosing 
the left side of a probabilistic choice with probability $1$ (so $0$ for the right side), and $\trop \model{M}_{\mathsf i}$ returns thus negative log-probabilities when computed on negative log-probabilities.
%$s:=\model{P\choice{X}Q:\Bool}_{\ONE}$ we have $\trop s(X:=0, \OV X:=+\infty)=0$ if there is at least one reduction-path $P\RED{X^n}{\mathsf i}$, and $=+\infty$ otherwise.
%The choice $X:=0, \OV X:=+\infty$ corresponds to consider the program which almost surely chooses the left-side of any $\choice{X}$.%, and then considering any of the most likely reduction paths to $\mathsf i$.
%For example, $\trop \model{\ONE\choice{X}\ZERO}_{\ONE}(X:=0, \OV X:=+\infty)=0$ and $\trop \model{\ONE\choice{X}\ZERO}_{\ZERO}(X:=0, \OV X:=+\infty)=+\infty$.
\end{example}
}

{
The situation of the previous example is not a coincidence. In fact,}
via tropicalisation, a program $M:\Bool$ is turned into two taf $\trop\model{M}_{\mathsf i}:\TS^{\mathbb X}\to \TS$ which compute the negative log-probability of most likely reductions of $M$, as the following proposition shows.
This is therefore a ``parametrization'' (allowing all choices of probabilities) of \cite[Corollary 10]{BarbaPist2024} or \cite[Theorem V.6]{Manzo2013}.

\begin{proposition}\label{prop:negproba}
Given $M:\Bool$, let $p\in[0,1]^{\mathbb X}$ any assignment of probabilities $p$ to the parameters.
Then (remark that $M[X:= p_{X}]$ is a PCF$^{[0,1]}$-term) \[(\trop\model{M}^{\mathbb X}_{\mathsf i})( X:=-\ln p_{{X}},{\OV X}:=-\ln p_{{\OV X}})=-\ln \, \sup\,\set{q\mid M[X:= p_{X}] \RED{q}{\mathsf i}}.\]
\end{proposition}
%\begin{proof}
%Remembering that $\BB X=\set{X_1,\OV X_1,\dots,X_n,\OV X_n}$, and identifying  $!\BB X$ with $\N^{2n}$, we have:
%    $\trop\model{M}_{\mathsf i}(-\ln p)=
%    \inf\limits_{\Vec{n}\in\mathrm{supp}(\model{M}_{\mathsf i})} \set{-\,\Vec{n}\cdot \ln p}=
%    - \sup\limits_{\Vec{n}\in\mathrm{supp}(\model{M}_{\mathsf i})} \sum_{i=1}^{2n} n_i\ln p_{X_i}=
%    - \sup\limits_{\Vec{n}\in\mathrm{supp}(\model{M}_{\mathsf i})} \ln \big(\prod_{i=1}^{2n} p_{X_i}^{n_i}\big)=
 %   - \ln \sup\limits_{\Vec{n}\in\mathrm{supp}(\model{M}_{\mathsf i})} \prod_{i=1}^{2n} p_{X_i}^{n_i}
%    $,
%    where in the last equality we used the fact that $\ln$ is non-decreasing.
%    Now simply remark that, by what said in Subsection \ref{subsec:PPCF_in_fps},  $\prod_{i=1}^{2n} p_{X_i}^{n_i}$ is precisely the weight (a probability, in this case) of a reduction of $M[X:= p_{X}]$ to $\mathsf i$, and $\mathrm{supp}(\model{M}_{\mathsf i})$ considers all possible such reductions.
%\end{proof}

The negative log-probabilities above are computed as an $\inf$ across %the (log)probabilities of 
\emph{all} trajectories leading to $\mathsf i$. 
Our goal is now to show that, \emph{independently of the parameters}, such an inf is always found within %the (neg-log)probabilities of 
a \emph{finite} set of trajectories (and is, therefore, a min).
The key result to get there is the following:

\begin{proposition}\label{prop:collapse}
Let $k\in\N$ and $\set{s_n\mid n\in\N^k}\subseteq \NINF$.
Then there exists a \emph{finite} set $S\subseteq \N^k$ such that $s_n<+\infty$ for all $n\in S$ and, for all $x\in \TS^n$,  
\[
\inf_{n\in \N^k}\{n x+s_n\}=\min_{n\in S}\{n x+s_n\}.
\]
\end{proposition}
\begin{proof}%[Sketch of the proof]
{Fix the well-founded order $m \prec n$ on $\N^k$ by saying that $m_i\leq n_i$ for all $1\leq i\leq k$ \emph{and} there is at least one $1\leq j\leq k$ such that $m_j<n_j$.
%the pointwise order $\preceq$ on $\N^k$ induced by $\leq$ on $\N$ (and remark that, then, $\prec$ is \emph{not} the pointwise order induced by $<$).
}
We claim that we can let
$S:=\set{n\in\N ^k \mid s_n<+\infty \textit{ and for all } m  \prec  n\textit{ one has }s_m>~s_n}$.
Indeed, if
$S$ is infinite {then one can easily see, using K\"onig's Lemma, that} there is a chain $m _0\prec  m_1\prec\cdots \subseteq S$. But by definition of $S$ this gives a chain $s_{  m_0}>s_{ m_1}>\cdots \subseteq \N$, which is absurd.
We have thus shown that $S$ is finite.
For the claimed equation, Wlog $S\neq\emptyset$ {(otherwise on the one hand $\min := +\infty$, and on the other hand one can easily see, by induction on $\prec$, that $\inf=+\infty$)}. 
Now fix $x\in\TS^n$.
We show by induction on $n$ wrt $\prec$, that $\forall n\in\N^k,\, \exists m\in S$ such that $s_m+ m   x  \leq s_n+ n   x$. Notice that this proves the desired equation.

Case $ n  =0$: Wlog $  n\notin S$. {By definition of $S$,} $s_n=+\infty$ and so any $m\in S\neq\emptyset$ works.

Case $ n {\succ} 0$: Wlog $  n\notin S$. {By definition of $S$, we} have two cases:
either $s_n=+\infty$, which is done as above.
Or there is $ n'\prec n  $ with $ s_{n'}\leq s_n$.
But then
$s_{n'}+ n  ' x   \leq s_n +  n' x   {\leq} s_n + ( n  - n  ') x  +  n  ' x   =  s_n+  n   x$.
If $n'\in S$, take $m:=n'$.
If $n'\notin S$, {take the $m  \in S$ with $s_m+ m   x  \leq s_{n'}+ n  ' x $ given by the IH on $n'$.}
\end{proof}

\begin{remark}[not all taf are polynomials]
An essential ingredient in the (proof of the) result above is that of considering fps with coefficients in a discrete set (like $\NINF$). 
In general, a fps with coefficients in $\TS$ needs not be equivalent to a polynomial: consider the fps $s=\sum_{n\in \N}\frac{1}{2^n}{x}^n\in\fps{\TS}{{x}}$; the corresponding tropical analytic function $s^!:\TS\to \TS$ is not a polynomial function, since $s^!(0)=\inf_n\{n\cdot 0+{1}/{2^n}\}=0$ is an $\inf$ that cannot be reduced to a $\min$.
\end{remark}

%As a corollary we have:

\begin{corollary}\label{thm:collapse}
For all terms $ M:\Bool^n\to \NAT$ (i.e.~$\Bool\to\dots\to \Bool\to \NAT$) of \pPCF and $i\in \BB N$ there {is} an all-one polynomial $s\in \fpp{\TS}{\mathbb X\cup\set{0,1}^n}$ {with} 
$\trop\model{M}_{\mathsf i}=s^!$ and $\mathrm{supp}(s)\subseteq \mathrm{supp}(\model{M}_{\mathsf i})$.
\end{corollary}
\begin{proof}
We have $\model{M}_{\mathsf i}\in\fps{\NINF}{\Sigma}$, for $\Sigma:=\set{X_1,\OV X_1,\dots,X_k,\OV X_k}\cup\set{0,1}^n$.
Then we can identify $!\Sigma\simeq\N^{2k+2^n}$.
Let $s_\mu:=0$ if $\mu\in\mathrm{supp}(\model{M}_{\mathsf i})$ and $:=+\infty$ otherwise. 
Then Proposition~\ref{prop:collapse} gives a finite $S\subseteq !\Sigma$ such that $\trop\model{M}_{\mathsf i}(x) = \inf_{\mu\in\mathrm{supp}(\model{M}_{\mathsf i})}\{\mu\cdot x\} = \inf_{\mu\in \N^k}\{\mu\cdot x+s_\mu\}=\min_{\mu\in S}\{\mu\cdot x\}=s^!(x)$, for the polynomial $s:=\sum_{\mu\in S} 1_{\TS}x^\mu_{\Sigma}\in\fpp{\TS}{\Sigma}$.
Moreover, remark that $\mathrm{supp}(s)=S$.
Then, for $\mu\in \mathrm{supp}(s)$, from Proposition~\ref{prop:collapse}, we have $s_\mu<+\infty$, so $\mu\in\mathrm{supp}(\model{M}_{\mathsf i})$ by definition of $s_\mu$.
\end{proof}

Intuitively, the polynomial $s$ takes into account only a finite number of the trajectories of $M$. Yet, the result above shows that the sup negative-log-probability across all trajectories is always found within the finite set $S$ selected by $s$ (and is, therefore, a max).
This leads to the following:

\begin{definition}
{Let} $M:\Bool^n\to \NAT$ in \pPCF. {For $i\in\N$}, the \emph{tropical degree $\F d_{i}(M)$ of $M$ at $i$} is the minimum degree of an all-one polynomial $s\in \fpp{\TS}{\mathbb X}$ {with} 
$\trop\model{M}_{\mathsf i}=s^!$ and $\mathrm{supp}(s)\subseteq ~\mathrm{supp}(\model{M}_{\mathsf i})$.
\end{definition}

The tropical degree expresses the fact that the sup of the probabilities of the execution paths of $M$ to $\mathsf i$ can be obtained by only looking at a finite number of execution paths whose degree is at most $\F d_{\mathsf i}(M)$.
For example,
let us expand the definition above for a closed term $M:\Bool$ with parameters $X,\overline X$ and let, say, $\mathsf i=\ZERO$.
Remember that $\model{M}_{\ZERO}^X=
\sum_{i,j\in \N}
\sharp(i,j) X^{i}\OV X^{j}$, with $\sharp(i,j)$ the number of reductions to $\ZERO$ of weight $X^{i}\OV X^{j}$.
Remembering Proposition \ref{prop:negproba} too, $\F d_{\ZERO}(M)$ is the \emph{smallest} $d\in\N$ such that there is a \emph{finite} $S\subseteq \N^2$~with 
%\[\begin{array}{c}
\begin{varenumerate}
%    \textit{1) }\ 
    \item 
    $\max_{(i,j)\in S} i+j = d$,
%%    
%    \qquad\qquad
%    \textit{2) }\ 
%%    
    \item for all $(i,j)\in S$ { there is a reduction } $M{\RED{}}\ZERO$ {of weight $X^i\OV X^j$}, 
%%    \\ \\
%    \textit{3) }\  
    \item  for all  $p\in[0,1] $,  the $\sup$ {of all probabilities }$P$ { across \emph{any} reduction} $M[X:=p]\RED{P}{\ZERO}${ equals the } $\max${ across the reductions %witnessed
    chosen in } $S$ {:}
$     \sup\{P\!\mid\! {M[X:=p]\RED{P}{\ZERO}} \} = \max_{(i,j)\in S} p^i(1\!-\!p)^j$. 
\end{varenumerate}
%\end{array}\]
%The tropical degree expresses then the fact that the sup of the probabilities of the execution paths of $M$ to $\mathsf i$ can be obtained by only looking at a finite number of execution paths whose degree is at most $\F d_{\mathsf i}(M)$.
%Observe again that $\F d_{\mathsf i}(M)$ is minimal wrt those properties.
%Concretely speaking, the tropical degree is obtained by \emph{guessing} the correct (and finite) subset $S$ of the support of the characteristic series of the interpretation of the term, which maximises the needed probabilities.
%In the example below, we will selected one trajectory %$M\RED{X}{\mathsf i}$
%among the potentially infinitely many in the support.
%

For example, one can easily see that the term $M_3$ from Section \ref{sec:inference} satisfies $\F d({M_3})=1$. 
%\davide{We {briefly} discuss a few less trivial examples below.}

\begin{example}
    Let $M:\Bool$ with parameters $X,\overline X$. Suppose that $M\RED{}{\mathsf i}$, for a fixed $i\in\set{0,1}$.
    Let us show that $\F d_{\mathsf i}(\mathsf i\choice{X}(M\choice{X}\Omega))=1$, where $\Omega := \YY I:\Bool$ is non-terminating.
    This means showing that $1$ is the smallest $d\in\N$ for which there is a finite $S\subseteq \N^2$ satisfying 1), 2), 3) from above.
    Observe that $d$ cannot be $0$, since there is no reduction $\mathsf i\choice{X}(M\choice{X}\Omega)\RED{X^0\OV X^0}{\mathsf i}$.
%
%    
%    
%    : indeed, if we were able to guess a set $S$ of reductions with maximal degree $0$ (this is point 1) above), then it means that we would have selected the reduction $\mathsf i\choice{X}(M\choice{X}\Omega)\RED{X^0\OV X^0}{\mathsf i}$.
%    But there is no such reduction, so the only choice violates 2).
    Let us show that $d=1$ satisfies all the required properties:
    the point is to guess the right finite set of reductions of maximal degree $1$. Take $S=\set{\tiny\begin{bmatrix}
        X=1\\ \OV X=0
    \end{bmatrix}}\subseteq \N^2\simeq !\set{X,\OV X}$, i.e.\ select the reduction $\mathsf i\choice{X}(M\choice{X}\Omega)\RED{X}{\mathsf i}$.
    This reduction exists, so 1) and 2) are satisfied.
    For 3), we need to show that the selected set maximises the sup of all the possible probabilities of all the possible reductions.
%    This requires to analyse the set of all the possible reductions, in other words, to consider the support $\mathrm{supp}(\model{\mathsf i\choice{X}(M\choice{X}\Omega)}^X_{\mathsf i})$ of the parametric interpretation of the term.
Since $\Omega$ does not terminate, the possible reductions to $\mathsf i$ are exactly those of weight either $X$ or $X^{i+1}\OV X^{j+1}$, for some $i,j\in\N$ such that $M\RED{X^i\OV X^j}{\mathsf i}$.
It is clear then that, for any choice of $p\in [0,1]$, the reduction corresponding to $\mathsf i\choice{X}(M\choice{X}\Omega)\RED{X}{\mathsf i}$ is the one with maximum weight.

%    
%    
%    For this, fix $p\in[0,1]$ and, since there is only the one reduction of weight $X$ in $S$, we have to show that $\sup\{ P\mid {\mathsf i\choice{X}(M[X:=p]\choice{X}\Omega)\RED{P}{\mathsf i}}\} =% \max_{(i,j)\in S} p^i(1-p)^j=
%    p$.
%            Therefore, we need to show that $\sup_{X^i\OV X^j\in \mathrm{supp}(\model{M}^X_{\mathsf i})}\set{p,p^{i+1}(1-p)^{j+1}}=p$, which in turn is easily seen, because both $p$ and $p-1$ are at most $1$.
%    %But since $p(1-p)p^{i}(1-p)^{j}\leq p(1-p)<p$ (the first inequality holds as $p^{i}(1-p)^{j}\leq 1$ and the second as $1-p<1$), the sup equals $p$ and we are done.
\end{example}

%{
%The example below shows that the tropical degree is very sensible to changes in the program, even if the structure of the program is the same and the only change is on the parameters.
%In fact, taking the two occurrences of $X$ in the example above as different parameters, already makes the tropical degree step from $1$ to at least $2$.
%This is because now there would be at least one reduction path of degree at least $2$ which can be given a probability higher than all the reductions of degree $1$.
%In other words, there is a choice of the parameters as probabilities such that the most likely one makes at least two random choices before stopping, instead of one as it was in the above example.}

\begin{example}\label{ex:tropdeg}
This example shows how the tropical degree is sensible to the number of parameters.
    Let $M:\Bool$, now with parameters $X_1,\overline{X_1},X_2,\overline{X_2}$. Suppose that $M\RED{}{\mathsf i}$, for a fixed $i\in\set{0,1}$.
    %Let us show that $\F d_{\mathsf i}(\mathsf i\choice{X_1}(M\choice{X_2}\Omega))=2+\F d_{\mathsf i}(M)$. We have to show that $2+\F d_{\mathsf i}(M)$ is the smallest degree $d\in\N$ of a all-one polynomial whose support $S\subseteq \N^4\simeq !\set{X_1,\OV X_1,X_2,\OV X_2}$ satisfies 1), 2), 3) from above.
    Let us show that $\F d_{\mathsf i}(\mathsf i\choice{X_1}(M\choice{X_2}\Omega))\geq 2$.
    It is easy, arguing as before, to see that the tropical degree cannot be $0$.
    But now, contrarily to the above, we can also exclude it to be $1$.
    %For this, we need to show that no matter the choice $S$ of reductions of degree $1$ we make, there will be one that can be given a higher probability and that makes at least $2$ probability choices in the reduction.
%    For this, let us study the set of all possible reductions.
    In this case the reductions to $\mathsf i$ are either that of weight $X_1$, or those of weight $X_1^{i_1}\OV X_1^{j_1+1}X_2^{i_2+1}\OV X_2^{j_2}$, for $M\RED{X_1^{i_1}\OV X_1^{j_1}X_2^{i_2}\OV X_2^{j_2}}{\mathsf i}$.
    Now, while the only possible choice of a set $S$ of reductions of maximal degree $1$, is the one of weight $X_1$, this choice does not maximises all the possible probabilities, i.e.\ it does not satisfy 3).
    For example, assigning $X_1,X_2$ with, respectively, probabilities $(p_1,p_2)\in[0,1]^2$ such that $p_1<\frac{1}{2}$ and $p_2>2p_1$ (for example, $p_1=\frac{1}{4}$ and $p_2=\frac{2}{3}$), 
    %Then we have $p_2(1-p_1)>\frac{p_2}{2}>p_1$ and, thus, $\sup_{\mathsf i\choice{X_1}(M[X_1:=p_1,X_2:=p_2]\choice{X_2}\Omega)\RED{P}{\mathsf i}} P > \sup_{X_1^{i_1}\OV X_1^{j_1}X_2^{i_2}\OV X_2^{j_2}\in\mathrm{supp}(\model{M}^{X_1,X_2}_{\mathsf i})}\set{p_1^{i_1}(1-p_1)^{j_1+1}p_2^{i_2+1}(1-p_2)^{j_2}}$.
    %But $(1-p_1)p_2\geq p_1^{i_1}(1-p_1)^{j_1+1}p_2^{i_2+1}(1-p_2)^{j_2}$, because those are all numbers in $[0,1]$, and so
    one can see that a reduction $\mathsf i\choice{X_1}(M\choice{X_2}\Omega)\RED{\OV{X_1}}M\choice{X_2}\Omega\RED{X_2}M\RED{P'}\mathsf i$, with $P'$ large enough, might yield a probability 
   $  (1-p_1)p_2 P' > p_1$, %= \max_{(i_1,i_2,j_1,j_2)\in S} p_1^{i_1}(1-p_1)^{j_1}p_2^{i_2}(1-p_2)^{j_2}$,
    thus violating 3).
    
    %Let us now finally show that $2+\F d_{\mathsf i}(M)$ satisfies all the required properties to make it the tropical degree:
    %let $s_M$ be the polynomial (given by Corollary \ref{thm:collapse}) realising the tropical degree $\F d_\mathsf i(M)$ of $M$.
    %Consider $s:=X_1+\OV X_1X_2s_M$.
    %Since $\deg s=2+\deg s_M$ and $\deg s_M=\F d_\mathsf i(M)$, we are done if we show that such $s$ satisfies the properties of Corollary \ref{thm:collapse} and its is of minimal degree among those.
    
    %If we call $S_M$ the support of $s_M$, the support $S$ of $s$ is $S=\set{\tiny\begin{bmatrix}
    %    X_1=1 \\ \OV X_1=0 \\ X_2=0 \\ \OV X_2=0 
    %\end{bmatrix}}\cup(
    %{\tiny\begin{bmatrix}
    %    X_1=0 \\ \OV X_1=1 \\ X_2=1 \\ \OV X_2=0 
    %\end{bmatrix}}+S_M)$, which is finite and clearly satisfies 1) and 2).
    %For 3), fix $p\in[0,1]$ and we have to show that $\sup_{\mathsf i\choice{X_1}(M[X_1:=p_1,X_2:=p_2]\choice{X_2}\Omega)\RED{P}{\mathsf i}} P =
    %\max_{(i_1,i_2,j_1,j_2)\in S} p_1^{i_1}(1-p_1)^{j_1}p_2^{i_2}(1-p_2)^{j_2}=
    %\max_{(i_1,i_2,j_1,j_2)\in S_M} \set{p_1\,,\,p_2(1-p_1)p_1^{i_1}(1-p_1)^{j_1}p_2^{i_2}(1-p_2)^{j_2}}$.
    %But this immediately follows from the fact that $S_M$ satisfies
    
    %$\sup_{M[X_1:=p_1,X_2:=p_2]\RED{P}{\mathsf i}} P =
    %\max_{(i_1,i_2,j_1,j_2)\in S_M} p_1^{i_1}(1-p_1)^{j_1}p_2^{i_2}(1-p_2)^{j_2}$.

    %\TODO{It only remains now to show the minimality of $2+\F d_{\mathsf i}(M)$. Not clear how}
    
\end{example}

Beyond these relatively simple cases, the general problem of finding the tropical degree of a \pPCF\ program  is not decidable (and indeed the proof of Corollary \ref{thm:collapse} is non-constructive).
%\davide{We can even say more:}{}

{
\begin{theorem}\label{thm:pi01}
%Finding the tropical degree $\F d_M$ for a term $M:\Bool$ is a $\Pi_1^0$-complete problem.The problem of input a number $d\in\N$ and a term $M:\Bool$ of \pPCFX{X_1,\dots,X_k}, with $k\geq 2$, and of output ``yes'' if $\F d_{\mathsf i}(M)=d$ and ``no'' otherwise, is \emph{not} RE, %it is co-RE 

%and %actually 
%$\Pi_1^0$-%complete
%hard.
Both the problems of input a number $d\in\N$ and a term $M:\Bool$ of \pPCFX{X_1,\dots,X_k} with $k\geq 2$, and of respective output ``yes'' if $\F d_{\mathsf i}(M)< d$ and ``no'' otherwise, and ``yes'' if $\F d_{\mathsf i}(M) = d$ and ``no'' otherwise, are \emph{not} RE, %it is co-RE 
and %actually 
$\Pi_1^0$-%complete
hard

%\TODO{The problem of input a number $d\in\N$ and a term $M:\Bool$ of \pPCFX{X_1,\dots,X_k}, with $k\geq 2$, and of output ``yes'' if $\F d_{\mathsf i}(M)\geq d$ and ``no'' otherwise, is RE but
%not co-RE.}
\end{theorem}
%\begin{proof}
%We reduce the computation of $\F d_M$ to the  $\Pi^0_1$-complete problem of knowing if a term $N:\Bool$ diverges.
%Take $M=\ONE\choice{X} (N\choice{Y}P)$, where $X\neq Y$, both do not appear in $N$ and $P$ is the diverging term
%$\YY (\lambda x.x)$.
%Since $N:\Bool$, $N$ may either diverge or reduce to either $\ZERO$ or $\ONE$. If $N$ reduces to $\ONE$, we must thus have $M\stackrel{\overline XY\mu}{\to}\ONE$ and, since $X,Y$ do not occur in $\mu$, $\overline XY\mu$ and $X$ are incomparable, so $\F d_M\geq 2$. A similar argument holds if $N$ reduces to $\ZERO$.
%Conversely, if $N$ does not reduce to either $\ONE$ or $\ZERO$, then the only converging reduction of $M$ is $M\stackrel{\mu}{\to}\ONE$, so $\F d_M=1$. 
%We conclude then that $\F d_M=1$ iff $N$ diverges. 
%\end{proof}
\begin{proof}
%We reduce the computation of $\F d_M$ to the  $\Pi^0_1$-complete problem \TODO{reference?} of knowing if a term $N:\Bool$ diverges.
%Take $M=\ONE\choice{X_1} (N\choice{X_2} (\YY (\lambda x.x)))$, where $X_1\neq X_2$ and both do not occur in $N$.
%Noticing that $\YY (\lambda x.x)$ can only diverge, we can see that $\F d_M=1$ iff $N$ diverges. 

%First, we sketch below a (Turing-)reduction from the problem of knowing if a term of type $\Bool$ is \emph{not} normalisable, to problems 1 and 2.Since the former is known to be not RE and $\Pi_1^0$-complete (it is equivalent to the complementary of the halting problem) \TODO{reference?}, we are done for 1 and 2.

%Given $N:\Bool$ in input, take $X_1\neq X_2\notin N$
We reduce the complementary of the halting problem (which is not RE and is $\Pi_1^0$-complete) to Problem 1 and to Problem 2.
Given in input a closed $M:\Bool$ of \pPCFX{\mathbb X}, take $X_1\neq X_2$ (does not matter whether they belong to $\mathbb X$ or not) and let $\widetilde M:=\mathsf i\choice{X_1} (M\oplus_{X_2} \Omega)$.
If $M$ is normalisable to $\mathsf i$ then, by arguing similarly to Example \ref{ex:tropdeg}, we see that $\F d_{\mathsf i}(\widetilde M)\geq2$.
If $M$ is not normalisable to $\mathsf i$ then, since $\Omega$ also is not normalisable to $\mathsf i$, we have $\model{\widetilde M}_{\mathsf i}=X_1$ and so $\F d_{\mathsf i}(\widetilde M)=1$.
Summing up, $\F d_{\mathsf i}(\widetilde M)=1$ iff $M$ is not normalisable to $\mathsf i$ iff $\F d_{\mathsf i}(\widetilde M)< 2$. Hence an oracle semideciding either problem 1 or 2 allows to semidecide the non-normalisability of $M$.
%, which is in turn equivalent to the complementary of the halting problem.
%Finally, Problem 3 is RE and not co-RE because it is the complementary of Problem 2.
%However, it is RE, since in order to semidecide it on input $(k,N)$ we can: \TODO{??}
%Conversely, we show that, with an oracle on the halting problem, we can design a brute force algorithm for Problem 2 (and thus Problem 1 as well): .
\end{proof}
}

{
The previous examples show that computing (or even estimating) the tropical degree by hand is a subtle task and, in general, exact values or even upper-estimations are not mechanisable.
This poses obvious limitations to the what can be achieved in general, algorithmically. However, 
in the next Sections we will show that it is still possible to design an algorithm that progressively computes estimations of the tropical degree eventually \emph{stabilizing} onto the correct value 
 $\F d_{\mathsf i}(M)$.
}

\section{%The Viterbi-Newton Algorithm
Convex Geometry and Newton Polytopes}\label{sec:geo}
% !TEX root = Tropical_II.tex

%In this and the following sections we show that, 
{As already mentioned, in this and the next section,} by combining the toolbox of tropical geometry with the one of programming language theory, we define an efficient procedure to solve the inference problems (I1) and (I2) for a term $M:\Bool/ \NAT$, that is, to compute the maximum a posteriori (log)probabilities of producing a given value, say $\ONE$, and to produce a most likely explanation for it. 

{
%That is, how to approximate the tropical degree of a program.
%Remember that the parametric interpretation $\model{M}^{\BB X}$ yields formal power series $s$ related to the reduction paths of $M$; 
While the finiteness of the tropical degree ensures that we may restrict ourselves to explore only a finite set of reductions $S$, the size of $S$
%those may be in an infinite number (Example~\ref{ex:33}), or in finite by
may still be exponentially large (cf.~Section~\ref{sec:unfeasable}).

In this section we show that one can compute some polynomially bounded subset $S'\subset S$ that still contains enough trajectories to track the most likely ones. This set $S'$ is obtained by associating a program (in fact, its associated fps) with a polytope, called the \emph{minimal Newton polytope}, which is a variant of the well-known Newton polytope of a polynomial. The same method will then be used to design an algorithm, similar to the Viterbi algorithm, to compute the tropical product of polynomials in an efficient way. This algorithm will be the key ingredient, in the next section, to design a compositional procedure to track the most likely trajectories of a \pPCF-program.

%
%In order to find the most likely reductions (which basically means computing $\trop s:\TS^n\to\TS$), the only hope is then to show that this set can be further \emph{shrinked} to one that is sufficiently small to be explored.
%%In Section \ref{sec:tropdeg} we showed that (for terms of a certain type) a finite space of trajectories exists via the notion of tropical degree; we are now going to approximate it by keeping it small.
%}
%\TODO{the goal of this section is to introduce an algo for the tropical product of polynomials. WHY DO WE DO THIS?}

\subsection{The Newton Polytope}

%We have seen in the previous section that, within the possibly infinite set $I$ of the trajectories of a probabilistic program, one can find a finite subset $J\subset I$ of such trajectories so that any trajectory in $I$ is \emph{dominated} in its probability by one in $J$. In algebraic terms this means finding, for some tps $f$, a polynomial $g$ so that $f$ and $g$ compute the same function, i.e.~$f^!=g^!$.

In this subsection we recall the standard definition of the Newton polytope of a polynomial.
%%%%
{
Let us fix some all-one polynomial $s=\sum_\mu %s_\mu
\mu \in \fpp{\OV{\BB R}}{\Sigma}$ in $n$ variables, where $\OV{\BB R}:=\BB R\cup\set{+\infty}$.
This is the standard tropical semiring in which tropical algebraic geometry is usually carried on.
}
It is well-known that the piece-wise linear function $f_s:\BB R^n\to \BB R$ defined by $s$ %over the reals (remember that $1_{\TS}=0_{\BB R}$)
by $f_s(x)=\min_{\mu\in\mathrm{supp}(s)}
%\{\mu\cdot x+s_\mu\} =
 %\min_{s_\mu=0_{\BB R}}
 \{ \mu\cdot x\}$ can be characterized via two, dual, geometric objects:
\begin{varitemize}
\item the \emph{tropical variety} $\gamma(f_s)$, i.e.\ the set of all \emph{tropical roots} of $s$, i.e.\ the $x\in \BB R^n$ such that the minimum $f_s(x)$ is reached by \emph{at least two} monomials (i.e., $f_s$ is not differentiable at $x$);

\item the \emph{Newton polytope} $NP(s)$, i.e.\ the convex hull in $\BB R_{\geq0}^n$ of the points $\mu\in \BB N^n$ %such that $s_\mu\neq 0_{\TS} \, (=+\infty)$
in $\mathrm{supp}(s)$.

\end{varitemize}

$\gamma(f_s)$ and $NP(s)$ describe two polyhedra in $\BB R^n$ with dual graphs (see \cite{Sturmfelds}), and any tropical root $a\in\gamma(f_s)$ of $s$ uniquely identifies a \emph{facet} $F_x$ of $NP(s)$: $a$ 
individuates $k\geq 2$ monomials $\mu_1,\dots, \mu_k\in\N^n$ such that $\mu_1\cdot a=\dots=\mu_k\cdot a=:b\in\BB R$, {so that $a$ is, by construction, normal to the hyperplane $H_a$ of $\BB R^n$ of equation (in $z$) $a\cdot z=b$ %(for any $i$, they are all equivalent),
} and $H_a$ is the supporting hyperplane of a unique facet $F_a$ of $NP(s)$, namely the one containing the points $\mu_1,\dots, \mu_k$.

A crucial remark now is that, even if we defined $NP(s)$ as the convex hull of the possibly very large set of points $\mathrm{supp}(s)$, it is uniquely determined by the set $\mathrm{Vert}(NP(s))\subseteq\mathrm{supp}(s)\subseteq NP(s)$ of its \emph{vertices} which is, in general, much smaller and in average efficiently computable:

\begin{theorem}\label{thm:newton}
Let $s\in\fpp{\OV{\BB R}}{x_1,\dots,x_n}$.
Then 
%For fixed $n$, i
{
then $\#\mathrm{Vert}(NP(s))=\C O(d^{2n-1})$ (\cite{Pachter2004, Pachter2004b}), where $d$ is the degree of $s$, and $\mathrm{Vert}(NP(s))$ can be computed with a randomised algorithm in expected time $\C O(|s|^{\lfloor\frac{n}{2}\rfloor})$ (\cite[p.~256 (line 4 from the bottom)]{Berg2008}), where $|s|:=\#\mathrm{supp}(s)$.}
\end{theorem}

A consequence of all this discussion is that, for a polynomial $s$ of degree $d$, we can always find a polynomial $s'$
formed by a \emph{subset} (namely, $\mathrm{Vert}(NP(s))$) of the monomials of $s$ of size polynomial in $d$ such that $NP(s)=NP({s'})$ and, {crucially, the functions $f_s$ and $f_{s'}$ coincide. Indeed:

\begin{lemma}\label{lm:supp_to_NP}
Let $s=\sum_\mu \mu \in \fpp{\OV{\BB R}}{\Sigma}$.
Then 
$\min_{\mu\in\mathrm{supp}(s)}\set{\mu\cdot x}=\min_{\mu\in \mathrm{Vert}(NP(s))}\set{\mu\cdot x}$. 
\end{lemma}
}
\begin{proof}
This is an immediate consequence of a well-known fact in linear optimisation (cfr. the appendix): for a polytope $P$, the $\inf$ on $P$ of a linear function $\mu\cdot x$ is found on the vertices of $P$. 
\end{proof}

\begin{example}\label{ex:freshman}
Consider the polynomial $s=\sum_{i+j+k=2}X_1^iX_2^jX_3^k$. Then $NP(s)$, illustrated in gray in Fig.~\ref{fig:newton}, is the convex hull of all the points $(i,j,k)\in \BB N^3$ such that $i+j+k=2$. $NP(s)$ is generated by its vertices, which are the three bold points $(2,0,0),(0,2,0),(0,0,2)$ in the figure. We deduce that $f_s$ coincides with $f_{s'}$, where $s'=X_1^2+X_2^2+X_3^2$.
What we have just described is in fact a geometric proof of the "old freshman dream" 
$
(x_1+x_2+x_3)^2=x_1^2+x_2^2+x_3^2
$
for tropical polynomial functions. 
\end{example}

\begin{figure}
\begin{subfigure}{0.31\textwidth}
$$
\begin{tikzpicture}[scale=0.4]
\draw[->, dashed] (0,0) to (0,3);
\draw[->, dashed] (0,0) to (3,0);
\draw[->, dashed] (0,0) to (1.5,-1.5);

\node (a) at (0,0) {};

%\node (a) at (1,0) {$\bullet$};

%\node (a) at (0,1) {$\bullet$};

\draw[fill=gray!20] (0,2) to (2,0) to (1,-1) to (0,2);
%\draw (2,0) to (1,-1);
%%\draw (0,0) to (2,0);
%%\draw (0,0) to (0,2);
%%\draw (0,0) to (1,-1);
%\draw (0,2) to (1,-1);
%\draw (2,0) to (1,-1);

\node (a) at (1,1) {\tiny$\bullet$};

%\node (a) at (0.5,-0.5) {$\bullet$};

\node (a) at (1.5,-0.5) {\tiny$\bullet$};

\node (a) at (0.5,0.5) {\tiny$\bullet$};
\node (a) at (0,2) {$\bullet$};
\node (a) at (2,0) {$\bullet$};
\node (a) at (1,-1) {$\bullet$};

\end{tikzpicture}
$$
\caption{Geometric proof of the "old freshman dream" $(X_1+X_2+X_3)^2=X_1^2+X_2^2+X_3^2$. The triangle is the convex hull of the points corresponding to the monomials in $(X_1+X_2+X_3)^2$. The triangle is spanned by the three vertices corresponding to $X_1^2,X_2^2,X_3^2$.}
\label{fig:newton}
\end{subfigure}
\hskip3mm
\begin{subfigure}{0.31\textwidth}
$$
\begin{tikzpicture}[line join=bevel,z=-5.5, scale=0.4]

\node(O) at (0,0,0) {\tiny$\bullet$};
\node(OO) at (0.3,-0.3,0) {\tiny$O$};
\draw[->, dashed] (0,0,0) to (3,0,0);
\draw[->, dashed] (0,0,0) to (0,3,0);
\draw[->, dashed] (0,0,0) to (0,0,4);

\coordinate (A1) at (2,3,2);
\coordinate (A2) at (3,2,2);
\coordinate (A3) at (1,1,3);
\coordinate (A4) at (3,0,3);
\coordinate (B1) at (5,3,4);
\coordinate (C1) at (0,1,0);

\node (a1) at (2,3.2,2) {\tiny$v_1$};
\node (a2) at (3,2.2,2) {\tiny$v_2$};
\node (a3) at (0.5,0.7,2) {\tiny$v_3$};
\node (a4) at (2.7,0,3) {\tiny$v_4$};
\node (b1) at (5.2,3.1,4) {\tiny$v_5$};

\draw (A1) -- (A2) -- (A4) -- (A3) -- cycle;
\draw (A1) -- (B1) -- (A2) -- cycle;
\draw (A1) -- (A3) -- (B1) -- cycle;
\draw (A2) -- (A4) -- (B1) -- cycle;
\draw (A3) -- (A4) -- (B1) -- cycle;

%
%\draw[->,dotted] (O) to (A1);
%
%\draw[->,dotted] (O) to (A2);
%
%\draw[->,dotted] (O) to (A3);
%
%\draw[->,dotted] (O) to (A4);
%\draw[->,dotted, thick] (O) to (B1);

%\draw (A4) -- (A1) -- (B1) -- cycle;
%\draw (A1) -- (A2) -- (C1) -- cycle;
%\draw (A4) -- (A1) -- (C1) -- cycle;
\draw [fill opacity=0.3,fill=gray!100!white]  (A1) -- (A2) -- (A4) -- (A3) -- cycle;
%\draw [fill opacity=0.3,fill=gray!40!white] (A2) -- (A4) -- (B1) -- cycle;
%
%\draw [fill opacity=0.7,fill=gray!30!black](A1) -- (B1) -- (A2) -- cycle;
%\draw [fill opacity=0.7,fill=gray!70!white] (A1) -- (A3) -- (B1) -- cycle;
%\draw [fill opacity=0.7,fill=gray!70!black] (A3) -- (A4) -- (B1) -- cycle;
%
\end{tikzpicture}
$$
\caption{Illustration of $NP_{\mathrm{min}}(s)\subset NP(s)$, for $s$ given in 
Example \ref{ex:visible}: $NP(s)$ is the
convex hull of $v_1,\dots,v_5$. %) is the Newton Polytope $NP(s)$. 
$NP_{\mathrm{min}}(s)$, coloured in grey, is the convex hull of the vertices $v_1,\dots,v_4$. Notice that $v_5\notin NP_{\mathrm{min}}(s)$, as it is not minimal (e.g.~$v_3\prec v_5$).
%%Visible vertices in the 
%\TODO{A minimal 
%Newton polytope}%: the point $v_5$ is not visible from $e_1$.   
}
\label{fig:visible}
\end{subfigure}
\hskip3mm
%
%\bigskip
%
%
%\bigskip
%
\begin{subfigure}{0.31\textwidth}
$$
\begin{tikzpicture}[scale=0.36]
\draw[->, dashed] (0,0) to (0,4);
\draw[->, dashed] (0,0) to (4,0);

\node (a) at (0,0) {};

\draw[fill=gray!20] (0,3) to (1,2) to (2,1) to (3,0);

\node (a) at (2.2,2.7) {\tiny$\sigma$};

\node (a) at (3,0) {$\bullet$};

\node (a) at (1,2) {\tiny$\bullet$};

\node (a) at (2,1) {\tiny$\bullet$};

\node (a) at (0,3) {$\bullet$};

\end{tikzpicture}
$$
\caption{
For the term $M_4$ of Example~\ref{ex:unfeasible}, 
both $NP_{\mathrm{min}}(M_4)$ and $NP(M_4)$ are given by the segment $\sigma$.
%
% is the polytope
%$NP_{\mathrm{min}}(M_4)$ in $\BB R^2$ (in this %particular 
%case it %even 
%coincides with $NP(M_4)$) associated with term $M_4$ 
Its vertices $(0,3),(3,0)$ contain, by Proposition~\ref{lm:NP_to_NPmin}, the %degrees 
weights of %two
at least one most likely reduction of $M_4$ (no matter how probabilities are assigned to the parameter $X$). 
%In standard linear optimisation terminology, they are %constitute 
%the $2$ \emph{basic feasible solutions}, while $\mathrm{supp}(\model{M_4}_{\ONE}^X)$ gives the $2^3$ \emph{feasible %solutions
%} ones.}
}
\label{fig:np_M4}
\end{subfigure}
\caption{\small Illustrations of the Newton polytope.}
\hskip8mm
\end{figure}

\subsection{{The minimal Newton polytope% and the modified ``Viterbi+Newton'' Algorithm
}}

{

%Let us consider a finitary variant of the problem discussed in Section~\ref{sec:tropdeg} (the general case is addressed in the next Section \ref{sec:inters_types}):
We now address the following question, indeed a finitary variant of the problem discussed in Section~\ref{sec:tropdeg}: given some very large, although finite, \emph{polynomial} $s\in\fpp{\TS}{\BB X}$%(wlog taken to be all-one, i.e.\ $s=\sum_\mu 1_{\TS}\mu \in \fpp{\TS}{\Sigma}$)
, can we find a sufficiently \emph{smaller}, and somehow \emph{minimal}, {all-one} polynomial $s'$ such that $\trop s=%\trop 
(s')^!$? 

%Equivalently, given a large set $I$ of trajectories, can we restrict our search for a most likely one to some sufficiently \emph{small} subset $J\subset I $?
%In this section, we fix an \NEW{all-one} polynomial $s=\sum_\mu %s_\mu
%1_{\TS}\mu \in \fpp{\TS}{\Sigma}$ in $n$ variables %with $s_\mu=1_{\TS} \, 
%for all $\mu\in !{\Sigma}\simeq \N^n$.
We have seen that the Newton polytope $NP(s)$ and its small number of vertices (Theorem \ref{thm:newton}) precisely serves this purpose... but for the fact that
$NP(s)$ characterizes the function $\BB R^n\to\BB R$ defined by tropical polynomials over the tropical semiring $\BB R\cup\set{+\infty}$, while we are interested in the function $\trop s:\TS^n\to\TS $ defined by tropical polynomials $s$ over the tropical semiring $\TS$ (with carrier set $[0,+\infty]$). 
To overcome this mismatch, we introduce the following:

\begin{definition}\label{def:npmin}
The \emph{minimal Newton Polytope} $NP_{\min}(s)$ of $s=\sum_\mu s_\mu\mu 
\in \fpp{\TS}{\Sigma}$ is the convex hull in $\BB R_{\geq0}^n$ of the following set, which is easily seen to be its set of vertices and non-empty:
\[
\mathrm{Vert}(NP_{\min}(s)):=\{%\mu\in \mathrm{Vert}(NP(s))\mid \lnot \exists \nu\in \mathrm{Vert}(NP(s)), \nu\prec \mu
\mu\in\N^n \mid \mu \textit{ minimal element of }(\mathrm{Vert}(NP(s)),\preceq)\}\subseteq \mathrm{supp}(s),
\]
where $\preceq$ is the pointwise (well-founded) order.
We also set $s_{\min}:=\sum_{\mu\in \mathrm{Vert}(NP_{\min}(s))}%s_\mu
\mu\in\fpp{\TS}{\Sigma}$.
%and we call an all-one polynomial $h\in \fpp{\TS}{\Sigma}$ \emph{minimal for $s$} whenever $\trop s_{\min}=h^!$. 
\end{definition}

The polytope $NP_{\min}(s)$ precisely captures the behavior of the function $\trop s(x):\TS^n\to\TS$: the latter coincides with the $\min$ computed over the monomials in $NP_{\min}(s)$:

{
\begin{proposition}\label{lm:NP_to_NPmin}
Let $s%=\sum_\mu s_\mu\mu 
\in \fpp{\TS}{\Sigma}$.
Then 
$\min_{\mu\in\mathrm{Vert}(NP(s))}\set{\mu\cdot x}=\min_{\mu\in \mathrm{Vert}(NP_{\min}(s))}\set{\mu\cdot x}$.
In particular, since the latter is $\trop s_{\min}(x)$, it follows from Lemma \ref{lm:supp_to_NP} that $\trop s(x)=\trop s_{\min}(x)$. %i.e., $s_{min}$ is minimal for $s$.
\end{proposition}}
\begin{proof}
Fix $x\in\BB R^n$.
The $(\leq)$ is trivial.
For $(\geq)$, we show, by induction on $\mu\in (\mathrm{Vert}(NP(s)),\preceq)$, that for all $\mu\in \mathrm{Vert}(NP(s))$, there is $\rho\in \mathrm{Vert}(NP_{\mathrm{min}}(s))$ such that $\rho\cdot x\preceq \mu\cdot x$.
If $\mu$ is minimal, then by definition $\mu\in\mathrm{Vert}(NP_{\mathrm{min}}(s))$, so we are done.
If $\mu$ is not, there is $\nu\in \mathrm{Vert}(NP(s))$ with $\nu\prec \mu$.
By IH there is $\rho\in \mathrm{Vert}(NP_{\mathrm{min}}(s))$ such that $\rho\cdot x\preceq \nu\cdot x \preceq \mu\cdot x$, and we are done.
\end{proof}

}

%{The crucial property is now that, even if the vertices of the minimal Newton Polytope of $s$ may seem to be defined in a non-constructive way, they can be computed.

The following shows that the set of vertices of $\mathrm{NP}_{\min}(s)$ can be computed fast wrt $s$.
Remembering that $|s|$ is the cardinality $\#(\mathrm{supp}(s))$ of the support of $s$, we have:}

\begin{theorem}\label{thm:np}
Let $s%=\sum_\mu s_\mu\mu 
\in \fpp{\TS}{\Sigma}$.
The set $\mathrm{Vert}(NP_{\min}(s))$ (i.e.\ $s_{\min}$) 
can be computed with a randomised algorithm in expected time $\C O(n|s|^{\max\{2,n\}})$.
\end{theorem}
\begin{proof}
%$NP_{\min}(s)$ is obtained by a quadratic check over $\mathrm{Vert}(NP(s))$, which can in turn be computed in time $\C O(|s|^{1+\lfloor\frac{n}{2}\rfloor})$ (cf.~\cite{Chan1995} and \NEW{\cite[p.~256 (line 4 from the bottom)]{Berg2008}}. \TODO{perché ``$+1$''? Io leggo senza. Inoltre, l'algo in [8] \`e probabilistico, quindi quello \`e l'expected time.}
First, compute $\mathrm{Vert}(NP(s))$.
For each $\mu\in \mathrm{Vert}(NP(s))$, the minimality check for $\mu$ can be done in time $n(\#\mathrm{Vert}(NP(s))-1)$ and, since we have $\#\mathrm{Vert}(NP(s))$ of them, we have an additional time $\sim n\,\#\mathrm{Vert}(NP(s))^2$.
By Theorem \ref{thm:newton}, we can compute $\mathrm{Vert}(NP(s))$ in expected time $\C O(|s|^{\lfloor\frac{n}{2}\rfloor})$.
Also $\#\mathrm{Vert}(NP(s))=\C O(|s|)$, whence the total time $\C O(|s|^{\lfloor\frac{n}{2}\rfloor})+n\C O(|s|^2)=O(n|s|^{\max\{2,n\}})$.
\end{proof}

\begin{example}\label{ex:visible}
%Let $s\in \fps{\TS }{\{X_1,X_2,X_3\}}$ be  
Let $s=X_1^2X_2^3X_3^2+X_1^3X_2^2X_3^2+X_1X_2X_3^3+X_1^3X_3^3+X_1^5X_2^3X_3^4+X_1^4X_2^2X_3^3\in\fps{\TS }{X_1,X_2,X_3}.$
Fig.~\ref{fig:visible} illustrates
$NP(s)$, the convex hull of the points $v_1=(2,3,2),v_2=(3,2,2),v_3=(1,1,3),v_4=(3,0,3),v_5=(5,3,4), v_6=(4,2,3)$, of which only $v_1,\dots,v_5$ are vertices, as $v_6$ is convex combination of $v_4,v_5$. $v_5%=(5,3,4)
$ is the only non minimal vertex, since e.g.~$v_1\prec v_5$, so 
%and nor are $v_1$ and $v_2$, as $v_3$ is smaller than both of them. 
$\mathrm{Vert}(NP_{\mathrm{min}}(s))=\{v_1,v_2,
v_3,v_4\}$. 
%
%: %intuitively, 
%the facet formed by the other four points ``cover'' the fifth. $NP_{\min}(s)$ (in gray% in the figure
%) is indeed formed by the other four points.

\end{example}

\begin{remark}\label{rmk:neg_or}
While the algorithm for $NP_{\min}(s)$ in Theorem \ref{thm:np} rests on a brute force minimality check on $NP(s)$ (still polynomial in $d$), a potential speed up may arise from geometric considerations. 
%
%However, one can characterise $NP_\mathrm{min}(s)$ in local and purely geometric terms as below. In practice, this may happen to give a more efficient computation of the minimal Newton Polytope.
In general, given a polytope $P$ in $\BB R^n$, any of its facets $f$ lies, by definition, on its supporting hyperplane $H_f$, and moreover $P$ is all contained inside one of the two closed halfspaces $H_f^+, H_f^-$ in which $H_f$ divides $\BB R^n$. Call $H_f^+$ the one containing $P$. 
Let us call $f$ \emph{negatively oriented} if the normal unit vector to $H_f$ towards $H_f^-$ has all strictly negative coordinates (in the canonical base). Intuitively, $f$ %is oriented \emph{towards the origin}
``sees the origin''.
Now, it can be easily proven (see the appendix) that a vertex 
$v$ belonging to some negatively oriented facet $f$ of $P$ is always minimal. The example in Fig.~\ref{fig:visible} illustrates this fact, since in this case $NP_{\min}(s)$ coincides with the unique negatively oriented facet of $NP(s)$.
This suggests that, to compute the minimal vertices, one could start by first selecting the negatively oriented facets, and restrict the brute force minimality check to the remaining ones. 
% (i.e.\ $v\in P_{\min}$) iff $v$ belongs to some negatively oriented facet $f$ of $P$}.
% 
%Therefore, instead of checking for the quadratic minimality check for a vertex $v$ of $P$ as in the proof of Theorem \ref{thm:np}, we can search for a negatively oriented facet containing $v$.
%Depending on the actual implementation of a polytope that one is using, and on the particular polytope under study, this may be more efficient.
%But, for the moment, we leave such investigations for future work.
However, the eventual speed up depends on the concrete representation of a polytope in use in the algorithm, so we leave such investigations for future work.
\end{remark}

{It is worth to rephrase and summarise what we got so far%in more geometric terms
:
given a \pPCFn\ term $M:\NAT$ and some fixed outcome $\mathsf i$, the parametric interpretation gives rise to a fps $%s:=
\model{M}_{\mathsf i}^{\BB X}\in\fps{\NINF}{\BB X}$.
If %$s$
the latter is a polynomial, %it gives rise to the Newton Polytope $NP(\model{M}_{\mathsf i}^{\BB X}\in\fpp{\NINF}{\BB X})$ in $\BB R_{\geq 0}^{2n}$, call it $NP(M)$.
then we have (cfr.\ Definition \ref{def:npmin}) a minimal polynomial 
$(\mathsf t %s
\model{M}_{\mathsf i}^{\BB X})_{\min}$
%$(\mathsf t\model{M}_{\mathsf i}^{\BB X})_{\mathrm{min}}\in\fpp{\TS}{\BB X}$
 or, equivalently, its Newton Polytope, call it $NP_{\mathrm{min}}^i(M)$.
By Proposition \ref{lm:NP_to_NPmin}%and Definition \ref{def:trop}
, this %the 
polytope %$NP_{\mathrm{min}}^i(M)$
\emph{approximates} the tropical degree $\F d_{\mathsf i}(M)$, in the sense that, for all probability assignment to the parameters of the program, the vertices of $NP_{\mathrm{min}}^i(M)$ (i.e.\ the minimal monomials in $\model{M}_{\mathsf i}^{\BB X}$) contain (the monomial of) \emph{at least one} most likely reduction.
This reads as $\trop\model{M}_{\mathsf i}^{\BB X}=(\trop \model{M}_{\mathsf i}^{\BB X})_{\min}$, whence $\F d_\mathsf i(M)\leq \deg (%s
(\mathsf t\model{M}_{\mathsf i}^{\BB X})_{\min})% \leq \deg (\mathsf t\model{M}_{\mathsf i}^{\BB X})=\deg (\model{M}_{\mathsf i}^{\BB X})
$.
Figure \ref{fig:np_M4} illustrates this discussion for the term $M_4$ from Example \ref{ex:unfeasible} (where the situation is more trivial: $\F d_\ONE(M_4)=\deg (\model{M_4}_{\ONE}^{\BB X})=3$, because all reductions of $M_4$ have degree $3$).

But we can actually do the same even if the fps $\model{M}_{\mathsf i}^{\BB X}$ is \emph{not} a polynomial (i.e.\ its support is infinite):
by arguing similarly to Proposition~\ref{prop:collapse}, one sees that the minimal monomials in its support are still in \emph{finite} number, therefore one still has a \emph{polynomial} $(\mathsf t %s
\model{M}_{\mathsf i}^{\BB X})_{\min}$ or, equivalently, its Newton Polytope, call it $NP^i_{\mathrm{min}}(M)$.
Moreover, one can follow the same proof of Proposition \ref{lm:NP_to_NPmin} in order to show that $NP^i_{\mathrm{min}}(M)$ still satisfies the exact same approximation property as in the finite case.

In conclusion, we can \emph{always} associate $M:\NAT$ with a minimal Polytope $NP_{\mathrm{min}}^i(M)$, i.e. a minimal all-one polynomial $(\mathsf t\model{M}_{\mathsf i}^{\BB X})_{\min}\in\fpp{\TS}{\BB X}$ such that $\trop\model{M}_{\mathsf i}^{\BB X}=(\trop \model{M}_{\mathsf i}^{\BB X})_{\min}$ and $\F d_\mathsf i(M)\leq \deg ((\mathsf t
\model{M}_{\mathsf i}^{\BB X})_{\min})$.

Notice that this is not in contradiction with Theorem \ref{thm:pi01}: the above upper-bound always holds, but in general we can only hope to compute \emph{all} the minimal monomials {for finite interpretations}.
%.. in which case we have access to %the full reduction graph 
%all the monomials of the reductions of the program and so we can well exactly compute the tropical degree.
%
%What if the fps $\model{M}_{\mathsf i}^{\BB X}\in\fpp{\NINF}{\BB X}$ is \emph{not} a polynomial (i.e.~its support is infinite)? On the one hand, by Proposition~\ref{prop:collapse}, the function $\trop s$ is captured by some polynomial $s'$ (which need not be unique), and we could define $NP^i_{\mathrm{min}}(M)$ as  $NP_{\min}(s')$. However, even without this fact, we can see (arguing similarly to Proposition~\ref{prop:collapse}) that the minimal monomials in $s$ are still in finite number. So they still give rise to a minimal polynomial $s_{\min}\in \fpp{\NINF}{\BB X}$. In other words, 
%%
%%(\mathsf t\model{M}_{\mathsf i}^{\BB X})_{\mathrm{min}}\in\fpp{\TS}{\BB X}$, 
%i.e.\ 
%%
%we can define $NP^i_{\mathrm{min}}(M):=NP^i_{\mathrm{min}}(s)$ even when $s$ is not a polynomial. \TODO{This satisfies the same property as in the finite case above. DAVVERO ? Senn\`e la somma di tutti quelli che realizzano il tropdeg funziona, ma meno elegante}.

%In addition to being %elegant and 
%standard in linear optimisation, the geometric understanding allows us to notice the following:

}

\subsection{The Viterbi-Newton Algorithm}

Recall that, from the discussion around \eqref{eq:mins} in Section \ref{sec:inference}, tracking the most likely runs of an application $MN$ requires to be able to compute (tropical) products of polynomials efficiently.
We will now use the results from the previous subsection to define an algorithm $\VN$ to compute, given $k$ polynomials $s_1,\dots, s_k$, a minimal polynomial $s$ capturing the tropical product of the $s_i$.

First observe that the number of monomials in $s_1\dots s_k$ grows exponentially in $k$. For instance, letting all $s_i$ be the same polynomial 
$X_1+\dots+X_n$, we have that $s_1\dots s_k=(X_1+\dots+X_n)^k$ contains 
$\binom{n+k-1}{k-1}\in \C O((n+k-1)^{k-1})$ distinct monomials. However, we have seen (cf.~Example \ref{ex:freshman}) that in the tropical setting we have $(X_1+\dots+X_n )^k=X_1^k+\dots+X_n^k$. This suggests that one may hope that a sufficiently small set of monomials is enough to capture the polynomial function $\trop(\prod_{i=1}^k s_i)$. This is precisely what we show below.

The main idea behind the algorithm described below is to compute the product as an operation performed directly over the minimal Newton polytopes $NP_{\min}(s_1),\dots, NP_{\min}(s_n)$, and producing the minimal polytope $NP_{\min}(s_1\dots s_n)$ (this is reminiscent of the \emph{polytope algebra} of \cite{McMullen1989}).
%
%Let $+$ indicate below the \emph{Minkowski sum}
%$A+B=\{v+w\mid v\in A, w\in B\}$ of two sets in $\BB R^n$.
The fundamental remark is that the product of polynomes translates into the \emph{Minkowski sum} of the corresponding polytopes, defined as $A+B=\{v+w\mid v\in A, w\in B\}$ for two sets in $\BB R^n$. 
The set $\mathrm{Vert}(A+B)$ can be computed in time $\mathcal O(nm)$, 
where $n=|\mathrm{Vert}(A)|,m=|\mathrm{Vert}(B)|$, cf.~\cite{Das2021}.
Using the well-known fact that $NP(s_1 s_2) = NP(s_1)+NP(s_2)$, we can prove:

%
%\davide{
%\begin{lemma}\label{lm:vertmin}
%Let $s_1,s_2\in \fps{\TS }{\Sigma}$.
%$\mathrm{Vert}(NP_{\mathrm{min}}(s_1s_2))\subseteq \mathrm{Vert}(NP_{\mathrm{min}}(s_1))+\mathrm{Vert}(NP_{\mathrm{min}}(s_2))$.
%\end{lemma}
%\begin{proof}
%It is well known that $NP(s_1 s_2) = NP(s_1)+NP(s_2)$.
%It is also well known that, for convex polytopes, $\mathrm{Vert}(\C P_1+\C P_2)\subseteq%=
%\mathrm{Vert}(\C P_1)+\mathrm{Vert}(\C P_2)$.
%Using those, it is easy to see that $NP_{\mathrm{min}}(s_1 s_2) = NP_{\mathrm{min}}(s_1)+NP_{\mathrm{min}}(s_2)$, whence the desired result.
%\end{proof}
%}
%{

%
%\begin{lemma}\label{lemma:submin}
%For all $A,B\subseteq \mathbb N^n$, if $A_{\min}\subseteq B\subseteq A$, then $A_{\min}=B_{\min}$.
%\end{lemma}
%\begin{proof}
%For $(\subseteq)$, let $a\in A_{\min}$ and $b\in B$, and suppose $b\preceq a$; since it must be $b\in A$, the minimality of $a$ in $A$ yields $b=a$. We conclude then $a\in B_{\min}$.
%For $(\supseteq)$, let $b\in B_{\min}\subseteq A$ and $a\in A$, and suppose $a\preceq b$; by the well-foundedness of $\prec$, we can find $a_0\in A_{\min}\subseteq B$ such that $a_0\preceq a\preceq b$. Now, the minimality of $b$ in $B$ yields $b= a_0= a$. We conclude then $b\in A_{\min}$.
%\end{proof}
%

\begin{lemma}\label{lm:vertmin}
Let $s_1,s_2\in \fps{\TS }{\Sigma}$.
$\mathrm{Vert}(NP_{\mathrm{min}}(s_1s_2)) 
= 
(\mathrm{Vert}(NP_{\min}(s_1))+\mathrm{Vert}(NP_{\min}(s_2)))_{\mathrm{min}}$.
\end{lemma}
%\begin{proof}
%It is well known that $NP(s_1 s_2) = NP(s_1)+NP(s_2)$. Moreover, given convex polygons $A=B+C$,
%Moreover, by Lemma \ref{lemma:submin} we have $(\mathrm{Vert}(A))_{\min}=((\mathrm{Vert}(B))_{\min}+(\mathrm{Vert}(C))_{\min})_{\min}$, so the claim follows with $A=NP(s_1 s_2), B=NP(s_1)$ and $C=NP(s_1)$.
%%% 
%
%%
%%For $(\subseteq)$, given $a=b+c\in \mathrm{Vert}(A)_{\min}$, if $b'\prec b$, then $a'=b'+c\prec a$, against the minimality of $a$. We conclude that $b\in \mathrm{Vert}(B)_{\min}$ and, similarly, that $c\in\mathrm{Vert}(C)_{\min}$. 
%%Moreover, for any $b'\in  \mathrm{Vert}(B)_{\min}$ and $c'\in \mathrm{Vert}(C)_{\min}$, since $b'+c'\in A$, it follows that $b'+c'\prec a$ cannot hold, and we can thus conclude that $a\in (\mathrm{Vert}(B)_{\min}+\mathrm{Vert}(C)_{\min})_{\min}$.
%%For $(\supseteq)$, given $b+c\in (\mathrm{Vert}(B)_{\min}+\mathrm{Vert}(C)_{\min})_{\min}\subseteq A$, suppose there is $a\in \mathrm{Vert}(A)$ such that $a\prec b+c$; 
%%since $A$ is finite, we can find $a_0\in \mathrm{Vert}(A)$ such that $a_0\preceq a\prec b+c$; but using $(\subseteq)$, $a_0\in  (\mathrm{Vert}(B)_{\min}+\mathrm{Vert}(C)_{\min})_{\min}$, so this contradicts the minimality of $b+c$. We conclude that $b+c\in \mathrm{Vert}(A)_{\min}$. 
%%%It is also well known that, for convex polytopes, $\mathrm{Vert}(\C P_1+\C P_2)\subseteq%=
%%%\mathrm{Vert}(\C P_1)+\mathrm{Vert}(\C P_2)$.
%%%Using those, it is easy to see that $NP_{\mathrm{min}}(s_1 s_2) = NP_{\mathrm{min}}(s_1)+NP_{\mathrm{min}}(s_2)$, whence the desired result.
%\end{proof}
%

We now show the existence an algorithm {(which we call $\VN$ -- for ``Viterbi+Newton'')} to compute the minimal polynomial $(\prod_{i=1}^k s_i)_{\min}$ efficiently from $(s_1)_{\min},\dots, (s_k)_{\min}$.

\begin{theorem}\label{thm:vn}%[``Viterbi+Newton'']
Let $k\geq 2$, $s_1,\dots, s_k\in \fps{\TS }{\Sigma}$ be minimal (i.e.~such that $s_i=(s_i)_{\min}$),  %$u=\max_i|s_i|$,
let $d:=\max_i\mathrm{deg}(s_i)$ and $n:=\#\Sigma$.
%Then $(\prod_{i=1}^k s_i)_{\min}$ has \TODO{$\C O(kd^{2n-1})$} monomials and 
There is an algorithm $%s:=
\VN(s_1,\dots, s_k)$ %minimal 
computing $%s:= \VN(s_1,\dots, s_k)$ %such that \NEW{which is minimal for} $%s=
(\prod_{i=1}^k s_i)_{\min}
$ in (expected) time 
{$\C O(nd^{k(2n-1)\max\{2,n\}})$} (when $n\leq k$),
and (deterministic) time
{$\C O((1+k)d^{k(2n-1)})$ (when $n>k$).}
%$\C O(nk^{\max\set{2,n}}d^{\max\set{2,n}(2n-1)})$.
Moreover, $(\prod_{i=1}^k s_i)_{\min}$ has {$\C O((kd)^{2n-1})$} monomials. 
\end{theorem}
\begin{proof}
 We could directly compute the product $\Pi_{i=1}^ks_i$ (in time $\C O(d^{k(2n-1)})$, since Theorem \ref{thm:newton} gives $|s_i|= \C O(d^{2n-1})$) and then extract its minimal Newton polytope, which gives, via Theorem \ref{thm:np}, 
the first bound.
%the claimed bound for $n\leq k$.expected time $\C O(n d^{k(2n-1)\max\{2,n\}})$.
When $n> k$, a speed-up is obtained by using Lemma \ref{lm:vertmin}: since $s_i=(s_i)_{\min}$, $NP(s_i)=NP_{\min}(s_i)$ and the vertices of $NP(s_i)$ are precisely the terms of $s_i$, so
we can compute in time $\C O(d^{k(2n-1)})$ the Minkowski sum $NP(\prod_i s_i)=\sum_iNP(s_i)$ (which, again by Theorem \ref{thm:newton}, has $\C O((kd)^{2n-1})$ terms) and then do a quadratic minimization. This gives the (deterministic) time $\C O(d^{k(2n-1)}+(kd)^{2(2n-1)})$, leading to the claimed bound. 
\end{proof}

Remark that $\trop (\prod_{i=1}^k s_i)_{\min}(x)=\sum_{i=1}^k (s_i)_{\mathrm{min}}(x)$.
By computing products via $\B{VN}$, once we fix the number of variables, the size of (tropical) products like \eqref{eq:mins} 
%(which corresponds to the particular case in which $s_i$ is the monomial $X_i\OV X_i$) 
grows polynomially in both $d$ and $k$.
% which was the announced aim of this Section \ref{sec:geo}.

\begin{example}\label{ex:unfeasible}
Consider 
%the term
{$
M_5=(\lambda x.x \choice{X%_1
}x)(\lambda x.x \choice{X%_2
}x)\dots
(\lambda x.x \choice{X%_n
}x)\ONE$ similar to $M_4$ from Section~\ref{sec:unfeasable}.}
%{\small
%$$M= \underbrace{(\lambda x. x\choice{X}x)\dots (\lambda x. x\choice{X}x)}_{n\text{ times}}\ONE.$$
%}
Each of the $2^{n}$ trajectories $M\RED{\mu}\ONE$ corresponds to a monomial $X^i\overline{X}^{n-i}$ and the sum of all such monomials produces the polynomial {(with the same support as)} %corresponding to
$(X+\overline{X})^{n}$.
{
Observe that,
since all reductions to $\mathsf i$ have degree $n$, the tropical degree $\F d_{\mathsf i}(M_4)=n$, and we can select \emph{all} the $2^n$ reductions to witness this fact.
But this tells nothing wrt the most likely reductions.
By contrast, by the old freshman dream {(Example \ref{ex:freshman})}, the Newton polytope of $(X+\overline{X})^{n}$ only selects the two monomials $X^{n}, \overline{X}^{n}$. 
One can see that, for all assignment of $X$ to probability $p\in[0,1]$, the probability of the most likely reductions is always found within the two selected ones (i.e., $\max_{i=0}^n p^i(1-p)^{n-i}=\max\set{p^n,(1-p)^n}$, as one can easily check).
In other words, the Newton Polytope (in fact, its minimal version) of (the parametric interpretation of) $M_4$, drastically reduces the search space for most likely reductions.
%ones maximal among all the possible reductions, it selects only $2$ most-likely reduction paths.
}
\end{example}

\section{Tropical Intersection Type System}\label{sec:inters_types}
% !TEX root = Tropical_II.tex
{We now put all the work of the previous section in use for the analysis of probabilistic programs of \pPCF:}
we introduce an intersection type system $\TIT$ that associates terms of \pPCF \ with \emph{minimal} all-one polynomials describing their most-likely reductions. After proving soundness and completeness of $\TIT$ wrt the parametric WRS semantics, we describe an algorithm that converges onto the Newton polynomial, thus producing an answer to the inference tasks (I1) and~(I2). 
%{Observe how, \emph{a priori}, having an algorithm doing that in a compositional way is far from being trivial.}

\subsection{The Type System $\TIT$}

Intersection type system have been largely used to capture the termination properties of higher-order programs. \emph{Non-idempotent} (n.~i.) intersection type systems, inspired from linear logic, have been shown to capture \emph{quantitative} properties like e.g.~the number of reduction steps \cite{decarvalho2018, Kesner2017, Kesner2018}. In a probabilistic setting, \cite{EhrhardTassonPagani14} have introduced a n.~i.~intersection type system $\IT$ for probabilistic PCF which precisely captures the probability that a program $M:\Bool$ reduces to, say, $\ONE$ in the following sense: for each reduction $M\stackrel{p}{\to}\ONE$ one can construct a derivation of the form $\vdash_{\IT}^p M:\ONE$ so that
\begin{equation}
\PROB\big(M\RED{}{\ONE}\big)=\sum
\left\{  w(\pi)\ \big \vert \ 
	\begin{matrix}
	 \pi\text{ is a derivation of }\vdash_{\IT}^p M:\ONE 
	\text{ of weight }w(\pi)=p
	\end{matrix}
\right\}.
\end{equation}
By replacing the positive real weights $p\in[0,1]$ in the system $\IT$ with the formal monomials of \pPCF\ one  obtains, in a straightforward way, a type system that produces all the monomials $\mu$ occurring in a reduction $M\RED{\mu}\ONE$. In other words, the type system explores \emph{all} possible reductions of $M$ and produces the associated monomial. This provides a way to fully reconstruct the parametric interpretation $\model{M}^{X_1,\dots, X_n}\in \fps{\NINF}{\mathbb X}$ of a term. 

Our goal, instead, is to design a type system that explores \emph{multiple} reductions at once, excluding those whose probability is dominated, so as to restrict to a finite set of most likely reductions. The goal is thus to capture a finite polynomial corresponding to the tropicalization $\trop\model{M}^{X_1,\dots, X_n}$ (in accordance with Theorem \ref{thm:collapse}).  A natural idea is to consider multiple $\IT$-derivations in parallel. Typically, while in the case of a choice $M\oplus_p N$ a derivation in $\IT$ chooses whether to look at $M$ or $N$ (that is, it chooses between the two reducts of $M\oplus_p N$), in our system the derivation branches so as to consider (and compare) both possible choices.

However, the feasibility of such a system is far from obvious: through reduction, even a term of small size may give rise to an exponentially large number of trajectories, as shown in the example below. Keeping track of all such trajectories through parallel branches in our type derivations can quickly become intractable (even for a computer-assisted formalization).

As already explained, this is where we exploit the minimal Newton polytope: while the rules of $\IT$ produce the probability by progressively multiplying the monomials obtained at each previous step, considering multiple $\IT$-derivations at once requires computing formal polynomials by repeatedly multiplying other formal polynomials produced at previous steps. By using the {results developed in Section \ref{sec:geo}, we are able to} keep the size of such polynomials under control. 

%Let us introduce the type system $\TIT$, that synchronizes multiple $\IT$-derivations using Viterbi-Newton to select the most likely ones. 

%%% EXAMPLE UNFEASABLE
\begin{definition}[{$\TIT$}]
The types of $\TIT$ are defined %, as in $\IT$,
by the grammar
$
a:= n\in\BB N \mid [a,\dots a]\multimap a
$.
A \emph{context} ${\gamma}$ is a function from variables to multisets of $\TIT$-types, {all empty except finitely many. We write it as a finite sequence of variable declarations $x:\gamma(x)$ such that $\gamma(y)=[]$ for all non-declared $y$}. Given contexts ${\gamma},{\delta}$, we indicate as ${\gamma}+{\delta}$ the context obtained by summing their image %variable by variable
variable-wise.
A \emph{pre-judgement} is an expression of the form

\vskip-2mm
{\small
\begin{align*}
M: \big \langle {\gamma}_{j}\vdash^{s_j} a_j  \big \rangle_{ j\in J}
\end{align*}
}

\noindent
and stands for a finite family of judgements ${\gamma}_j\vdash^{s_j}M:a_j$, where $s_j$ indicates a formal polynomial in $\fpp{\TS}{\BB X}$. A pre-judgement %as above
is a \emph{judgement} when the pairs $({\gamma}_j,a_j)_{j\in J}$ are \emph{pairwise distinct} and the polynomials $s_j$ are \emph{minimal}.
Given a pre-judgement as above, we can always produce a judgement 
$M: \SELECT\big \langle {\gamma}_{j}\vdash^{s_j} a_j  \big \rangle_{ j\in J}$ 
by merging equal typings (e.g.~turning $\langle {\gamma}\vdash^s a\ \vert \ {\gamma}\vdash^{s'} a\rangle$ into $ \langle {\gamma}\vdash^{s+s'} a\rangle$) and minimizing {each obtained} polynomial $s$ {via $\VN(s)$}.
%Notice that the $+$ there is the one of $\fpp{\TS}{\BB X}$ and $s+s'$ corresponds to taking the union of their support.
%
% For instance, we will identify the two elements family 
%$\langle \Gamma\vdash^X a, \Gamma\vdash^{\overline X} a\rangle$ with the single one element family
%$\langle \Gamma\vdash^{X+\overline X} a\rangle$.
%
%
%ADD RULES FOR THE NUMERALS AND FOR THE EMPTY FAMILY
%
%CONTEXTS ARE MULTISETS, DEFINITION OF + ETC.
%
%
%where $\Big \langle \Gamma_{j}\vdash^{s_j} a_j  \Big \rangle_{J\to I}$ is a $ I$-indexed family of expressions of the form
%$$
%\Gamma\vdash^s a,
%$$
%where $\Gamma$ is a set declarations of the form $x:[a_1,\dots, a_n]$, $a$ is a type and $s$ is ftm. 
%\TODO{expand informal explanation of rules}
The rules of $\TIT$ are illustrated in Fig.~\ref{fig:tITrules}. 
\end{definition}

{Crucially, we design the rules so as to precisely keep track of the reductions selected by the \emph{minimal} Newton Polytope of a compound term, by combining those of its constituent.}

Except for the rule $(\emptyset)$, that introduces an empty family of judgements, each rule of $\TIT$ results from a corresponding rule of $\IT$ by extending it to families of judgements. %While 
The rules (n), (id), (S), (P), $(\lambda)$ are self-explanatory{: they correspond to rules that create no new parametric reduction.} 
The rules $(\mathrm{ifz})$, $(\oplus)$, $(@)$ and $(\YY)$ deserve some discussion. The rule $(\oplus)$ collects a family of typings of $M$ with polynomials $s_i$, and a family of typings of $N$ with polynomials $s'_j$, to produce a family of typings of $M\choice{X}N$, with polynomials {$s_i\cdot X$} and {$s'_j\cdot \overline X$}, that is successively merged.
{Observe that this precisely corresponds to keeping track of the reductions selected by the minimal Newton Polytope of $M\choice{X}N$, by combining those of $M$ and $N$.}
The rule $(\mathrm{ifz})$ works in a similar way, but uses $\VN(-)$ also \emph{before} merging, since it needs to compute the possibly non-trivial tropical products $s_0\cdot {t_j}$ and $s_{i+1}\cdot {t'_j}$.
The application rule $(@)$ collects, on the one hand, a family of typing ${m_i}\multimap b_i$ of $M$ with polynomials $s_i$, where $m_i=[m_{i1},\dots, m_{ip_i}]$; on the other hand, for each typing ${m_i}\multimap b_i$, and each
type $m_{ij}$ inside $m_i$, it collects a typing $N:m_{ij}$ with polynomials $s'_{ij}$. 
The conclusion of the rule {computes minimal polynomials for the types $b_i$ by calling $\VN(s_i,s'_{i1},\dots,s'_{ip_i})$ to minimise the
tropical multiplication $s_i\cdot\prod_{j}s'_{ij}$.} The rule $(\YY)$ works in a very similar way.
%
%The most involved rules are $(@)$ and the similar rule (FIX). We suppose here that the multiset $m_i$ is of the form $[m_{i1},\dots, m_{i\sharp J_i}]$; moreover, in the $K\to J\to I$-indexed family of left premisses we note the contexts $\Delta_{ijk}$,as well as the types $m_{ijk}$ with only the indices $i,j$ su indicate that they do not actually depend on $k\in K$, that is, that they are constant on the fibers of the map $K\to J$. 
%In order to compute the reduced ftp for the conclusion we exploit the Viterbi-Newton algorithm. 
%
%

%
%Notice also that in the conclusion we obtain a reduced family of derivation starting from one indexed on $\boldsymbol\Gamma(K\to J)\to I$: this means that, for each $i\in I$, we are taking all possible combinations of sums over the indexes $jf(j)$, for $f$ a section of $K\to J$. 
%Algebraically, at the level of weights we are here computing for each $i\in I$ a product of tropical polynomials, corresponding to:
%$$
%s_i+\sum_{j\in J_i}\min_{k\in K_j}s'_{ijk}= s_i+\min_{f\in \boldsymbol\Gamma(K\to J)}\sum_{j\in J_i}s'_{ijf(j)}=
% \min_{f\in \boldsymbol\Gamma(K\to J)}s_i+\sum_{j\in J_i}s'_{ijf(j)}.
%$$
%These rules are then possibly responsible for a large growth in the number of parallel derivations considered at each step. However, Proposition \ref{prop:countftp} ensures us that, for a fixed number of parameters $V$, a derivation of depth $D$ cannot produce more than $\binom{V+D-1}{D-1}\in \C O((V+D)^{D-1})$.

\begin{figure*}
%\begin{subfigure}{0.48\textwidth}
\fbox{
\begin{minipage}{0.99\textwidth}

\centering
\adjustbox{scale=0.7, center}
{
$
\AXC{\phantom{$\Big\langle\Big\rangle$}}
\RL{$\emptyset$}
\UIC{$M: \emptyset$}
\DP
\qquad
\AXC{\phantom{$\Big\langle\Big\rangle$}}
\RL{id}
\UIC{$x: \Big \langle x:[a_i]\vdash^1 a_i\Big \rangle_{i\in I}$}
\DP
\qquad
\AXC{\phantom{$\Big\langle\Big\rangle$}}
\RL{n}
\UIC{$\mathsf n: \Big \langle \vdash^1 n\Big \rangle_{\{\star\}}$}
\DP
\qquad
\AXC{$M:\Big\langle \gamma_i\vdash^{s_i}  n_i\Big\rangle_{i\in I}$}
\RL{S}
\UIC{$\SUCC M:\Big\langle \gamma_i\vdash^{s_i}  n_i+1\Big\rangle_{i\in I}$}
\DP
\qquad
\AXC{$M:\Big\langle \gamma_i\vdash^{s_i}  n_i\Big\rangle_{i\in I}$}
\RL{P}
\UIC{$\PRED M:\Big\langle \gamma_i\vdash^{s_i}  n_i\dotdiv 1\Big\rangle_{i\in I}$}
\DP
$
}

\bigskip

\adjustbox{scale=0.7, center}
{
\AXC{$M: \Big \langle \gamma_0\vdash^{s_0} 0 \ \Big \vert \ 
 \gamma_{i+1}\vdash^{s_{i+1}} i+1 \Big \rangle_{i\in I\subset \BB N}$}
\AXC{$N:   \Big \langle \delta_{j}\vdash^{{t}_{j}} a_j\Big \rangle_{j\in J_0}$}
\AXC{$P:   \Big \langle \delta'_{j}\vdash^{{t'}_{j}} a'_j\Big \rangle_{j\in J_1}$}
\RL{ifz}
\TIC{$\ITE{M}{N}{P}: \SELECT\Big \langle\gamma_0 +\delta_{j }\vdash^{{\VN(s_0\,,\,{t}_j)}} a_j
\ \Big \vert \ 
\gamma_{i+1} +\delta'_{j }\vdash^{{\VN(s_{i+1}\,,\,{t'}_j)}} a'_j\
\Big \rangle_{i\in I, j\in J_0+J_1 }$}
\DP
}

\bigskip

\adjustbox{scale=0.7, center}
{
$
\AXC{$M: \Big \langle \gamma_i\vdash^{s_i}a_i \Big \rangle_{i\in I}$}
\AXC{$N: \Big \langle \gamma_{j}\vdash^{s'_{j}} a_{ j}\Big \rangle_{ j\in J}$}
\RL{$\oplus$}
%\BIC{$M\choice{X}N: \Big \langle
%\Gamma_\ell\vdash^{\VN(s_i\cdot X,s'_j\cdot \overline X)_\ell}a_\ell
%\Big \rangle_{\ell\in \VN(I+J)}
%$}
\BIC{$M\choice{X}N: \SELECT\Big \langle
\gamma_i\vdash^{s_i\cdot X}a_i \ \Big \vert 
\gamma_j\vdash^{s'_j\cdot \overline X}a_j
\Big \rangle_{i\in I,j\in J}$}
\DP
\qquad 
\AXC{$M: \Big \langle\gamma_i,x:m_i\vdash^{s_i} b_i\Big \rangle_{ i\in I}$}
\RL{$\lambda$}
\UIC{$\lambda x.M: \Big \langle \gamma_i\vdash^{s_i}m_i\multimap b_i\Big \rangle_{ i\in I}$}
\DP
$
}

\bigskip

\adjustbox{scale=0.7, center}
{
\AXC{$M: \Big \langle \gamma_i\vdash^{s_i}m_i\multimap b_i\Big \rangle_{ i\in I}$}
\AXC{$N:  \Big \langle \Big \langle \delta_{ij}\vdash^{s'_{ij}} m_{ ij}\Big \rangle_{j\in J_{{i}} }\Big \rangle_{i\in I}$}
\RL{$@$}
\BIC{$MN: \SELECT\Big \langle\gamma_i+\sum_j\delta_{ij }\vdash^{{\VN(s_i\,,\, s'_{i1},\dots,s'_{ip_i})}} b_i\Big \rangle_{i\in I}$}
\DP
\qquad
\AXC{$M: \Big \langle \gamma_i\vdash^{s_i}m_i\multimap b_i\Big \rangle_{i\in I}$}
\AXC{$\YY M:  \Big \langle \Big \langle \delta_{ij}\vdash^{s'_{ij}} m_{ ij}\Big \rangle_{j\in J_i}\Big \rangle_{i\in I}$}
\RL{$\YY$}
\BIC{$\YY M:\SELECT\Big \langle\gamma_i+\sum_j\delta_{ij }\vdash^{{\VN(s_i\,,\, s'_{i1},\dots,s'_{ip_i})}} b_i\Big \rangle_{i\in I}$}
\DP
}

\medskip

\
\end{minipage}
}

\caption{\small Typing Rules of $\TIT$. {In rules $@$ and $\YY$, $m_i = [m_{i1},\dots, m_{ip_i}]$ and $p_i= \mathrm{Card}(J_i)$}.}
\label{fig:tITrules}
\end{figure*}

\begin{figure}
\fbox{
\begin{minipage}{0.99\textwidth}
\centering
\adjustbox{scale=0.55, center}{
\begin{tabular}{c l c c l}
$\pi_0:$ & 
\AXC{$x:\emptyset$}
\AXC{$\ONE:\Big \langle \vdash^{{1}} 1\Big \rangle$}
\BIC{$x\choice{X} {\ONE}: \Big \langle\vdash^{\overline X} 1\Big \rangle$}
\UIC{$\lambda x.x\choice{X} \ONE:\Big \langle \vdash^{\overline X}\emptyset\multimap 1\Big \rangle$}
%\AXC{$\YY (\lambda x.x\choice{X} \ONE):\emptyset$}
\UIC{$\YY (\lambda x.x\choice{X} \ONE): \Big \langle\vdash^{\overline X} 1\Big \rangle$}
\DP
& \hskip1cm  & 
$\pi_{n+1}:$ &
\AXC{$\ONE: \Big \langle\vdash^{{1}} 1\Big \rangle$}
\AXC{$x: \Big \langle x:[1]\vdash^{{1}} 1\Big \rangle$}
\BIC{$x\choice{X} \ONE: \Big \langle\vdash^{\overline X} 1\Big \vert x:[1]\vdash^X 1\Big \rangle$}
\UIC{$\lambda x.x\choice{X} T:\Big \langle \vdash^{\overline X}{\emptyset}\multimap 1\Big\vert \vdash^X[1]\multimap 1\Big \rangle$}
\AXC{$\emptyset\Big\vert\pi_n$}
\noLine
\UIC{$\YY (\lambda x.x\choice{X} \ONE):\Big \langle\emptyset\Big\vert\vdash^{\overline X} 1\Big \rangle$}
\BIC{$\YY (\lambda x.x\choice{X} \ONE): \Big \langle\vdash^{\overline X} 1\Big \rangle$}
\DP
\end{tabular}
}

%
%\resizebox{\textwidth}{!}{
%\begin{tikzpicture}
%\node(a) at (-2,0) 
%{
%\AXC{$x:\emptyset$}
%\AXC{$\ONE:\Big \langle \vdash^{\emptyset} 1\Big \rangle$}
%\BIC{$x\choice{X} v: \Big \langle\vdash^{\overline X} 1\Big \rangle$}
%\UIC{$\lambda x.x\choice{X} \ONE:\Big \langle \vdash^{\overline X}\emptyset\multimap 1\Big \rangle$}
%%\AXC{$\YY (\lambda x.x\choice{X} \ONE):\emptyset$}
%\UIC{$\YY (\lambda x.x\choice{X} \ONE): \Big \langle\vdash^{\overline X} 1\Big \rangle$}
%\DP
%};
%\node[rectangle, draw, thick, gray, rounded corners, minimum width=4cm, minimum height=4cm](a) at (-2,0) {};
%\node(p) at (-2,2.3) {$\Pi_0$};
%
%\node(b) at (5.4,0) {
%\AXC{$\ONE: \Big \langle\vdash^{0} 1\Big \rangle$}
%\AXC{$x: \Big \langle x:[1]\vdash^{0} 1\Big \rangle$}
%\BIC{$x\choice{X} \ONE: \Big \langle\vdash^{\overline X} 1\Big \vert x:[1]\vdash^X 1\Big \rangle$}
%\UIC{$\lambda x.x\choice{X} T:\Big \langle \vdash^{\overline X}\emptyset\multimap 1\Big\vert \vdash^X[1]\multimap 1\Big \rangle$}
%\AXC{$\emptyset\Big\vert\Pi_n$}
%\noLine
%\UIC{$\YY (\lambda x.x\choice{X} \ONE):\Big \langle\emptyset\Big\vert\vdash^{\overline X} 1\Big \rangle$}
%\BIC{$\YY (\lambda x.x\choice{X} \ONE): \Big \langle\vdash^{\overline X} 1\Big \rangle$}
%\DP
%};
%\node[rectangle, draw, thick, gray, rounded corners, minimum width=10.3cm, minimum height=4cm](a) at (5.4,0) {};
%\node(p) at (5.4,2.3) {$\Pi_{n+1}$};
%
%\end{tikzpicture}
%}
\end{minipage}
}
\caption{\small Derivations from Example \ref{example62}.}
\label{fig:example62}
\end{figure}

\begin{figure*}
\fbox{
\begin{minipage}{0.99\textwidth}
\centering
%{\footnotesize
%$$
%\AXC{$x:\Big \langle x:[1]\vdash^0 1\Big \rangle$}
%\AXC{$y:\Big \langle y:[1]\vdash^{0} 1\Big \rangle$}
%\BIC{$x\choice{X} y: \Big \langle x:[1]\vdash^X 1\Big\vert y:[1]\vdash^{\overline X}1  \Big \rangle$}
%\doubleLine
%\UIC{$\lambda xy.x\choice{X} y: \Big \langle\vdash^X [1]\multimap \emptyset \multimap 1\Big\vert \vdash^{\overline X}\emptyset \multimap [1] \multimap1  \Big \rangle$}
%\AXC{$(\ONE\choice{X}\ZERO):\Big \langle\vdash^X [1]\Big\vert \vdash^0 \emptyset \Big \rangle$}
%\BIC{$(\lambda xy.x\choice{X} y)(\ONE\choice{X}\ZERO):\Big \langle\vdash^{2X} \emptyset\multimap 1\Big\vert  \vdash^{\overline X}[1]\multimap 1\Big \rangle$}
%\AXC{$((\ONE\choice{X}\ZERO)\choice{X}\ONE): \Big \langle\emptyset\Big\vert \vdash^{2X,\overline{X}}1\Big \rangle$}
%\BIC{$M: \Big \langle\vdash^{2X,2\overline X} 1\Big \rangle$}
%\DP
%$$
%}
\resizebox{\textwidth}{!}{
$$
\AXC{}
\UIC{$x:\LEFT x:[[a]\multimap a]\vdash^1 [a]\multimap a\RIGHT$}
\AXC{}
\UIC{$x:\LEFT x:[[a]\multimap a]\vdash^1 [a]\multimap a\RIGHT$}
\BIC{$x\choice{X}x:\LEFT  x:[[a]\multimap a]\vdash^{X+\overline{X}} [a]\multimap a \RIGHT $}
\UIC{$\lambda x.x\choice{X}x:\LEFT \vdash^{X+\overline X}[[a]\multimap a]\multimap [a]\multimap a\RIGHT $}
\AXC{}
\UIC{$x:\LEFT x:[[1]\multimap 1]\vdash^1 [1]\multimap 1\RIGHT$}
\AXC{}
\UIC{$x:\LEFT x:[[1]\multimap 1]\vdash^1 [1]\multimap 1\RIGHT$}
\BIC{$x\choice{X}x:\LEFT  x:[[1]\multimap 1]\vdash^{X+\overline{X}} [1]\multimap 1 \RIGHT $}
\UIC{$\lambda x.x\choice{X}x:\LEFT \vdash^{X+\overline X}[[1]\multimap 1]\multimap [1]\multimap 1\RIGHT $}
\BIC{$(\lambda x.x\choice{X}x)\lambda x.x\choice{X}x:\LEFT \vdash^{X^2+\overline{X}^2}[1]\multimap 1 \RIGHT $}
\AXC{}
\UIC{$x:\LEFT x:[1]\vdash^1 1\RIGHT$}
\AXC{}
\UIC{$x:\LEFT x:[1]\vdash^1 1\RIGHT$}
\BIC{$x\choice{X}x:\LEFT  x:[1]\vdash^{X+\overline{X}} 1 \RIGHT $}
\UIC{$\lambda x.x\choice{X}x:\LEFT \vdash^{X+\overline{X}}[1]\multimap 1 \RIGHT $}
\BIC{$(\lambda x.x\choice{X}x)(\lambda x.x\choice{X}x)\lambda x.x\choice{X}x:\LEFT \vdash^{X^3+\overline{X}^3}[1]\multimap 1 \RIGHT$ }
\AXC{}
\UIC{$\ONE:\LEFT \vdash^1 1\RIGHT$}
%\AXC{}
%\UIC{$\ONE:\LEFT \vdash^1 1\RIGHT$}
%\BIC{$\ONE\choice{X}\ONE:\LEFT \vdash^{X+\overline{X}} 1 \RIGHT $}
\BIC{$(\lambda x.x\choice{X}x)(\lambda x.x\choice{X}x)(\lambda x.x\choice{X}x)\ONE:\LEFT \vdash^{X^3+\overline{X}^3} 1 \RIGHT $}
\DP
$$
}
\end{minipage}
}
\caption{\small Derivation from Example \ref{example61}, where $a=[1]\multimap 1$.}
\label{fig:example61}
\end{figure*}

%
%In the following examples, we represent a family \emph{of families} of entities (e.g.~the family $\langle \langle x_{ij}\rangle_{j\in \{0,1\}}\rangle_{i\in \{0,1\}}$), using "$ \vert$" and "$;$", as in 
%$\langle x_{00};x_{01}     \vert x_{10};x_{11}\rangle$.
%
% For example, we represent
% 
%  derivations $\Big \langle\Big \langle \Gamma_{ijk}\vdash^{s_{ij}} a_{ij}  \Big \rangle_{j\in J_i}\Big \rangle_{ i\in I}$
%
% the notation 
%{\footnotesize
%$$
% \Big \langle \dots 
% \Big\vert\dots;\dots ; 
% \Gamma_{ijk_1}\vdash^{s_{ijk_1} a_{ijk_1}},
% \dots,
%  \Gamma_{ijk_{\sharp K_j}}\vdash^{s_{ijk_{\sharp K_j}}} a_{ijk_{\sharp K_j}}
%  ,\dots
%  ;\dots;
%  \dots \Big\vert\dots   \Big \rangle
%$$
%}
%that is, we use "$\Big \vert$" to separate distinct $I$-fibers, "$;$" to separate distinct $J_i$-fibers and "," to separate distinct elements $k\in K_j$. 
%

\begin{example}\label{example62}
In Fig.~\ref{fig:example62} we illustrate a family $\pi_n$ of derivations for the term $M_3$ from Section 2. 
$M_3$ admits arbitrary long reductions, the first one being the most likely. 
$\pi_0$ computes the weight of the most likely derivation $M_3\RED{X}\ONE$; $\pi_{n+1}$ compares the weights from all $\pi_i$, for $i\leq n$ with the weight of the $n+1$th reduction, but ends up selecting in each case only the weight from $\pi_0$, since $(\sum_n X\overline{X}^n)_{\min}= X$.
%(
% (in the last step of $\Pi_{n+1}$ we use the fact that the visible Newton polytope of 
%$
%\{\overline X, X+\overline X\}$ is the singleton $\{\overline X\}
%$
%)).
Hence, all $\pi_n$ correctly compute the minimal polynomial, providing a correct estimation of the tropical degree 
 $\F d_1({M_3})= 1$ of $M_3$. \end{example}

\begin{example}\label{example61}
In Fig.~\ref{fig:example61} we illustrate a derivation for the term $M_4$ {from} {Section~\ref{sec:unfeasable}, choosing} $n=2$ {and $X_1=X_2$}. {It computes} the reduced polynomial $X^3+\overline X^3$, thus correctly estimating $\F d_1({M_4})=3$.
%
%The typing derivation for $M=(\lambda xy.x\choice{X} y)(\ONE\choice{X}\ZERO)((\ONE\choice{X} \ZERO)\choice{X}\ONE)$ shown in Fig.~\ref{fig:example61} captures its tropical polynomial $\F p_M(X,\overline X)=X^2+\overline X^2$.
%
%
%In the last rule we computed $\mathrm{Vit}([2X, \overline X]\Join [\emptyset; 2X, \overline X])$ corresponding to 
%$$
%\min\{2X,\overline X+\min\{2X, \overline X\}\}= \min\{2X, 2X+\overline X, 2\overline X\}=\min\{2X, 2\overline X\}. 
%$$
%The system correctly predicts the minimum weights in the paths leading to $T$ in the normal form of $M$:
%$$
%(\ONE\choice{X}\ZERO)\choice{X}((\ONE\choice{X}\ZERO)\choice{X}\ONE)
%$$
\end{example}

%
%
%
%The following result, proved by induction on derivations, states that derivations in $\TIT$ correspond to families of derivations in $\IT$.
%\begin{proposition}\label{prop:ITtIT}
%$M:\Big \langle\Gamma_{i}\vdash^{s_{i}}a_{i}\Big )_{i\in I}$ is derivable iff for all $k\in K$, 
%$\Gamma_{ijk}\vdash_{\mathbf{IT}}^{s_{ijk}}M:a_{ijk}$ is derivable.
%\end{proposition}

The number of families explored in parallel in a derivation is a parameter controlled by the user. For example, in a term $M\choice{X}N$ we can decide whether to explore both branches or only one, and this choice affects the size of the derivation $|\pi|$, that is, the number of rules.  Instead, the size of the polynomials obtained through the derivation is not controlled by the user. Thanks to the %use of the %Viterbi-Newton 
%$\VN$ algorithm
estimation from Theorem \ref{thm:vn}, though, their size remains polynomial in $|\pi|$. Indeed, the following can be proved by induction on $\pi$:
%To state this result precisely, define the \emph{size} of a judgement $ \gamma=M: \langle\Gamma_i\vdash^{s_i}a_i \rangle_{i\in I}$ as $|\gamma|=\sum_{i\in I}|s_i|$, where $|s_i|$ is the number of monomials in $s_i$. Intuitively, 
%$|\gamma|$ counts the number of distinct reductions accounted for in $\gamma$. 

\begin{proposition}
%For all derivation 
Let $\pi$ be a derivation of $ M: \langle\Gamma_i\vdash^{s_i}a_i \rangle_{i\in I}$.
{Then $\max_i\set{\deg s_i}=O(|\pi|)$.
As a consequence, }
$
|s_i|
\in \C O( |\pi |^{2n-1})$ {for all $i\in I$.} 
\end{proposition}
%As a consequence, if $M:\langle\Gamma\vdash^{s}a \rangle$ contains a single judgement, then the polynomial $s$ has size at most $\C O( |\pi |^{2n-1})$.

{
%A $\TIT$-derivation can be seen as a way of collecting a finite number of $\IT$-derivations, that is, of reductions of the underlying term, and of selecting the most likely ones. 
Given a derivation $\pi$ of $M:\langle \vdash^s\mathsf i\rangle$, by replacing, in each rule ($\mathsf{ifz}$), ($@$) and ($\YY$), the minimized products $(\Pi_i s_i)_{\min}$ obtained by $\B{VN}(s_1,\dots, s_n)$ with the full products $\Pi_i s_i$, we obtain in the end a \emph{larger} polynomial, that we call $\mathrm{traj}^i(\pi)$.
Using Lemma \ref{lm:vertmin} one can easily check that:

%
%
%
%
%
%As Theorem ?? above shows, one $\IT$-derivation of $\vdash^\mu M:\mathsf i$ captures one point in $\mathrm{supp}(\model{M})$, that is, one reduction $M\RED{\mu}\mathsf i$, while 
%If we did \emph{not} use the Viterbi-Newton algorithm to compute products in the rules ($\mathsf{ifz}$), ($@$) and ($\YY$), and we took instead the standard product, 
%
%
%TRAJECTORIES

\begin{proposition}\label{prop:trajmin}
For any derivation $\pi$ of $M:\langle\vdash^{s}\mathsf i\rangle$, $s=(\mathrm{traj}^{\mathsf i}(\pi))_{\min}$.
\end{proposition}
%\begin{proof}
%Should just be the fact that $(st)_{\min}=(s_{\min}\cdot t_{\min})_{\min}$.
%\end{proof}

Intuitively, the polynomial $\mathrm{traj}^{\mathsf i}(\pi)$ tracks \emph{all} reductions of $M$ that $\pi$ explored and out of which it selected the most likely ones. This claim will be justified by the soundness theorem below.
%
%
%Using the properties of the minimal Newton polytope and reasoning similarly as in Theorem~\ref{th:completeness}, we have:
%
%
% be the polynomial obtained at the end. 
%
%
% we can define, by induction on $\pi$, a polynomial $\mathrm{traj}^i(\pi)$ intuitively tracking \emph{all} reductions of $M$ collected (and possibly discarded) by $\pi$: for this it is enough to replace, in the rules ($\mathsf{ifz}$), ($@$) and ($\YY$), the optimized products $\B{VN}(s_1,\dots, s_n)$ obtained by the Viterbi-Newton algorithm, with the standard products $\Pi_i s_i$, and let $\mathrm{traj}^i(\pi)$ be the polynomial obtained at the end. 
%Crucially, while $ \mathrm{Traj}^I(\pi)$ may contain a number of trajectories that is \emph{exponential} in the size of $\pi$ (cf.~Example \ref{example61} and Example \ref{example76} below), the number of monomials in $s$ remains polynomial.
For instance, consider Example \ref{example61}, if we replace, in the derivation of Fig.~\ref{fig:example61}, products computed by $\B{VN}$ by standard products, we obtain $\mathrm{traj}^{\ONE}(\pi)= \sum_{i+j=3}X^i\overline{X}^j$, while $X^3+\overline{X}^3=(\mathrm{traj}^{\ONE}(\pi))_{\min}$.

}

\subsection{Soundness and Completeness of $\TIT$ for the parametric WRS}

%The two examples above show that $\TIT$ is in a sense sound wrt the parametric WRS. 
Intuitively, a $\TIT$-derivation is an optimized way to collect multiple $\IT$-derivations, which, in turn, encode the reductions of the underlying term. More precisely,
%As usual with intersection types, this can be made precise.
%the idea is that, 
for any choice of probabilities $p\in [0,1]^{\mathbb X}$, for any derivation of 
$M: \langle\Gamma_{i}\vdash^{s_{i}}a_{i} \rangle_{i\in I}$, for each $i\in I$ and for each monomial $\mu$ in $s_i$, there is a %exists a corresponding 
derivation of 
$\Gamma_{i}\vdash^{{p^\mu}}M[X:=p]:a_{i}$ in $\IT$, where
% {(the already mentioned n.i.\ system introduced in \cite{Pagani2018})}, 
%where $M[X:=p]$ is the pPCF term obtained by replacing the parameters $X_i$ by the $p_i$, 
{$p^\mu=\Pi_{V\in\BB X} p_V^{\mu(V)}$ (cfr.\ Section \ref{subsec:fps}) is the probability of the reduction of $M[X:=p]$ (to some normal form) corresponding to $\mu$.
This suggests then that soundness and completeness can be \emph{lifted} from $\IT$ \cite[Lemma 20 and Equation 11]{EhrhardTassonPagani14} to $\TIT$.

The fundamental ingredient is the notion of a $\TIT$-derivation \emph{refining} a \pPCF-derivation.
%(this is also adapted from \cite[Lemma 20]{EhrhardTassonPagani14}).
{First, given a simple type $A$, a $\TIT$-type $a$, a simple context $\Gamma$ and a $\TIT$-context $\gamma$, we say that $(\gamma,a)$ refines $(\Gamma,A)$ whenever $a\in \model{A}$ (so e.g.~$\ZERO,\ONE$ refine $\Bool$ and $m\multimap b$ refines $A\to B$ whenever $m\in \ !\model{A}$ and $b\in \model{B}$) and $\gamma(x)\in!\model{\Gamma(x)}$ for all $x$ declared in $\Gamma$, and $\gamma(x)=[]$ otherwise.
}
Now, given a \pPCF-derivation $\Pi$ of $\Gamma\vdash M:A$ %, a $\TIT$-type $a\in \model A$ and a $\TIT$-context $\gamma$ such that $\gamma(x)\in!\model{\Gamma(x)}$ for all $x\in\mathrm{dom}(\Gamma)$ and $\gamma_k(x)=[]$ otherwise 
and a $\TIT$-derivation $\pi$ of $M: \big \langle \gamma_k\vdash^{s_k} a_k \big \rangle_k$ {with $(\gamma,a)$ refining $(\Gamma,A)$}, $\pi$ refines $\Pi$ when, intuitively, its rules match the corresponding rules in $\Pi$: for instance, if $\Pi$ ends with the abstraction rule of conclusion $\Gamma\vdash M:A\to B$, $\pi$ ends with the rule $(\lambda)$ of conclusion
$M:\langle\gamma_i\vdash M:m_i\multimap b_i\rangle_{i\in I}$, 
%\davide{where $\gamma_i\in\ !\model{\Gamma}$, $m_i\in !\model{A}$ and $b_i\in \model{B}$}{
where $(\gamma_i,m_i\multimap b_i)$ refines $(\Gamma,A\to B)$}.
Two exceptions are the cast rule (which is skipped%, as we did not include it in $\TIT$
) and the $\YY$-rule. For the latter, 
the intuition is that a $\TIT$-derivation $\pi$ for $\YY M$ actually determines one possible \emph{finite unfolding} of $\YY M$. More precisely, $\pi$ must end with a \emph{cluster} of $h$ consecutive $\YY$-rules of which the last one {has} premise of type $[]\multimap b_j$. By replacing each such $(\YY)$ with $(@)$ one obtains then a $\TIT$-derivation $\mathbf{unfold}(\pi)$ in which $\YY M$ has been replaced by the unfolded term $M^{(h)} y$ ($y$ is a fresh variable). Correspondingly, $\Pi$ can also be transformed into a \pPCF-derivation $\mathbf{unfold}^h(\Pi)$ of $M^{(h)}y$ by replacing the fixpoint rule with a cluster of $h$ application rules; we say that $\pi$ refines $\Pi$ when $\mathbf{unfold}(\pi)$ refines $\mathbf{unfold}^h(\Pi)$. For the definition to make sense, a transfinite induction is required, by treating the $\YY$-rule as a $\omega$-rule.

By adapting the argument for $\IT$ \cite[Lemma 20]{EhrhardTassonPagani14}, we obtain (by induction on a \pPCF-derivation):

\begin{theorem}[Soundness of $\TIT$ wrt WRS]\label{thm:soundwrtwrs2}
    Let $\Pi$ a \pPCF-derivation of $\Gamma\vdash M:A$ %    \davide{, $a\in \model A$ a $\TIT$-type and $\gamma$ a $\TIT$-context such that $\gamma(x)\in!\model{\Gamma(x)}$ for all $x\in\mathrm{dom}(\Gamma)$ and $\gamma_k(x)=[]$ otherwise. %for all $x\notin\mathrm{dom}(\Gamma)$.
%    }
   {and $(\gamma,a)$ refining $(\Gamma,A)$.}
    For all $\TIT$-derivation $\pi$ of $M:  \langle \gamma\vdash^{s} a  \rangle$ that refines $\Pi$, $\mathrm{supp}(s) 
    \subseteq
    \mathrm{supp}(\model{M}^{\mathbb X}_{\gamma, a})$.
    %Here for a derivation $\pi$ with polynomial $s_\pi$, we set $NP(\pi):=\mathrm{supp}(s_\pi)$.
    \end{theorem}

It follows that, in particular, the minimal polynomials produced by typing derivations for a closed ground-type term $M$ produce an over-approximation of the tropicalisation of $M$:
\begin{corollary}\label{thm:correct}
Let $\Pi$ derive in \pPCF $\vdash M:\NAT$, $n\in \mathbb N$. For all $\pi$ derivation in $\TIT$ of $M: \langle\vdash^{s}n \rangle$ that refines $\Pi$, we have $\trop\model{M}_{\mathsf n}(q)\leq s^!(q)$ for all $q\in\mathbb T^{2n}$ ($n$ being the number of parameters).
\end{corollary}
\begin{proof}
It follows from taking $\Gamma=\emptyset$, $A=\NAT$ in the soundness, which gives: for all $n\in\N$ and $q\in\mathbb T^{2n}$,
$\trop\model{ M}^{\mathbb X}_{\mathsf n}(q)
\leq
\inf\set{\mu\cdot q \mid \mu\in\mathrm{supp}(s) \textit{ for some $\pi$ of } M:  \langle \vdash^{s} n   \rangle}$.
\end{proof}

In fact, we can say more: $\TIT$ captures at least the minimal part of the parametric WRS.
Reasoning in a similar way as for Theorem~\ref{thm:soundwrtwrs2}, one can show:

\begin{theorem}[Completeness of $\TIT$ wrt \emph{minimal} WRS]\label{thm:completeness2}
Let $\Pi$ be a \pPCF-derivation of $\Gamma\vdash M:A$ and $(\gamma,a)$ {refining $(\Gamma,A)$}. %for all $x\in\mathrm{dom}(\Gamma)$ and $\gamma(x)=[]$ for all $x\notin\mathrm{dom}(\Gamma)$.
For all $\mu\in\mathrm{supp}(\model{M}^{\mathbb X}_{\gamma, a})_{\mathrm{min}}$ there is a $\TIT$-derivation $\pi$ of $M:  \langle \gamma\vdash^{s} a  \rangle$ that refines $\Pi$ and with $\mu\in\mathrm{supp}(s)$.
\end{theorem}

Notice that, together, Theorems \ref{thm:soundwrtwrs2} and \ref{thm:completeness2} give the equality
\begin{equation}
\trop\model{ M}_{\mathsf n}(q)
=
\inf\set{\mu\cdot q \mid \mu\in\mathrm{supp}(s) \textit{ for some $\pi$ of } M:  \langle \vdash^{s} n   \rangle},
\end{equation}
showing that the taf interpreting $M$ can be in principle \emph{approximated} {via} larger and larger %$\TIT$-
derivations, 
%\davide{Actually, as the $\inf$ above is always a min (by Corollary \ref{thm:collapse}), we can hope to construct \emph{a single} $\TIT$-derivation that reaches it. This is what we do below.}
{which we construct below. Actually, coherently with Corollary~\ref{thm:collapse}, we show in Theorem~\ref{thm:stabilizza} that there is a \emph{single} %$\TIT$-
derivation reaching the $\inf$, which is thus a (non computable, in general) min.}

%
%
%\begin{remark}
%Remember the definition of tropicalisation and that in general $((\model{\_}^{\mathbb X}_a)_{\mathrm{min}})^!=\trop\model{\_}^{\mathbb X}_a$ as discussed at the end of Section 6.2.
%Therefore, Theorem~\ref{thm:completeness} implies that we actually have an equality in the inequality in the proof of Theorem~\ref{thm:correct}.
%Notice that the search space in the $\inf$ above is still potentially infinite.
%However, now we can hope to approximate it via $\TIT$, in order to give an approximative answer to (I1), (I2): this is what we do in the remainder of the paper.
%Actually, via a further subtle argument for the case $\Gamma=\emptyset$ and $A=\NAT$ we will show that the $\inf$ above is always a min (this is the second statement of Theorem \ref{thm:stabilizza}). Of course, a non computable one.
%\end{remark}
%

%
%Finally, thanks to Theorem \ref{thm:collapse} theorem , which states that $\model{M}(\vec X)_n$ is equivalent to a tropical polynomial, we obtain the following:
%\begin{theorem}
%For all closed term $M$ and $n\in \mathbb N$, there exists a typing derivation of $M:\Big \langle\vdash^{s_{i}}n\Big \rangle_{ I}$ such that $\model{M}(\vec X)_n= \min_Is_i$.
%\end{theorem}

% !TEX root = Tropical_II.tex
\subsection{Constructing a Solution to Tasks (I1) and (I2)}\label{subsec:inference_algo}

We now exploit the type system $\TIT$ to design an algorithm that, by reconstructing the Newton polynomial of a term, converges onto a correct solution to the inference tasks (I1) and (I2). % We will not provide a precise implementation, as this would go beyond the scope of this paper, but we illustrate, through a couple of examples, the main potential challenges. 
Given a \pPCF-term $M:\Bool$ with parameters $\vec X$ and a chosen output value $i\in \{0, 1\}$, the goal is to produce in output a finite set $\TT{selected\_ trajs}^i(M)$ of tuples $(\mu,w_\mu, S)$, where $w_\mu\in\{0,1\}^*$ is a sequence describing a candidate most likely trajectory $M\RED{\mu}\mathsf i$ and $S\subseteq [0,+\infty]^n$ is the set of $(-\log)$-values of the parameters $\vec X$ that make $\mu$ minimum across \emph{all} $\mu\in\TT{selected\_ trajs}^i(M)$. 
The algorithm is divided in two phases: the construction of a suitable $\TIT$-derivation of $M:\langle\vdash^s \mathsf i\rangle$, and the extraction of the set $\TT{selected\_ trajs}^i(M)$ from $\pi$, described in the following two subsections.

\subsubsection{Constructing a Stable Derivation}

The main idea for the construction of a derivation exploring the reductions of $M$ was already suggested in Example \ref{example62}: we progressively make $\pi$ grow so as to explore \emph{more and more} reductions, until the produced polynomial $s$ \emph{stabilizes}: anyhow the derivation may still grow, the polynomial $s$ does not change.
% As shown below, at this point $s$ will indeed coincide with $NP_{\min}^{\mathsf i}(M)$. 

%
%the produced polynomial $s$ thus progressively grows, but the whole process terminates when, thanks to the finiteness of $NP^i(M)$, $s$ finally \emph{stabilizes}: adding further $\IT$-derivations no more makes it grow. 

As a first simple example, suppose $M$ is a closed first-order term built using only $\ZERO,\ONE,\oplus_X$ and $\ITE{-}{-}{-}$, so that its typing in \pPCF\  only contains the type $\Bool$. 
It is not difficult then to construct, by induction on $M$, a derivation $\pi:\langle \vdash^{s_0} \ZERO\ \vert \ \vdash^{s_1}\ONE\rangle$ that explores \emph{all} trajectories of $M$ (i.e.~for which $\mathrm{supp}(\mathrm{traj}^{\mathsf i}(\pi))=\mathrm{supp}(\model{M}_i)$). Notice that the derivation has size linear in the size of $M$, since the index sets $I$ for any judgment can have at most two elements (the only two refinements $\ZERO,\ONE$ of the type $\Bool$).
In this simplified setting, as illustrated in the example below, the construction of $s$ can then be seen as a parametric variant of the Viterbi algorithm.

\begin{example}\label{example76}
Consider closed terms $S_0,S^1_i,\dots, S^n_i :\Bool$, $i=0,1$, $n>0$ and let $P=P_n:\Bool$,  
where $P_0=S_0$ and 
$P_{j+1}=\ITE{P_j}{S^{j+1}_0}{S^{j+1}_1}$.
 The term $P:\Bool$ encodes the graphical model in Fig.~\ref{fig:dag73}. This model has $2^n$ trajectories leading to either $S^n_0,S^n_1$. There is a derivation $\pi_n$ of $P:\langle \vdash^{s_0}\ZERO\ \mid\ \vdash^{s_1}\ONE\rangle$ (of size linear in $P$) capturing all such trajectories, made of a chain of $\mathsf{ifz}$-rules (the derivations $\pi_{j+1}$ for the terms $P_j$ are illustrated in Fig.~\ref{fig:td73}, supposing given derivations for $S_0$ and the $S^j_i$). 
% $$
% \AXC{$P_j:\langle \vdash^{s^j_0}\ZERO\ \mid\  \vdash^{s^j_1}\ONE\rangle$}
% \AXC{$S^{j+1}_0:\langle \vdash^{u^{j+1}_0}\ZERO\ \mid\  \vdash^{u^{j+1}_1}\ONE\rangle$}
% \AXC{$S^{j+1}_1:\langle \vdash^{v^{j+1}_0}\ZERO\ \mid\  \vdash^{v^{j+1}_1}\ONE\rangle$}
%\RL{$\mathsf{ifz}$}
%\TIC{$P_{j+1}:\big\langle \vdash^{\B{VN}(s^j_0\cdot u^{j+1}_0+s^j_1\cdot v^{j+1}_0)} \ZERO\ \big \vert \ 
% \vdash^{\B{VN}(s^j_0\cdot u^{j+1}_1+s^j_1\cdot v^{j+1}_1)} \ONE\big \rangle$}
%\DP
% $$
The polynomials $s_0:=s^n_0$ and $s_1:=s^n_1$ are constructed by choosing, at each step, the best way to expand either $s^j_0$ or $s^j_1$ with the monomials $u^{j}_{l},v^j_l$ coming from either $S^{j}_0$ or $S^{j}_1$. If the $u^{j}_l,v^j_l$ were scalars, this would indeed correspond to computing a Viterbi sequence for the model in Fig.~\ref{fig:dag73}.
\end{example}

\begin{figure}
\begin{subfigure}{0.3\textwidth}
\fbox{
\begin{minipage}{.99\textwidth}
\adjustbox{scale=0.6,center}{
\begin{tikzpicture}
\node[draw, circle, fill=gray!20, minimum width=0.8cm](s0) at (-0.2,-0.75) {$S_0$};
\node[draw, circle, fill=gray!20, minimum width=0.8cm](s1) at (1,0) {$S^1_0$};
\node[draw, circle, fill=gray!20, minimum width=0.8cm](o1) at (1,-1.5) {$S^1_1$};
\node[draw, circle, fill=gray!20, minimum width=0.8cm](s2) at (2.2,0) {$S^2_0$};
\node[draw, circle, fill=gray!20, minimum width=0.8cm](o2) at (2.2,-1.5) {$S^2_1$};
\node[ minimum width=0.8cm](sf3) at (3.4,0) {};
\node[ minimum width=0.8cm](of3) at (3.4,-1.5) {};
\node[ minimum width=0.8cm](sf4) at (3.6,0) {};
\node[ minimum width=0.8cm](of4) at (3.6,-1.5) {};
\node[draw, circle, fill=gray!20, minimum width=0.8cm](s3) at (5,0) {$S^n_0$};
\node[draw, circle, fill=gray!20, minimum width=0.8cm](o3) at (5,-1.5) {$S^n_1$};
\node(o5) at (3.6,-0.75) {$\dots$};

\node(o5) at (4.2,-2.2) {\ };

\draw[->] (s0) to (s1);
\draw[->] (s0) to (o1);
\draw[->] (s1) to (s2);
\draw[->] (s1) to (o2);
\draw[->] (o1) to (s2);
\draw[->] (o1) to (o2);
\draw[->] (s2) to (sf3);
\draw[->] (s2) to (of3);
\draw[->] (o2) to (sf3);
\draw[->] (o2) to (of3);

\draw[->] (sf4) to (s3);
\draw[->] (sf4) to (o3);
\draw[->] (of4) to (s3);
\draw[->] (of4) to (o3);

\end{tikzpicture}
}
\end{minipage}
}

\caption{Bayesian Network. }
\label{fig:dag73}
\end{subfigure} \hspace{3mm}
\begin{subfigure}{0.65\textwidth}
\fbox{
\begin{minipage}{\textwidth}
\vskip2mm

\adjustbox{scale=0.75,center}{
$
 \AXC{$\pi_j$}
 \noLine
 \UIC{$P_j:\langle \vdash^{s^j_0}\ZERO\ \mid\  \vdash^{s^j_1}\ONE\rangle$}
 \AXC{$\pi_{S,j+1,0}$}\noLine\UIC{$S^{j+1}_0:\langle \vdash^{u^{j+1}_0}\ZERO\ \mid\  \vdash^{u^{j+1}_1}\ONE\rangle$}
  \AXC{$\pi_{S,j+1,1}$}\noLine\UIC{$S^{j+1}_1:\langle \vdash^{v^{j+1}_0}\ZERO\ \mid\  \vdash^{v^{j+1}_1}\ONE\rangle$}
\RL{$\mathsf{ifz}$}
\TIC{$P_{j+1}:\big\langle \vdash^{\B{VN}(s^j_0\cdot u^{j+1}_0+s^j_1\cdot v^{j+1}_0)} \ZERO\ \big \vert \ 
 \vdash^{\B{VN}(s^j_0\cdot u^{j+1}_1+s^j_1\cdot v^{j+1}_1)} \ONE\big \rangle$}
\DP
 $
 }
 
 \vskip1.4mm
 
 \
 \end{minipage}
}
\caption{Typing derivation $\pi_{j+1}$ for $P_{j+1}$. }
\label{fig:td73}
\end{subfigure}
\caption{Bayesian Network and typing derivation for Example \ref{example76}.}
\end{figure}

While in the case above it was possible to generate a single derivation encompassing all trajectories of the term, this is not possible in general, because the term may have \emph{infinitely} many trajectories, which is typically the case if the term contains a fixpoint.
Another problem is that, if the term has some \emph{higher-order} applications $PQ$, the number of possible intersection types $m\multimap b$ refining the type $A\to B$ of $P$ is also infinite, since $m\in \ !\model{A}$ may be arbitrarily long.
%) or, in any case, too large to be explored in practice.

This is why we construct our derivation incrementally. 
For two fixed parameters $n,p\in \mathbb N$, we can always construct a derivation $\pi_{n,p}^{\mathsf i}(M)$ of $M:\langle \vdash^s \mathsf i\rangle$ that collects \emph{all} $\IT$-derivations of $\vdash M:\mathsf i$ in which 
the rule $(\YY)$ is used at most $n$ times and, for all intersection types $m\multimap b$, $m$ has at most $p$ elements and does not contain atoms $\mathsf i> p$.
This corresponds to looking at reductions of $M$ in which fixpoints are \emph{unfolded at most $n$ times} and, and in each $\beta$-reduction $(\lambda x.P)Q$, the term $Q$ may be \emph{duplicated at most $p$ times}}.
In our algorithm, we initially set $n=p=1$ and then progressively increment either $n$ or $p$ until reaching stability. 

\begin{remark}[complexity of $\pi_{n,p}^{\mathsf i}(M)$]
In general, $\pi_{n,p}^{\mathsf i}(M)$ may not be constructible efficiently (i.e.~in polynomial time wrt $|M|$): the rules $(@)$ and $(\YY)$ have arity at most $ p+1$, while the corresponding operations on terms are at most binary, yielding a derivation much \emph{larger} than the corresponding \pPCF-typing, as well as a \emph{longer} computation time due to $p+1$-ary products (as the time for computing multiplications - optimized or not - grows exponentially in the number of factors, cf.~Theorem \ref{thm:vn}); moreover, also the index sets $I$ may grow very large, as they vary over all  
refinings $[a_1,\dots, a_k]\multimap b$, with $k\leq p$, of the corresponding types $A\to B$.
However, when $p=1$, $\pi_{n,1}^{\mathsf i}(M)$, corresponding to looking to \emph{affine} reductions (i.e.~in which terms may be deleted but not duplicated) with at most $n$ $\YY$-unfolding, only requires \emph{binary} multiplications $\B{VN}(s_1 s_2)$ (computed in time $\C O(d^{4n+2})$ by Theorem \ref{thm:vn}) and has size $\mathcal O(n|M| 3^t)$, where $t$ is the maximum size of a simple type in $M$. This bound is obtained by unfolding the subterms $\YY P$ as $P^{i}y$ (so that the sum of all such $i$ does not exceed $n$), yielding a term of size $\C O(n|M|)$, observing that the arity of $(@),(\YY)$ is at most $2$, and noticing that the index sets $I$ cannot exceed the number of affine refinings of the corresponding type (which are obtained by replacing each atom $\Bool,\NAT$ by any of $[],[\ZERO],[\ONE]$), which is bounded by $3^t$. 
\end{remark}

%If we can reach stability by only incrementing $p$ (i.e.~if all most likely reductions are affine), then we are able to construct each derivation $\pi_{n,1}^{\mathsf i}(M)$ in time linear in both $n$ and the size of $M$ (although exponential in the size of the largest type in the derivation of $M$).

\begin{example}\label{example78}
Let us illustrate our algorithm for the term $M_2$ from Example \ref{example2}.
%Fig.~\ref{fig:redutree} illustrates the reduction tree of $M_2$. 
Let $\pi_0$ be a derivation of $ND: 
\langle
\vdash^{u_0}
\ZERO \ \vert  \
 \vdash^{u_1}\ONE\rangle$, capturing the four reductions of $ND$ to either $\ZERO$ or $\ONE$, where, letting $X^0=X$ and $X^1=\overline X$, $u_i=X_0X_{1}^i+\overline{X_0}X_2^i$. One can similarly construct a derivation $\pi_1$ of  
 $B:\langle \vdash^{v_i} []\multimap [i]\multimap \ONE \rangle_{b\in \{0,1\}}$, capturing the two reductions $(\YY B)\mathsf i\RED{}B(\YY B)\mathsf i \RED{}\ONE$ that unfold $\YY$ only once, where $v_{i}=\overline{X_{3+i}}$. 
We construct $\pi_{1,1}^{\ONE}(M)$ by combining $\pi_0$ and $\pi_1$ via the $\YY$-rule yielding $M_2:\langle \vdash^s \ONE\rangle$, where
$s=u_0v_0+u_1v_1$ captures all reductions of $M_2$ with one $\YY$-unfolding.
To construct $\pi_{n,2}^{\ONE}(M)$, we must construct a derivation $\pi_2$ of
 $B:\langle \vdash^{t_{ij}} [[j]\multimap \ONE]\multimap [i]\multimap \ONE
\rangle_{i,j\in \{0,1\}}$, where $t_{ij}=X_{3+i}X_{1+i}^j$, 
which tracks all pairs of reductions $(\YY B)\mathsf i\RED{}{}B(\YY B)\mathsf i \RED{}(\YY B)(N\mathsf i)$, of weight $X_{3+i}$, and 
$N\mathsf i\RED{}\mathsf j$, of weight $X_{1+i}^j$. The derivation $\pi_{2,1}^{\ONE}(M)$ of 
 of $M_2:\langle \vdash^s \ONE\rangle$, is illustrated in Fig.~\ref{fig:m2piall}, where now
$s=u_0v_0+u_1v_1+u_0t_{00}v_0+u_0t_{01}v_1+u_1t_{10}v_0+u_1t_{11}v_1$ captures all reductions of $M_2$ with at most 2 $\YY$-unfoldings.
With $n=3$, we can iterate the process adding a new $\YY$-rule. Intuitively, this should lead to add to $s$ all monomials $w_{abc}:=u_{a}t_{ab}t_{bc}v_{c}$, for $a,b,c\in\{0,1\}$, corresponding to 3 iterations.
However, each $w_{abc}$ is dominated by one monomial in $s$: we have that either $a=b,b=c$ or $a=c$; if $a=b$ holds, then $w_{abc}$ is dominated by $u_at_{ac}v_c$, and similarly for the other cases. Hence $\pi_{3,1}^{\ONE}(M)$ yields \emph{the same} polynomial as $\pi_{2,1}^{\ONE}(M)$.
Moreover, further incrementing either $n$ or $p$ does not make the polynomial change. In fact, we can easily see that the minimal polynomial $s$ produced by $\pi_{2,1}^{\ONE}(M)$ coincides $NP_{\min}^{\ONE}(M_2)$: an arbitrary trajectory of $M_2$ 
%(see again Fig.~\ref{fig:redutree}) 
yields a monomial of the form $w_{a_1\dots a_{n+1}}=u_{a_1}t_{a_1a_2}\dots t_{a_{n}a_{n+1}}v_{a_{n+1}}$, and if $n\geq 2$, then the monomial 
$w_{a_1\dots a_{n+1}}$ is dominated by some monomial in $s$. 
As anticipated in Section 3, it follows then that $\F d^{\ONE}(M_2)=\mathrm{deg}(s)=5$.

%
%\begin{figure}
%\fbox{
%\begin{minipage}{.95\textwidth}
%\centering
%\adjustbox{scale=0.65}{
%$
%\AXC{$x:\langle x:[\ZERO]\vdash^{[]} \ZERO \ \vert \ x:[\ZERO]\vdash^{[]} \ONE\rangle$}
%\AXC{$M:\langle \vdash^{X_1} \ZERO \ \vert \ \vdash^{\overline{X_1}} \ONE\rangle$}
%\AXC{$N:\langle \vdash^{X_2} \ZERO \ \vert \ \vdash^{\overline{X_2}} \ONE\rangle$}
%\RL{$\mathsf{ifz}$}
%\TIC{$\ITE{x}{M}{N}:
%\big\langle
% x:[\ZERO]\vdash^{X_1} \ZERO \ \big\vert  \
%  x:[\ZERO]\vdash^{\overline{X_1}} \ONE \ \big\vert  \
%    x:[\ONE]\vdash^{X_2} \ZERO \ \big\vert \ 
%      x:[\ZERO]\vdash^{\overline{X_2}} \ONE
%\big\rangle
%$}
%\RL{$\lambda$}
%\UIC{$N:
%\big\langle
%\vdash^{X_1}
% [\ZERO] \multimap\ZERO \ \big\vert  \
% \vdash^{\overline{X_1}}[\ZERO] \multimap\ONE \ \big\vert  \
%    \vdash^{X_2}[\ONE]\multimap \ZERO \ \big\vert \ 
%      \vdash^{\overline{X_2}}[\ZERO] \multimap\ONE
%\big\rangle
%$}
%\AXC{$D:\langle \vdash^{X_0}\ZERO\ \vert \ \vdash^{\overline{X_0}}\ONE\rangle$}
%\RL{$@$}
%\BIC{$
%ND: 
%\big\langle
%\vdash^{X_0X_1+\overline{X_0}X_2}
%\ZERO \ \big\vert  \
% \vdash^{\overline{X_0}\overline{X_1}+\overline{X_0}\overline{X_2}}\ONE\big\rangle
%$}
%\DP
%$
%}
%\end{minipage}
%}
%\caption{Typing of $ND$.}
%\label{fig:m2pi0}
%\end{figure}
\begin{figure}
%\fbox{
%\begin{minipage}{.99\textwidth}
%\begin{subfigure}{0.16\textwidth}
%\fbox
%{
%\begin{minipage}{.95\textwidth}
%\adjustbox{scale=0.6,center}{
%$
%\begin{tikzcd}[ampersand replacement=\&,cramped]
%	\& D \\
%	{N\ZERO} \&\& {N\ONE} \\
%	{O\ZERO} \&\& {O\ONE} \\
%	\& \ONE
%	\arrow["{X_0}"'{pos=0.2}, from=1-2, to=2-1]
%	\arrow["{\OV X_0}"{pos=0.2}, from=1-2, to=2-3]
%	\arrow["{\OV X_1}"{pos=0.467}, curve={height=-6pt}, from=2-1, to=3-1]
%	\arrow["{X_1}"{pos=0.12}, curve={height=-6pt}, from=2-1, to=3-3]
%	\arrow["{X_2}"'{pos=0.12}, curve={height=6pt}, from=2-3, to=3-1]
%	\arrow["{\OV X_2}"'{pos=0.469}, curve={height=6pt}, from=2-3, to=3-3]
%	\arrow["{X_3}"{pos=0.4}, curve={height=-6pt}, from=3-1, to=2-1]
%	\arrow["{\OV X_3}"'{pos=0.2}, from=3-1, to=4-2]
%	\arrow["{X_4}"'{pos=0.4}, curve={height=6pt}, from=3-3, to=2-3]
%	\arrow["{\OV X_4}"{pos=0.2}, from=3-3, to=4-2]
%\end{tikzcd}
%$
%}
% \end{minipage}
%}
%\caption{Reduction tree}
%\label{fig:redutree}
%\end{subfigure}
%\ \ \ 
%\begin{subfigure}{.81\textwidth}
\fbox
{
\begin{minipage}{.99\textwidth}
%\vskip.34cm
\begin{center}
\adjustbox{scale=0.72}{
$
\AXC{$\pi_1+\pi_2$}
\noLine
\UIC{$B:\big\langle \vdash^{v_i} []\multimap [i]\multimap \ONE \ \big\vert 
\ \vdash^{t_{ij}} [[j]\multimap \ONE]\multimap [i]\multimap \ONE
\big\rangle_{i,j\in \{0,1\}}
 $}
\AXC{$\pi_1$}
\noLine
\UIC{$B:\big\langle \vdash^{v_j} []\multimap [j]\multimap \ONE 
\big\rangle_{j\in \{0,1\}}
 $}
 \RL{$\YY$}
 \UIC{$
 \YY B: \big\langle \emptyset \ \big\vert \ \vdash^{v_j}[ j]\multimap \ONE\big\rangle_{j\in\{\ZERO,\ONE\}}
 $}
 \RL{$\YY$}
 \BIC{
 $
 \YY B: 
 \big\langle
 \vdash^{v_i+t_{i0}v_0+t_{i1}v_1}
 [\mathsf i]\multimap \ONE \big\rangle_{i\in\{0,1\}}
 $
 }
 \AXC{$\pi_0$}
 \noLine
 \UIC
 {$
ND: 
\big\langle
\vdash^{u_0}
\ZERO \ \big\vert  \
 \vdash^{u_1}\ONE\big\rangle
$}
\RL{$@$}
\BIC{
$
M_2:\big\langle \vdash^{u_0v_0+u_1v_1+u_0t_{00}v_0+u_0t_{01}v_1+u_1t_{10}v_0+u_1t_{11}v_1} \ONE\big\rangle
$
}
 \DP
 $
 }
 \end{center}
  \end{minipage}
}
\caption{Derivation ${\pi}^{\mathrm{stab}}=\pi_{2,1}^{\ONE}(M_2)$, where 
%, where $s=u_0v_0+u_1v_1+u_0t_{00}v_0+u_1t_{10}v_0+u_1t_{10}v_0+u_1t_{11}v_1$.
$u_i=X_0X_{1}^i+\overline{X_0}X_2^i$, $v_i=\overline{X}_{3+i}$ and $t_{ij}=X_{3+i}\overline{X}_{1+i}^j$.
}
\label{fig:m2piall}
%\label{fig:redutree}
%\end{subfigure}
%%\end{minipage}
%%}
%
%\caption{Reduction and typing of $M_2$.}
\end{figure}

\end{example}

The result below shows that the algorithm above {stabilizes} onto the minimal Newton polynomial.

\begin{theorem}\label{thm:stabilizza}
%For all $(n,\tau)$, the derivation $\pi=\pi_{n,\tau}^{\mathsf i}(M)$ proves $M:\langle\vdash^{s}\mathsf i\rangle$, where $s=(\sum_{\mu\in\mathrm{Traj}^{\mathsf i}(\pi)}\mu)_{\mathrm{min}}$.
For any \pPCF-term $M:\Bool$ and $\mathsf i\in \{\ZERO,\ONE\}$, there exists $n,p\in \mathbb N$ such that the derivation $\pi^{\mathrm{stab}}:=\pi_{n,p}^{\mathsf i}(M)$ is stable and proves $M:\langle\vdash^{s}\mathsf i\rangle$, where $s=(\mathsf t\model{M}_{\mathsf i}^{\BB X})_{\min}=NP_{\min}^{\mathsf i}(M)$.
\end{theorem}
\begin{proof}
 By Corollary \ref{thm:collapse} $\mathrm{supp}(\model{M}^{\BB X}_{\mathsf i})_{\min}$ is finite and by Theorem \ref{thm:completeness2} all its points are reached by some $\TIT$-derivations. We can then set $n$ and $p$ to be larger than the corresponding numbers of $\YY$-unfoldings and 
 multiset sizes in all such (finitely many) derivations.
  \end{proof}

%The result above implies that the procedure sketched above stabilizes, after a finite number of steps, over the minimal Newton polynomial. 

%While the result above shows that one can always in principle reach the Newton polynomial of a term, recall that 

Stabilization at $n,p$ is a $\Pi^0_1$ (i.e.~non-recursive) property: it means that \emph{for any} $n',p'$ larger than $n,p$, the produced polynomial does not change. Therefore we might not be able to tell \emph{when} the algorithm did actually stabilize.
Recall that, by Theorem \ref{thm:pi01}, we cannot hope to compute $NP_{\min}^{\mathsf i}(M)$ in all situations (even though we always compute $(\mathrm{traj}^i(\pi))_{\min}$, by Proposition \ref{prop:trajmin}). 
 In practice, we can let the proof-search algorithm terminate after the polynomial has remained stable for some fixed number of iterations (for instance, one stable iteration was enough to reach $NP_{\min}^{\mathsf i}(M)$ in Examples \ref{example62} and \ref{example78}). Let $\tilde{\pi}^{\mathrm{stab}}$ be the derivation obtained after a finite number of stable iterations.

 {
 \begin{remark}[reaching efficiency in the affine case]
%Even once the parameters $n,p$ are fixed, the size of the derivation $\pi_{n,p}^{\mathsf i}(M)$ may be very large, since
%any application rule (of arity $2$ in the \pPCF-derivation of $M$) may have arity at most $p$ and, moreover, one has to potentially consider, for any type $A\to B$ in $M$, any possible intersection type $m\multimap b$, with $|m|\leq p$, that refines $A\to B$.

As we observed, when the most-likely reductions of $M$ are affine, we can reach $\pi^{\mathrm{stab}}$ via the derivations $\pi_{n,1}^{\mathsf i}(M)$, which have size $\mathcal O(n|M| 3^t)$. This is indeed the situation underlying our examples \ref{example62}, \ref{example76} and \ref{example78}.
%, where $t$ is the maximum size of a type in the \pPCF-derivation of $M$. To see this, just replace 
%all subterms of the form $\YY P$ of $M$ by $P^{i}(\lambda x.x)$ (with the sum of all such $i$ not exceeding $n$), and observe that in a $\TIT$-derivation of $P:\langle \vdash^s a_i\rangle_{i\in I}$, where the $a_i$ refine some type $A$, the index-set $I$ cannot exceed the number of affine refinings of $A$ (obtained by replacing each atom $\Bool$ by any of $[],[\ZERO],[\ONE]$).
By contrast, any affine reduction of $M$ can make at most $|M|$ probabilistic choices (since any choice strictly decreases the size of the term), so the trajectory space explored by each such derivation has in general size $\C O(2^{|M|})$.

 \end{remark}
 }

\subsubsection{Extracting a (Partial) Solution from $\tilde{\pi}^{\mathrm{stab}}$}

We now show how the set $\TT{selected\_ trajs}^i(M)$ is extracted from a (candidate) stable derivation.

By a straightforward adaptation of the algorithm $\VN$ (cf.~Remark \ref{rem:monomial}) and of the $\TIT$-rules we can perform a \emph{traceback} of $s$, i.e.~keep track, at each step of the construction of $\tilde{\pi}^{\mathrm{stab}}$, of a word $w_\mu\in\{0,1\}^*$, describing the sequence of probabilistic choices made during a reduction $M\RED{\mu}\mathsf i$; the word $w_\mu$ then traces back one \emph{most likely explanation}.
Moreover, for any monomial $\mu$ in $s$, we can compute the \emph{normal cone} of (the convex polytope generated by) $s$ at vertex $\mu$, defined as 
$\C N(\mu;s)=\{ x\mid  x\cdot\mu =\min_{\nu\in s}x\cdot\nu\}$, see \cite[p.~193]{Gunter}.
Computing $\C N(\mu;s)$ corresponds to solving the linear system of inequalities $\{(\mu-\nu)x\leq 0\}_{\nu\in s}$, which can be done via linear programming (cf.~\cite{Winston1988}).
Observe that $\C N(\mu;s)$ precisely captures the set of ($-\ln$-)values that maximize $\mu$ across all vertices in $s$, that is, that make the pair $(\mu,w_{\mu})$ a most likely explanation across those in $s$.

We can then let $\TT{selected\_ trajs}^i(M)$ $=\{(\mu,w_\mu,\C N(\mu;s))\mid \mu \text{ in }s\}$. The 
 following holds:

\begin{theorem}\label{thm:stats}
\begin{enumerate}

\item the pairs $(\mu,w_\mu)$ in $\TT{selected\_ trajs}^i(M)$ identify the most likely trajectories in $\mathrm{traj}^i(\tilde{\pi}^{\mathrm{stab}})$ (resp.~the most likely trajectories $M\RED{}\mathsf i$ in case $\tilde{\pi}^{\mathrm{stab}}={\pi}^{\mathrm{stab}}$).

\item the sets $\C N(\mu;s)$ in $\TT{selected\_ trajs}^i(M)$ identify all ($-\ln$)-values of the parameters $\vec X$ minimizing $\mu$ across all $\mathrm{traj}^i(\tilde{\pi}^{\mathrm{stab}})$ (resp.~across all trajectories $M\RED{}\mathsf i$ in case $\tilde{\pi}^{\mathrm{stab}}={\pi}^{\mathrm{stab}}$).

\end{enumerate}
\end{theorem}

Whenever $\tilde{\pi}^{\mathrm{stab}}={\pi}^{\mathrm{stab}}$, i.e.~the proof search reached the Newton polynomial, the result above states that $\TT{selected\_ trajs}^i(M)$ provides a correct answer to both (I1) and (I2). Otherwise, it provides an answer to (I1) and (I2) only \emph{relatively} to the trajectories explored by $\tilde{\pi}^{\mathrm{stab}}=\pi_{n,p}^{\mathsf i}(M)$. This means that any selected trajectory could still be dominated by some intuitively \emph{larger} one, i.e.~one requiring either more than $n$ $\YY$-unfoldings or $\beta$-reductions performing more than $p$ duplications.

\section{Conclusion}\label{sec:conc}
% !TEX root = Tropical_II.tex

\paragraph{Related Work}

A growing literature has explored foundational approaches to graphical probabilistic models and higher-order %programming 
languages for them, both from a categorical \cite{Jacobs_Zanasi_2020, Dahlqvist2018, Heunen2017, Staton2017, Staton2017} and from a more type-theoretic perspective \cite{Vanoni2024}.
Methods for statistical inference based on tropical polynomials and the Newton polytope, in the line of Section 4, have been 
%recently 
explored for several types of graphical probabilistic models, including HMM and Boltzmann machines \cite{Pachter2004, Pachter2004b, Cueto2010, Maragos2018, Maragos2021}.
Tropical geometry has also been applied to the study of deep neural networks %operating 
with ReLU activation functions \cite{Zhang2018, Maragos2017, Maragos2021}, as well as to piecewise linear regression \cite{Maragos2020}.
The interpretation of probabilistic PCF in the weighted relational model of linear logic is well-studied. The fully abstract model of probabilistic coherent spaces \cite{Pagani2018} relies on it. %this semantics. 
Tropical variants of this semantics are studied first in \cite{Manzo2013}, and more recently in \cite{BarbaPist2024}.
Beyond the one from \cite{Pagani2018}, many other kinds of intersection type systems to capture probabilistic properties have been proposed, e.g.~\cite{Breuvart2018, PistoneLICS2022, Heij2025}.
\DEL{Finally, the literature on programming languages for differential privacy, revolving around languages like FUZZ \cite{Reed2010} and type theories for relational reasoning \cite{Barthe2017} has grown vast \cite{Barthe_2012, Gaboardi2013, Gaboardi2017, Gabo2019, Gabo2019b}. We are not aware of applications of tropical methods in this area.}

\paragraph{Future Work}

In this paper we demonstrated the %possibility
%interest
potential of combining methods from programming language theory (especially, weighted relational semantics) and %from
tropical geometry, %and we applied 
by applying them to study %the
most likely behaviours of probabilistic higher-order programs.
We are convinced that this %combination
interaction %may 
leads to many other %intriguing
interesting 
applications in 
programming language~theory.

For instance, 
%A promising research direction is the following: while 
in this paper we only %described
manipulated the tropicalization %via
by means of the \emph{trivial} valuation $\mathrm{val}_0:\NINF\to\TS$ that sends $0$ to $\infty$ and all %(discrete) 
coefficients $n>0$ to $0$.
This was enough to study the most probable outcomes, but considering other valuations may lead to capture different properties of probabilistic programs. For example, $\mathrm{val}_c:\BB R_{\geq 0}\to \overline{\BB R}$ defined by $\mathrm{val}_c(x)=-c\ln x$ yields  
an exciting connection with \emph{differential privacy} \cite{Dwork2014} (already studied from the perspective of programming languages \cite{Gaboardi2017,Gabo2019,Gaboardi2013}), that we are currently exploring: 
for a function $f:\mathrm{db}\to \mathrm{Dist}(X)$ (corresponding to some probabilistic protocol), it is not difficult to see that $f$ is $\epsilon$-differentially private precisely when the function $\mathrm{val}_c\circ f:\mathrm{db}\to[X\to \overline{\BB R}]$ 
is $\epsilon$-Lipschitz continuous, taking the Euclidean distance on $\overline{\BB R}$.
As tropical polynomials are always Lipschitz-continuous (%as 
they are piecewise linear functions), this suggests that a program with \emph{finite tropical degree} (w.r.t.~the valuation $\mathrm{val}_c$, %and 
not $\mathrm{val}_0$ as in this work) could be proved differentially private.

%
%As an application of our results, we explore a connection with differential privacy.
%A key factor to show that a program may not extract private information is that the program need not be too \emph{sensitive} to 
%small changes in the input. When a program $M$ has a low sensitivity, well-established probabilistic methods (e.g.~the Laplace mechanism \cite{Dwork2014}) can be applied to turn $M$ into a differentially-private program.
%
%By exploiting the fact that tropical polynomials are always Lipschitz continuous, we show that the tropical degree of a probabilistic program can be used to produce an estimation of its differential privacy. Since the tropical degree of a complex probabilistic program may be quite high in general, it is not obvious that the produced estimations are of practical use, beyond simple explanatory cases. 
%At the same time, our results highlight a surprising conceptual connection between these two areas that we think it might be worth to explore further.

Another important future work is to %actually
\emph{implement} our type system, in order to mechanise, in interaction with the user, some inference tasks over programs as explained in Section~\ref{sec:inters_types}.
Related to that, we notice that the crucial notion of minimal monomial in a (Newton) polytope, that we develop in Section~\ref{sec:geo}, is reminiscent of that of Gr\"obner basis in computational algebra.
Similarly, we wonder whether the algebra of generating functions -- essentially, the formal power series given by our parametric interpretation -- reflects computational properties of %the respective
programs.

%
%Beyond exploring further the suggestive connections with differential privacy, we can think of 
Beyond that, there are other natural areas of applications. For instance,
 \cite{BarbaPist2024} illustrated a notion of differentiation for tropical power series, relying on the theory of cartesian differential categories \cite{Blute2009, Manzo2012}, that aligns with existing notions in the literature on tropical differential equations \cite{Grigoriev2017}.
 Finally, the growing interest towards higher-order frameworks for automatic differentiation \cite{Mazza2021, Nunes2023} suggests to look at the tropical methods currently employed for ReLU neural networks~\cite{Maragos2021, Giansiracusa:2024aa}.

\begin{acks}
%The authors would like to greatly thank the anonymous reviewers for their careful reading and for several questions that helped us significantly improve the first version of the paper.

Davide Barbarossa has been funded by the EPSRC grant n.\ EP/W035847/1. Paolo Pistone has been funded by the ANR grant n.\
ANR-23-CPJ1-0054-01. For the purpose of Open Access the authors have applied a CC BY public copyright licence to any Author Accepted Manuscript version arising from this submission.
\end{acks}

\bibliographystyle{plain}
\bibliography{tropical.bib}

\newpage

\appendix

% !TEX root = Tropical_II.tex
\section{Proofs from Section 4}

\subsection{$\fps{\NINF}{\Sigma}$ is the Free Continuous Commutative Semiring over $\Sigma$}
We only prove Proposition \ref{prop:free_ccs}, the other results are either known in the literature (as referenced in the paper, or immediately obtained by them.

We will use the following result, which is obtained by a straightforward adaptation to the commutative case of the statement (and the proof) of \cite[Theorem 2.1]{EsiK2017}:

\begin{proposition}\label{prop:literature} 
    Let $S$ be a continuous commutative semring and $\Sigma$ a (finite) set.
    For any continuous commutative semring $Q$, $q\in Q^\Sigma$ and homomorphism of continuous commutative semirings $h:S\to Q$, the $\widetilde h_q$ below is the unique homomorphism of continuous commutative semirings which makes the following diagram commute:
\[\begin{tikzcd}[ampersand replacement=\&,cramped]
	\&\&\&\& \Sigma \\
	{} \&\& {} \&\& {\fps{S}{\Sigma}} \&\& Q \& {\widetilde{h}_q(\sum_\mu s_\mu x^\mu):=\sum_\mu h(s_\mu)q^\mu} \\
	\&\&\&\& S
	\arrow[hook', from=1-5, to=2-5]
	\arrow["q", from=1-5, to=2-7]
	\arrow["{\widetilde{h}_q}"{description}, dashed, from=2-5, to=2-7]
	\arrow[hook, from=3-5, to=2-5]
	\arrow["h"', from=3-5, to=2-7]
\end{tikzcd}\]
\end{proposition}

Moreover, we have:

\begin{lemma}\label{lm:char}
    Let $Q$ be a continuous commutative semring. The map $(\_)_Q:\NINF\to Q$ by $n_Q:=\sum_{i=1}^n 1$ is a continuous commutative semiring homomorphism.
\end{lemma}
\begin{proof}
    It is clearly well defined because $Q$ is continuous.
    By definition of sums in $Q$ we trivially have $0_Q=0$ and $1_Q=1$.
    It is easy to see that $(n+m)_Q=n_Q+m_Q$.
    Finally, let us show that it preserves products and supremas:
    \[\begin{array}{rclcccrcl}
        (nm)_Q & 
        = & \sum\limits^{nm}1 &&&&(\bigvee\limits_i n_i)_Q & 
        = & \bigvee\limits_i \sum\limits^{n_i} 1 \\
        & = & \bigvee\limits_{t\leq_{\textit{fin}}nm} \sum\limits^t 1&&&&& = & \bigvee\limits_i\bigvee\limits_{k\leq_{\textit{fin}} n_i} \sum\limits^k 1 \\
        & = & \bigvee\limits_{k\leq_{\textit{fin}}n}\bigvee\limits_{r\leq_{\textit{fin}}m} \sum\limits^{kr} 1&&&&& = & \bigvee\limits_{k\leq_{\textit{fin}} \bigvee_i n_i} \sum\limits^{k} 1  \\
        & = & \bigvee\limits_{k\leq_{\textit{fin}}n}\bigvee\limits_{r\leq_{\textit{fin}}m} \left(\sum\limits^{k} 1\right)\left(\sum\limits^{r} 1\right)&&&&& = & \sum\limits^{\bigvee_i n_i} 1 \\
        & = & \bigvee\limits_{k\leq_{\textit{fin}}n} \left(\left(\sum\limits^{k} 1\right)\left(\bigvee\limits_{r\leq_{\textit{fin}}m} \sum\limits^{r} 1\right)\right)&&&&& = & \left(\bigvee\limits_i n_i\right)_Q \\
        & = & \left(\bigvee\limits_{r\leq_{\textit{fin}}n} \sum\limits^{k} 1\right)\left(\bigvee\limits_{r\leq_{\textit{fin}}m} \sum\limits^{r} 1\right)&&&&& & \\
        & = & \left(\sum\limits^{n} 1\right)\left(\sum\limits^{m} 1\right) \\
        & = & n_Qm_Q &&&&& & 
        \end{array}\]
       In the third equality at the right column we used that if $k\leq_{\textit{fin}} \bigvee_i n_i$ then $k\leq n_j$ for some $j$.
\end{proof}

\begin{lemma}\label{lm:aux}
    Remember that for $Q$ a continuous commutative semring, $q\in Q$ and $n\in\NINF$, we defined in the paper $np:=\sum^n p$.
    We have $n(pq)=(np)q$ for all $n\in\NINF$ and $p,q\in Q$.
\end{lemma}
\begin{proof}
    $(np)q=(\sum^n p)q=(\bigvee_{k\leq_{\textit{fin}} n}\sum^k p)q =\bigvee_{k\leq_{\textit{fin}} n}\sum^k (pq) = \sum^n pq = n(pq)$.
\end{proof}

Now we can give the:

\begin{proof}[Proof of Proposition \ref{prop:free_ccs}]
We are give a finite set $\Sigma$, and we have to show that for any continuous commutative semiring $Q$ and $q\in Q^\Sigma$, the map $\texttt{ev}_q$ defined in the statement of the proposition is the unique homomorphism of continuous commutative semirings $\fps{\NINF}{\Sigma}\to Q$ which sends $X$ in $q_X$ for all $X\in\Sigma$.

Applying Proposition \ref{prop:literature} to $q$ and to $h:=(\_)_Q$ of Lemma \ref{lm:char},  we obtain the map $(\widetilde{(\_)_Q})_q$ which, by looking at its definition and using Lemma \ref{lm:aux}, is exactly the desired map $\texttt{ev}_q$ of the statement.
Thus, in particular, $\texttt{ev}_q$ is a homomorphism of continuous commutative semirings such that $\texttt{ev}_q(X)=q_X$ for all $X\in \Sigma$ and it only remains to show that it is uniquely determined by $q$.
For this, let $h':\fps{\NINF}{\Sigma}\to Q$ a homomorphism of continuous commutative semirings such that $h'(X)=q_X$ for all $X\in\Sigma$.
Then for all $n\in \NINF$ we have $h'(n)=h'(\sum^n 1)=\sum^n h'(1)=\sum^n 1=n_Q$, i.e.\ $h'$ extends $(\_)_Q$.
But then by the uniqueness of $(\widetilde{(\_)_Q})_q$ we have $\texttt{ev}_q=(\widetilde{(\_)_Q})_q=h'$.
\end{proof}

\section{The Category $\an Q$}

Following \cite{Martini1992}, a cartesian category $\B C$ is a \emph{weak cartesian closed category} (wCCC) if for every objects $a,b$ there exists an object $ b^a$ together with natural transformations
\begin{align*}
\mathrm{ev}_{-,a,b}& :  \B C(-,  b^a) \To \B C(-\times a, b) \\
\Lambda_{-,a,b}& :  \B C(-\times a, b) \To  \B C(-,  b^a) \\
\end{align*}
satisfying 
\begin{equation}
\Lambda_{-,a,b}\circ \mathrm{ev}_{-,a,b}=\mathrm{id}_{\B C(-\times a, b)} \tag{$\beta$}
\end{equation} 
Observe that a wCCC is a CCC precisely when the converse, $\eta$, equation also holds
\begin{equation}
\mathrm{ev}_{-,a,b} \circ \Lambda_{-,a,b}=\mathrm{id}_{\B C(-, b^a)} \tag{$\eta$}
\end{equation} 
A wCCC is thus an \emph{intensional} model of the simply typed $\lambda$-calculus, that is, one in which the rule $\beta$ is valid but the rule $\eta$ needs not be valid.

The category $\rel{Q}_!$ is cartesian closed, with exponential $!X\times Y$. The maps $\mathrm{ev}_{-,X,Y}$ and 
$\Lambda_{-,a,b}$ are defined, for $\mu\in !X, \nu\in !Y$ and $y\in Y$, by
\begin{align*}
\mathrm{ev}_{-,X,Y}(t)_{\mu\oplus\nu, y}&= (t_{\mu})_{\nu, y},\\
((\Lambda_{-,X,Y}(t))_\mu)_{\nu,y}&=t_{\mu\oplus\nu , y},
\end{align*}
where we used the fact that, via the natural isomorphism $\sigma_{A,B}: !A\times !B\to !(A+B)$, an element of $!(A+B)$ can be uniquely written as $\mu\oplus\nu$, where $\mu\in !A$ and $\nu\in !B$.

For the category $\an Q$ the following holds:
\begin{proposition}
If $Q$ is a complete lattice with sums and products commuting with arbitrary joins, 
then the category $\an Q$ is a wCCC.
\end{proposition}
\begin{proof}
$\an Q$ inherits the cartesian product $X+Y$ from $\rel{Q}_!$. We show that $!X\times Y$ is a weak exponential. 

We will exploit the natural isomorphism $\langle-,-\rangle,Q^A\times Q^B \to Q^{A+B}$.

%Observe that we have a natural isomorphism $\tau_{A,B}:  Q^{!(A+B)}\to Q^{!A\times!B}$ given by 
%$\tau_{A,B}(x)_{(\mu,\nu)}=x_{S(\mu,\nu)}$. 

Since $Q$ is a complete lattice, the semirings $\rel{Q}_!(!A,B)=Q^{!A\times B}$ are complete lattices as well (for the pointwise order), with sums and products commuting with joins. We will exploit this fact to define the natural family of maps $\Lambda_{-,X,Y}$. 
Let us first define the set of power series representations of $f$:
$$
\mathrm{PS}_{X,Y}(f):= \left \{ t\in Q^{!X,Y}\ \Big \vert \  \forall x\in Q^X\ \forall y\in Y, f(x)=\sum_{\mu\in !X}t_{\mu,y}x^\mu\right \}.
$$
Observe that the sets $\mathrm{PS}_{X,Y}(f)$ are non-empty: since $f\in \an Q(X,Y)$ is analytic, there exists a matrix $\widehat f\in \rel{Q}_!(X,Y)$ such that $f(x)_y=\sum_{\mu\in !X}\widehat f_{\mu,y}x^\mu$, that is, $\widehat f\in \mathrm{PS}_{X,Y}(f)$.

We define the operators $\mathrm{ev}_{-,X,Y}$ and $\Lambda_{-,X,Y}$ as follows, for $f\in \an Q(-\times X, Y)$ and $g\in \an Q(-,!X\times Y)$:
\begin{align*}
\mathrm{ev}_{-,X,Y}(f)(z)(x)_y&= \sum_{\nu \in !X}f(z)_{\nu,y} x^\nu,\\
\Lambda_{-,X,Y}(g)(z)_{\nu,y} &= \bigvee_{s\in \mathrm{PS}_{X,Y}(g(\langle z,-\rangle))}s_{ \nu,y}.
\end{align*}
Intuitively, $\Lambda_{-,X,Y}$ chooses the \emph{largest} among all power series representations of $f$. 

Let us first check equation $(\beta)$: given any $s\in \mathrm{PS}_{Z+ X,Y}(g)$ and $\langle z,x\rangle\in Q^{X+Y}$ we have by definition that 
\begin{align*}
\sum_{\mu\oplus\nu\in !(Z+X)}s_{\mu\oplus\nu, y} z^\mu x^\nu=
\sum_{\mu\oplus\nu\in !(Z+X)}s_{\mu\oplus\nu, y} \langle z,x\rangle^{\mu\oplus\nu}= g(\langle z,x\rangle) \tag{$\star$}
\end{align*}
Using the fact that infinite sums and finite products commute with joins, we deduce then that 
\begin{align*}
\mathrm{ev}_{Z,X,Y}(\Lambda_{Z,X,Y}(f))(z)(x)_y&=
\sum_{\nu \in !X}
\Big( \Lambda_{Z,X,Y}(f)(z)\Big)_{\nu,y} x^\nu\\
&=
\sum_{\nu \in !X}
\bigvee_{s\in \mathrm{PS}_{Z+X,Y}(f(z))}
s_{\nu,y}x^\nu \\
%&=
%\sum_{\nu \in !X}
%\bigvee_{s\in \mathrm{PS}_{Z+X,Y}}
%\sum_{\mu\in !Z}
%s_{\mu\oplus\nu,y}z^\mu x^\nu\\
&=
\bigvee_{s\in \mathrm{PS}_{Z+X,Y}(f(z))}
\sum_{\nu \in !X}
s_{\nu,y} x^\nu
\stackrel{{\tiny(\star)}}{=}f(\langle z,x\rangle).
\end{align*}

Let us check that the operations $\mathrm{ev}_{-,X,Y}$ and $\Lambda_{-,X,Y}$ are natural.

Let $f\in \an Q(Z+X,Y)$, $g\in \an Q(Z,!Y\times X)$ and $h\in \an Q(Z',Z)$. 
\begin{align*}
\mathrm{ev}_{Z',X,Y}(f\circ h)(z')(x)_y&=
\sum_{\nu \in !X} (f\circ h)(z')_{\nu,y}x^\nu\\
&=
\sum_{\nu\in !X}f(h(z'))_{\nu,y}x^\nu
=
\mathrm{ev}_{Z,X,Y}(f)(h(z'))(x)_y.
\end{align*}
On the other hand we have
\begin{align*}
\Lambda_{Z,X,Y}(g)(h(z'))_{\nu,y}=
\bigvee_{s\in \mathrm{PS}_{X,Y}(g(h(z')))}s_{\nu,y}=
\bigvee_{s\in \mathrm{PS}_{X,Y}(g\circ (h\times\mathrm{id}_X)(z))}s_{\nu,y}
=
\Lambda_{Z',X,Y}(g\circ (h\times\mathrm{id}_X))(z')_{\nu,y}
\end{align*}
\end{proof}

All the continuous semirings $\BS,\NINF,\RS,\TS$ satisfy the hypothesis of the theorem, so their respective categories of analytic functions are wCCC.
\cite{Ehrhard2005}, p.~20 furthermore shows that $\an \RS$ is even CCC.% cartesian closed. This is not true, as we saw, for $\an \TS$. 

% !TEX root = Tropical_II.tex

\section{Proofs from Section 5}

\subsection{Full details of Proposition \ref{prop:collapse}}

When $\Sigma$ has $k$ elements, the set $!\Sigma$ can be identified with $\N^k$.  

\begin{definition}
 Let $\preceq$ be the product order on $\N^k$ (i.e.\ for all $ m  , n  \in \N^{k}$, $ m  \preceq  n  $ iff $m_{i}\leq n_{i}$ for all $1\leq i\leq k$).
 Of course $ m  \prec  n  $ holds exactly when $ m  \preceq  n  $ and $m_{i}<n_{i}$ for at least one $1\leq i\leq k$.
 Finally, we set $ m  \prec_{1} n  $  iff
$ m  \prec  n  $ and $\sum_{i=1}^{k}n_{i}-m_{i}=1$ (i.e.\ they differ on exactly one coordinate).
\end{definition}

\begin{remark}\label{rmk:AC2}
If $U\subseteq \N^{k}$ is infinite, then $U$ contains an infinite ascending chain $ m  _{0}\prec  m  _{1} \prec  m  _{2} \prec \dots$.
This is a consequence of K\"onig Lemma (KL): consider the directed acyclic graph $(U,\prec_{1})$, indeed a $k$-branching tree; if there is no infinite ascending chain $  m  _{0}\prec  m  _{1} \prec  m  _{2} \prec \dots$, then in particular there is no infinite ascending chain $  m  _{0}\prec_{1}  m  _{1} \prec_{1}  m  _{2} \prec_{1} \dots$ so the tree $U$ has no infinite ascending chain; then by KL it is finite, contradicting the assumption. 
\end{remark}

Now we can give the

%\begin{lemma}\label{lemma:fps}
%Let $\Sigma$ be a finite set and $s\in \fps{\TS}{\Sigma}$. 
%If $s_\mu\in\NINF$ (as real numbers) for all $\mu\in !\Sigma$, then there exists a finite set $P(s)\subseteq \ !\Sigma$ such that, for all $x\in \TS^\Sigma$,  
%$$
%s^!(x)=\inf_{\mu\in !\Sigma}\{\mu\cdot x+s_\mu\}=\min_{\mu\in P(s)}\{\mu\cdot x+s_\mu\}.
%$$
%\end{lemma}
\begin{proof}[Proof of Proposition \ref{prop:collapse}]
Let $F(x)=\inf_{\mu\in !\Sigma}\{\mu\cdot x+s_\mu\}$.
Let $k$ be the cardinality of $\Sigma$. Observe then that $!\Sigma$ can be identified with $\N^k$ (and we write $n$ instead of $\mu$). 
We will actually show the existence of $P(s) \subseteq_{\mathrm{fin}} \N^k$ such that:
\begin{enumerate}
% \item $\C F_\epsilon$ is finite
 \item if $P(s)= \emptyset$ then $F( x ) = +\infty$ for all $ x \in \TS^{\Sigma} $;
 \item if $F( x _0) = +\infty$ for some $ x _0\in [0,+\infty)^\Sigma$ then $P(s)= \emptyset$;
 \item $F(x)=\min\limits_{n\in P(s)}\set{n x+s_n}$.
%where $nx:=\sum_{i=1}^k n_ix_i$.
\end{enumerate}
Let $P(s)$ be the complementary in $\N^k$ of:
$
 \set{ n  \in\N^{k} \mid \textit{either } s_n=+\infty \textit{ or there is }  m  \prec  n  \textit{ s.t.\ } s_m \leq s_n}.
$ 
In other words, $ n  \in P(s)$ iff $s_n<+\infty$ and for all $ m  \prec  n  $, one has $s_m> s_n$.
Suppose that $P(s)$ is infinite; then, using Remark~\ref{rmk:AC2}, it contains an infinite ascending chain $\set{ m  _0\prec  m  _1\prec\cdots}$.  
By definition of $P(s)$ we have then an infinite \emph{descending} chain
$+\infty>s_{  m  _0}>s_{ m  _1}>s_{ m  _2}>\cdots$ in $\N$, which is impossible.
We conclude thus that $P(s)$ is finite.

\begin{enumerate}
\item
We show that if $P(s)=\emptyset$, then $s_n=+\infty$ for all $ n  \in\N^{k}$.
This immediately entails the desired result.
We go by induction on the well-founded order $\prec$ over $ n  \in\N^{k}$:
\begin{itemize}
\item if $ n  =0^{k}\notin P(s)$, then $s_n=+\infty$, because there is no $ m  \prec n  $.

\item if $ n  \notin P(s)$, with $ n  \neq 0^{k}$ then suppose there is $ m  \prec  n  $ s.t.\ $s_m\leq  s_n$.
By induction $s_m=+\infty$ and we obtain $s_m=+\infty \leq s_n$ so $s_n=+\infty$.
\end{itemize}
\item If $F( x _0)=+\infty$ for some finite $x_0\in [0,+\infty)^\Sigma$, then necessarily $s_n=+\infty$ for all $ n  \in\N^{k}$.
Therefore, no $ n  \in\N^{k}$ belongs to $P(s)$.

\item We have to show that $F( x )=\min_{n\in P(s)}\set{n x+s_n}$.
By 1), it suffices to show that we can compute $F( x )$ by taking the $\inf$, that is therefore a $\min$, only in $P(s)$ (instead of all $\N^{k}$).
If $P(s)=\emptyset$ then by 1) we are done (as $\min\emptyset := +\infty$).
If $P(s)\neq\emptyset$, we show that for all $ n  \in\N^{k}$, if $ n   \notin P(s)$, then there is $ m  \in S$ s.t.\ $s_m+ m   x  \leq s_n+ n   x $ .
We do it by induction on $\prec%_{1}
$:
\begin{itemize}
\item if $ n  =0^{k}$, then from $  n\notin P(s)$, by definition of $S$, we have $s_n=+\infty$ (because there is no $ n  '\prec n  $).
So any element of $P(s)\neq\emptyset$ works.

\item if $ n  \neq 0^{k}$, then we have two cases:
either $s_n=+\infty$, in which case we are done as before by taking any element of $P(s)\neq\emptyset$.
Or $s_n<+\infty$, in which case (again by definition of $P(s)$) there is $ n  '\prec n  $ such that $ s_{n'}\leq s_n$  ($\star$).
Therefore we have (remark that the following inequalities hold also for the case $x=+\infty$):
\[\begin{array}{rclr}
 s_{n'}+ n  ' x  & \leq & s_n +  n' x  & \textit{by }(\star) \\
 & \leq & s_n + ( n  - n  ') x  +  n  ' x  & \text{since $n'\prec n$ (the $\leq$ is a $=$ is $x=0$)}\\
 & = & s_n+  n   x . &
\end{array}\]
Now, if $ n  '\in P(s)$ we are done.
Otherwise $ n  '\notin P(s)$ and we can apply the induction hypothesis on it, obtaining an $ m  \in P(s)$ s.t.\ $s_m+ m   x  \leq s_{n'}+ n  ' x $.
Therefore this $ m  $ works.
\end{itemize}
\end{enumerate}
\end{proof}

\section{Proofs from Section 6}

{
\subsection{Lemma \ref{lm:supp_to_NP}}
This lemma is actually an instance of the well-known minimum principle at the heart of linear optimisation (e.g.\ in the simplex algorithm).
Let us prove it below for the sake of clarity:

\begin{lemma}\label{lm:opt}
Let $w_1,\dots,w_h\in\BB R^n$ ($h\geq1$), let $\C P$ be the polytope given by the convex hull of $w_1,\dots,w_h$ and let $v_1,\dots,v_k$ ($1\leq k\leq h$) be the vertices of $\C P$.
For all $x\in\BB R^n$,
\[\inf_{y\in\C P}\set{y\cdot x}=\min\set{v_1\cdot x,\dots,v_k\cdot x}.\] 
\end{lemma}
\begin{proof}
Fix $x\in\BB R^n$.
The $(\leq)$ is trivial.
For $(\geq)$, we show that $y\cdot x\geq \min\set{v_1\cdot x,\dots,v_k\cdot x}$ for all $y\in\C P$.
Since $y\in\C P$ then, by definition of vertices, $y= \alpha_1v_1+\dots+\alpha_kv_k$ for some $\alpha_i\in[0,1]$ such that $\sum_{i=1}^k\alpha_i=1$.
But now:
\[\begin{array}{rcl}
    \min\set{v_1\cdot x,\dots,v_k\cdot x} & = & (\sum_{i=1}^k\alpha_i) \min\set{v_1\cdot x,\dots,v_k\cdot x} \\
     & = & \alpha_1\min\set{v_1\cdot x,\dots,v_k\cdot x}+\dots+\alpha_k\min\set{v_1\cdot x,\dots,v_k\cdot x} \\
     & \leq & \alpha_1v_1\cdot x+\dots+\alpha_kv_k\cdot x \\
     & = & (\alpha_1v_1+\dots+\alpha_kv_k)\cdot x \\
     & = & y\cdot x.
\end{array}\]
\end{proof}

\begin{proof}[Proof of Lemma \ref{lm:supp_to_NP}]
Remembering that $\mathrm{Vert}(NP(s))\subseteq \mathrm{supp}(s)\subseteq NP(s)$, we have:
\[\min_{\mu\in\mathrm{Vert}(NP(s))} \mu\cdot x=\min_{\mu\in NP(s)} \mu\cdot x\leq \min_{\mu\in\mathrm{supp}(s)} \mu\cdot x\leq \min_{\mu\in\mathrm{Vert}(NP(s))} \mu\cdot x,\]
where in the first equality we used Lemma \ref{lm:opt}.
\end{proof}

%Let us now prove Lemma \ref{lm:NP_to_NPmin}.
%
%\begin{proof}[Proof of Lemma \ref{lm:NP_to_NPmin}]
%Fix $x\in\BB R^n$.
%The $(\leq)$ is trivial.
%For $(\geq)$, we show that for all $\mu\in \mathrm{Vert}(NP(s))$, there is $\rho\in \mathrm{Vert}(NP_{\mathrm{min}}(s))$ such that $\rho\cdot x\preceq \mu\cdot x$.
%We do it by induction on $\mu\in (\mathrm{Vert}(NP(s)),\preceq)$.
%
%Case $\mu$ minimal: then by definition $\mu\in\mathrm{Vert}(NP_{\mathrm{min}}(s))$, so we are done.

%Case $\mu$ not minimal:
%this means that there is $\nu\in \mathrm{Vert}(NP(s))$ with $\nu\prec \mu$.
%By IH there is $\rho\in \mathrm{Vert}(NP_{\mathrm{min}}(s))$ such that $\rho\cdot x\preceq \nu\cdot x \preceq \mu\cdot x$.
%\end{proof}
%}

%To conclude the proof of Proposition \ref{prop:subset} we need the following lemma:
\subsection{Statement contained in Remark \ref{rmk:neg_or}%Vertices Contained in Negatively Oriented facets
}

We prove in Lemma \ref{lm:neg_or_to_min} below the statement contained in Remark \ref{rmk:neg_or}.

\begin{definition}
Let $P$ be a polytope in $\BB R^n$.
We call a face $f$ of $P$ \emph{negatively oriented} iff $\hat n_f^-$ has coordinates (in the canonic base) all strictly negative (i.e.\ $\hat n_f^-\cdot e_i<0$ for all $i$, where $\set{e_1,\dots,e_n}$ is the canonic base).
\end{definition}

\begin{lemma}\label{lm:neg_or_to_min}
Let $P$ be a polytope in $\BB R^n$ and $v$ a vertex of $P$.
If $v$ belongs to some facet $f$ of $P$ which is negatively oriented, then $v\in P_\mathrm{min}$.
\end{lemma}
\begin{proof}
We have to show that there is no $w$ vertex of $P$ such that $w\prec v$.
Given $w\in\BB R^n$, if $w\prec v$ then by definition there is a coordinate $i_0$ such that $w_{i_0}<v_{i_0}$.
Therefore, $\hat n_f^-\cdot (w-v)=\sum_{i\neq {i_0}} (\hat n_f^-)_i(w_i-v_i)+(\hat n_f^-)_{i_0}(w_{i_0}-v_{i_0})\geq (\hat n_f^-)_{i_0}(w_{i_0}-v_{i_0})>0$,
where the two inequalities hold because $(\hat n_f^-)_i<0$ and $(w_i-v_i)\leq 0$ for all $i$ and, for $i=i_0$, the latter is strict.
But remember that, by definition of facet, there is a hyperplane $H_f$ that contains $f$ and such that $P\subseteq H_f^+$, where (remembering that $v\in f$) $H_f^+$ has equation $\hat n_f^-\cdot(x-v)\leq 0$.
What we found above precisely means that $w\notin H_f^+$ and thus, a fortiori, $w\notin P$. 
\end{proof}

\subsection{Lemma \ref{lm:vertmin}}

\begin{lemma}\label{lemma:submin}
For all $A,B,C\subseteq \mathbb N^n$, if $A_{\min}\subseteq B+C\subseteq A$, then $A_{\min}=(B_{\min}+C_{\min})_{\min}$.
\end{lemma}
\begin{proof}
For $(\subseteq)$, let $a\in A_{\min}$; since $A_{\min}\subseteq B+C$, we can write $a=b+c$, with $b\in B,c\in C$.
Let $b'\in B$ and suppose $b'\preceq b$; then $b'+c\in B+C\subseteq A$ and $b'+c\leq b+c=a$; by the minimality of $a$ we deduce $b'+c=b+c=a$, which in turn implies $b'=b$. We conclude thus that $b\in B_{\min}$. A similar argument shows that $c\in C_{\min}$. We have thus proved $A_{\min}\subseteq B_{\min}+C_{\min}$. 
Let now $b'+c'\in B_{\min}+C_{\min}$, and suppose $b'+c'\preceq a$;
 by $B_{\min}+C_{\min}\subseteq B+C\subseteq A$ we deduce $b'+c'\in A$, and by the minimality of $a$ in $A$ we deduce then $b'+c'=a$. This shows then that $a\in (B_{\min}+C_{\min})_{\min}$.
 
 For $(\supseteq)$, let $d\in (B_{\min}+ C_{\min})_{\min}\subseteq B+C\subseteq A$ and $a\in A$, and suppose $a\preceq d$; by the well-foundedness of $\prec$, we can find $a_0\in A_{\min}\subseteq B_{\min}+C_{\min}$ such that $a_0\preceq a\preceq d$. Now, the minimality of $d$ in $B_{\min}+ C_{\min}$ yields $d= a_0= a$. We conclude then $d\in A_{\min}$.
\end{proof}

%\begin{lemma}\label{lm:vertmin}
%Let $s_1,s_2\in \fps{\TS }{\Sigma}$.
%$\mathrm{Vert}(NP_{\mathrm{min}}(s_1s_2)) 
%= 
%(\mathrm{Vert}(NP_{\min}(s_1))+\mathrm{Vert}(NP_{\min}(s_2)))_{\mathrm{min}}$.
%\end{lemma}
\begin{proof}[Proof of Lemma \ref{lm:vertmin}]
It is well known that $NP(s_1 s_2) = NP(s_1)+NP(s_2)$. 
We thus have $NP_{\min}(s_1s_2)\subseteq NP(s_1)+NP(s_2)\subseteq NP(s_1s_2)$, so we can apply 
Lemma \ref{lemma:submin} which gives the result.
\end{proof}

\section{Details and Proofs for Section~\ref{sec:inters_types}}

Conceptually speaking, the notion of refinement that we present below, the soundness and completeness theorems that follow, together with their proofs are, in a sense an adaptation of the proof of \cite[Lemma 20]{EhrhardTassonPagani14}.
However, the arguments there are left partial and very implicit, while here we give more precise and explicit definitions and proofs. That said, there are a few differences: they work with the n.i.i.-type system $\mathbf P$, while we here deal with $\TIT$, which intuitively speaking considers several (but finitely many) $\mathbf P$-derivations in parallel.
In fact, we are here in a tropical setting for a probabilistic language, whence the additional (and brand new) treatment of the notion of minimality, Viterby-Newton and inference task partial algorithm of Section~\ref{subsec:inference_algo}, while in their case it is standard probabilistic weighted relational semantics.
Ignoring these tropical aspects, our constructions and proofs could, however, be adapted for having a clean parametric version of \cite[Lemma 20]{EhrhardTassonPagani14}.    

\subsection{Definition of \emph{refinement}}

\begin{definition}\label{def:adequate}
Let $\Gamma,A$ be, respectively, a \pPCF-context and a \pPCF-type.
We say that a pair $(\gamma,a)$ of, respectively, a $\TIT$-context and a $\TIT$-type is \emph{adequate for $(\Gamma,A)$} iff\footnote{To be pedantic, for making sense of the condition $a\in\model A$ we are secretly implementing $\TIT$-types as set theoretic entities via the map $\hat{(\_)}$ defined in the obvious way: $n\in\N\mapsto n\in\N$, $[e_1,\dots,e_h]\multimap f \mapsto ([\hat e_1,\dots,\hat e_h],\hat f)$. Similarly, for making sense of the condition $\gamma(x)\in!\model{\Gamma(x)}$ we secretly really mean $\hat\gamma(x)\in!\model{\Gamma(x)}$, where if $\gamma(x)=[a_1,\dots,a_h]$ we set $\hat{\gamma}(x):=[\hat a_1,\dots,\hat a_h]$.} $a\in \model A$ and $\gamma(x)\in!\model{\Gamma(x)}$ for all $x\in\mathrm{dom}(\Gamma)$ and $\gamma(x)=[]$ for all $x\notin\mathrm{dom}(\Gamma)$.

Let $\Gamma\vdash M:A$ be a (Curry) well-typed \pPCF-term.
We say that a $\TIT$-derivation $\deriv{\pi}{M:\big\langle \gamma_k\vdash^{s_k} a_k \rangle_{k\in K}}$ is \emph{adequate for $\Gamma\vdash M:A$} iff for all $k\in K$, $(a_k,\gamma_k)$ is adequate for $(\Gamma,A)$.
\end{definition}

The above notation $\deriv{\pi}{M\langle\_\rangle}$, and the similar $\deriv{\Pi}{\_\vdash\_}$, which will be both used in the following, are shortcuts for saying that $\pi$ (resp.\ $\Pi$) are derivations in $\TIT$ (resp.\ \pPCF) of the respective judgments.

In order to define the fundamental notion of refinement, it is crucial to carefully handle the nature of a fixed point: we introduce for this the following definitions of unfolding of a derivation, and an ordinal size of well-typing derivations.

\begin{definition}\label{def:unfold_Pi}
Let %$\deriv{\Pi}{\Gamma\vdash \YY M:A}$ obtained via a $\YY$-rule on 
$\deriv{\Pi}{\Gamma\vdash M:A\to A}$. For a variable $y$ non-declared in $\Gamma$ we define derivations $\deriv{\mathbf{unfold}^{(n)}.\Pi}{\Gamma,y:A\vdash M^{(n)}y:A}$ by induction on $n\in\N$ in Figure~\ref{fig:unfold_Pi}.
\end{definition}

%%%%%

\newsavebox\unfoldzero
\sbox\unfoldzero{\small
        \AXC{}
        \UIC{$\Gamma,y:A\vdash y:A$}
    \DisplayProof
}

\newsavebox\unfoldsucc
\sbox\unfoldsucc{\small
        \AXC{$\vdots \ \mathbf{wkn}^{(y:A)}.\Pi$}
        \noLine\UIC{$\Gamma,y:A\vdash M:A\to A$}
        \AXC{$\vdots \ \mathbf{unfold}^{(n)}.\Pi$}
        \noLine\UIC{$\Gamma,y:A\vdash M^{(n)}y:A$}
        \BIC{$\Gamma,y:A\vdash M^{(n+1)}y:A$}
    \DisplayProof
}

%%%%%

\begin{figure}
\fbox{\begin{minipage}{.99\textwidth}
\[\begin{array}{ccc}
    \mathbf{unfold}^{(0)}.\Pi & := & \usebox\unfoldzero  \\ \\
    \mathbf{unfold}^{(n+1)}.\Pi & := & \usebox\unfoldsucc
\end{array}\]
\end{minipage}}
\caption{Unfolding of $\deriv{\Pi}{\Gamma\vdash M:A\to A}$. Here $\mathbf{wkn}^{(y:A)}.\Pi$ is the \emph{weakening of $\Pi$}, obtained from $\Pi$ by adding $y:A$ in all axiom rules' contexts, and propagating it top-down in all judgments.}
\label{fig:unfold_Pi}
\end{figure}

\begin{definition}\label{def:unfold_pi}
Let $\deriv{\pi}{\YY M:\big\langle \gamma_k\vdash^{s_k} a_k\big\rangle_k}$.
Employing the same terminology as \cite[Lemma 20(iii)]{EhrhardTassonPagani14}, we remark that any $\TIT$-derivation of $YM:\big\langle \cdots\big\rangle_k$ ends with a finite cluster of $\YY$ rules whose hypotheses judgments are always $M$ and $YM$, and the highest $\YY$ rule necessarily has hypotheses the judgments $M:\big\langle \gamma_j\vdash^{u_j} [\,]\multimap a_j \big\rangle_{j\in J}$ (for some set $J$) and $YM:\big\langle\big\rangle$ (proven via the $\TIT$ rule $\emptyset$).
Let $\mathbf{cl}(\pi)$ be the height of the cluster of $\pi$.
Now it is easy to see that, given any $\TIT$-derivation of $YM:\big\langle \cdots\big\rangle_k$, one can chose a fresh variable $y$ and replace each $\YY$ rule at height $h\in\N$ in the cluster (counting bottom-up), by a $@$ rule, now with hypotheses judgments the same exact one for $M$ and, instead of $YM:\langle \cdots\rangle$, a judgment $M^{(\mathbf{cl}(\pi)-h)}y:\langle \cdots\rangle$, so that the obtained derivation for $M^{(\mathbf{cl}(\pi)-h)}y$ has the following properties:
the variable $y$ is declared in the context as $y:[]$ (so it does not occur in the derivation, according to our notation which hides empty multiset context declarations); the rest of the contexts, the polynomials and the output types are exactly as in the original derivation for $YM$ at that step.
Therefore, we obtain in this way a derivation with conclusion judgment $M^{(\mathbf{cl}(\pi))}y\,:\big\langle \gamma_k \vdash^{s_k} a_k\big\rangle_k$.
Notice that it has exactly the same contexts, polynomials and output types as $\pi$.
\end{definition}

%%%%%%%%%%%%%%%%%%%%%%%%%%

\newsavebox\pisucc
\sbox\pisucc{\tiny
        \AXC{$\vdots\ \pi$}
        \noLine\UIC{$M:\big\langle \cdots\rangle$}
        \UIC{$\SUCC M:\big\langle \cdots\rangle$}
    \DisplayProof
}

\newsavebox\Pisucc
\sbox\Pisucc{\tiny
        \AXC{$\vdots\ \Pi$}
        \noLine\UIC{$\Gamma\vdash M:A$}
        \UIC{$\Gamma\vdash \SUCC M:A$}
    \DisplayProof
}
%%%%%%%%%   succ

\newsavebox\pipred
\sbox\pipred{\tiny
        \AXC{$\vdots\ \pi$}
        \noLine\UIC{$M:\big\langle \cdots\rangle$}
        \UIC{$\PRED M:\big\langle \cdots\rangle$}
    \DisplayProof
}

\newsavebox\Pipred
\sbox\Pipred{\tiny
        \AXC{$\vdots\ \Pi$}
        \noLine\UIC{$\Gamma\vdash M:A$}
        \UIC{$\Gamma\vdash \PRED M:A$}
    \DisplayProof
}
%%%%%%%%%   pred

\newsavebox\picast
\sbox\picast{\tiny
        \AXC{$\vdots\ \pi$}
        \noLine\UIC{$M:\langle \cdots\rangle$}
    \DisplayProof
}

\newsavebox\Picast
\sbox\Picast{\tiny
        \AXC{$\vdots\ \Pi$}
        \noLine\UIC{$\Gamma\vdash M:\Bool$}
        \UIC{$\Gamma\vdash M:\NAT$}
    \DisplayProof
}
%%%%%%%%%   cast

\newsavebox\pilam
\sbox\pilam{\tiny
        \AXC{$\vdots\ \pi$}
        \noLine\UIC{$M: \big \langle\cdots\big \rangle$}
        \UIC{$\lambda x.M: \big \langle \cdots\big \rangle$}
    \DisplayProof
}

\newsavebox\Pilam
\sbox\Pilam{\tiny
        \AXC{$\vdots\ \Pi$}
        \noLine\UIC{$\Gamma,x:B\vdash M:A$}
        \UIC{$\Gamma\vdash \lambda x.M:B\to A$}
    \DisplayProof
}
%%%%%%%%%   lam

\newsavebox\piplus
\sbox\piplus{\tiny
        \AXC{$\vdots \pi_1$}
        \noLine\UIC{$M: \big \langle \cdots \big \rangle$}
        \AXC{$\vdots \pi_2$}
        \noLine\UIC{$N: \big \langle \cdots\big \rangle$}
        \BIC{$M\choice{X}N: \big \langle\cdots\big \rangle$}
    \DisplayProof
}

\newsavebox\Piplus
\sbox\Piplus{\tiny
        \AXC{$\vdots\ \Pi_1$}
        \noLine\UIC{$\Gamma\vdash M:A$}
        \AXC{$\vdots\ \Pi_2$}
        \noLine\UIC{$\Gamma\vdash N:A$}
        \BIC{$\Gamma\vdash M\oplus_X N:A$}
    \DisplayProof
}
%%%%%%%%%   plus

\newsavebox\piifz
\sbox\piifz{\tiny
        \AXC{$\vdots\ \pi_1$}
        \noLine\UIC{$M : \big\langle \cdots\big\rangle$}
        \AXC{$\vdots\ \pi_2$}
        \noLine\UIC{$N : \big\langle \cdots\big\rangle$}
        \AXC{$\vdots\ \pi_3$}
        \noLine\UIC{$P : \big\langle \cdots\big\rangle$}
        \TIC{$\ITE M N P : \big\langle \cdots\big\rangle$}
    \DisplayProof
}

\newsavebox\Piifz
\sbox\Piifz{\tiny
        \AXC{$\vdots\ \Pi_1$}
        \noLine\UIC{$\Gamma\vdash M:\NAT$}
        \AXC{$\vdots\ \Pi_2$}
        \noLine\UIC{$\Gamma\vdash N:A$}
        \AXC{$\vdots\ \Pi_3$}
        \noLine\UIC{$\Gamma\vdash P:A$}
        \TIC{$\Gamma\vdash \ITE M N P:A$}
    \DisplayProof
}
%%%%%%%%%   ifz

\newsavebox\piapp
\sbox\piapp{\tiny
        \AXC{$\vdots \pi_1$}
        \noLine\UIC{$M: \big \langle \cdots \big \rangle$}
        \AXC{$\vdots \pi_2$}
        \noLine\UIC{$N: \big \langle \cdots\big \rangle$}
        \BIC{$MN: \big \langle\cdots\big \rangle$}
    \DisplayProof
}

\newsavebox\Piapp
\sbox\Piapp{\tiny
        \AXC{$\vdots\ \Pi_1$}
        \noLine\UIC{$\Gamma\vdash M:B\to A$}
        \AXC{$\vdots\ \Pi_2$}
        \noLine\UIC{$\Gamma\vdash N:B$}
        \BIC{$\Gamma\vdash MN:A$}
    \DisplayProof
}
%%%%%%%%%   app

\newsavebox\pifix
%\sbox\pifix{\tiny
%        \AXC{$\vdots \pi_0$}
%        \noLine\UIC{$M: \big \langle \cdots \big \rangle$}
%        \AXC{$\vdots \pi_1$}
%        \noLine\UIC{$\YY M: \big \langle \cdots\big \rangle$}
%       \BIC{$\YY M: \big \langle\cdots\big \rangle$}
%   \DisplayProof
%}
\sbox\pifix{\tiny
        \AXC{$\vdots\ \pi$}
        \noLine\UIC{$\YY M:\langle\cdots\rangle$}
    \DisplayProof
}

\newsavebox\Pifix
\sbox\Pifix{\tiny
        \AXC{$\vdots\ \Pi$}
        \noLine\UIC{$\Gamma\vdash M:A\to A$}
        \UIC{$\Gamma\vdash \YY M:A$}
    \DisplayProof
}
%%%%%%%%%   fix

\newsavebox\pigeneric
\sbox\pigeneric{\tiny
        \AXC{$\vdots\ \pi_1$}
        \noLine\UIC{$M_1: \big \langle\cdots\big \rangle$}
        \AXC{$\cdots$}
        \AXC{$\vdots\ \pi_h$}
        \noLine\UIC{$M_h: \big \langle\cdots\big \rangle$}
        \TIC{$\odot(M_1,\dots,M_h): \big \langle \cdots\big \rangle$}
    \DisplayProof
}

\newsavebox\Pigeneric
\sbox\Pigeneric{\tiny
        \AXC{$\vdots\ \Pi_1$}
        \noLine\UIC{$\Gamma_1\vdash M_1: A_1$}
        \AXC{$\cdots$}
        \AXC{$\vdots\ \Pi_h$}
        \noLine\UIC{$\Gamma_h\vdash M_h: A_h$}
        \TIC{$\odot(\Gamma_1,\dots,\Gamma_h)\vdash \odot(M_1,\dots,M_h): \odot(A_1,\dots,A_h)$}
    \DisplayProof
}
%%%%%%%%%   generic rule

%%%%%%%%%%%%%%%%%%%%%%%%%%

\begin{figure}
\fbox{\begin{minipage}{.99\textwidth}

Minimal size: 

\vspace{1em}

\begin{minipage}{0.01\textwidth}\scriptsize
\begin{prooftree}
\AXC{}
\dashedLine
\UIC{$\appr{\dfrac{}{\ZERO:\langle \cdots \rangle}}{\dfrac{}{\Gamma\vdash \ZERO:\NAT}}$}
\end{prooftree}
\end{minipage} %%%%%%%%%   0 : nat 
\hfill
\begin{minipage}{.01\textwidth}\scriptsize
\begin{prooftree}
\AXC{}
\dashedLine
\UIC{$\appr{\dfrac{}{\ZERO:\langle \cdots \rangle}}{\dfrac{}{\Gamma\vdash \ZERO:\Bool}}$}
\end{prooftree}
\end{minipage} %%%%%%%%%   0 : bool
\hfill
\begin{minipage}{.01\textwidth}\scriptsize
\begin{prooftree}
\AXC{}
\dashedLine
\UIC{$\appr{\dfrac{}{\ONE:\langle \vdash^1 1 \rangle}}{\dfrac{}{\Gamma\vdash \ONE:\Bool}}$}
\end{prooftree}
\end{minipage} %%%%%%%%%   1 : bool
\hfill
\begin{minipage}{.25\textwidth}\scriptsize
\begin{prooftree}
\AXC{}
\dashedLine
\UIC{$\appr{\dfrac{}{x: \langle \cdots \rangle}}{\dfrac{}{\Gamma,x:A\vdash x:A}}$}
\end{prooftree}
\end{minipage} %%%%%%%%%   var

\vspace{1em}

Successor size

%\begin{minipage}{.29\textwidth}\scriptsize
%\begin{prooftree}
%\AXC{$\appr \pi \Pi$}
%\dashedLine
%\UIC{$\appr{\usebox\pisucc}{\usebox\Pisucc}$}
%\end{prooftree}
%\end{minipage} %%%%%%%%%   succ
%\hfill
%\begin{minipage}{.29\textwidth}\scriptsize
%\begin{prooftree}
%\AXC{$\appr \pi \Pi$}
%\dashedLine
%\UIC{$\appr{\usebox\pipred}{\usebox\Pipred}$}
%\end{prooftree}
%\end{minipage} %%%%%%%%%   pred
%\hfill
\begin{minipage}{.5\textwidth}\scriptsize
\begin{prooftree}
\AXC{$\appr{\pi_1}{\Pi_1}$}
\AXC{$\cdots$}
\AXC{$\appr{\pi_h}{\Pi_h}$}
\dashedLine
\TIC{$\appr{\usebox\pigeneric}{\usebox\Pigeneric}$}
\end{prooftree}
\end{minipage} %%%%%%%%%   generic rule
\hfill
\begin{minipage}{0.25\textwidth}\scriptsize
\begin{prooftree}
\AXC{$\appr \pi \Pi$}
\dashedLine
\UIC{$\appr{\usebox\picast}{\usebox\Picast}$}
\end{prooftree}
\end{minipage} %%%%%%%%%   cast
%
%\vspace{1em}
%
%\begin{minipage}{.39\textwidth}\scriptsize
%\begin{prooftree}
%\AXC{$\appr \pi \Pi$}
%\dashedLine
%\UIC{$\appr{\usebox\pilam}{\usebox\Pilam}$}
%\end{prooftree}
%\end{minipage} %%%%%%%%%   lam
%\hfill
%\begin{minipage}{.59\textwidth}\scriptsize
%\begin{prooftree}
%\AXC{$\appr{\pi_1}{\Pi_1}$}
%\AXC{$\appr{\pi_2}{\Pi_2}$}
%\dashedLine
%\BIC{$\appr{\usebox\piplus}{\usebox\Piplus}$}
%\end{prooftree}
%\end{minipage} %%%%%%%%%   plus
%
%\vspace{1em}
%
%\begin{minipage}{.99\textwidth}\scriptsize
%\begin{prooftree}
%\AXC{$\appr{\pi_1}{\Pi_1}$}
%\AXC{$\appr{\pi_2}{\Pi_2}$}
%\AXC{$\appr{\pi_3}{\Pi_3}$}
%\dashedLine
%\TIC{$\appr{\usebox\piifz}{\usebox\Piifz}$}
%\end{prooftree}
%\end{minipage} %%%%%%%%%   ifz
%
%\vspace{1em}
%
%\begin{minipage}{.99\textwidth}\scriptsize
%\begin{prooftree}
%\AXC{$\appr{\pi_1}{\Pi_1}$}
%\AXC{$\appr{\pi_2}{\Pi_2}$}
%\dashedLine
%\BIC{$\appr{\usebox\piapp}{\usebox\Piapp}$}
%\end{prooftree}
%\end{minipage} %%%%%%%%%   app

\vspace{1em}

Limit size

\begin{minipage}{.99\textwidth}\scriptsize
\begin{prooftree}
\AXC{$\appr{\mathbf{unfold}.\pi}{\mathbf{unfold}^{(\mathbf{cl}(\pi))}.\Pi}$}
\dashedLine
\UIC{$\appr{\usebox\pifix}{\usebox\Pifix}$}
\end{prooftree}
\end{minipage} %%%%%%%%%   fix

\end{minipage}}
\caption{Definition of the refinement $\appr\pi\Pi$ for adequate $\pi\neq\emptyset$. For the case $\pi$ obtained via a $\emptyset$-rule, we set, in each case of the transfinite induction on $\Pi$, that $\appr{\frac{}{M:\emptyset}}{\Pi}$, where  $\deriv{\Pi}{\Gamma\vdash M:A}$ (notice the same $M$).\\ 
Three more explanations: i) we use dashed lines to indicate the inductive step of the definition, in order to distinguish it from the lines building the derivation trees. ii) In the bottom left rule, $\odot$ is a $h$-ary rule of \pPCF\ among $\SUCC,\PRED,\oplus_X,\mathsf{ifz},@,\lambda x$ (for $x$ any variable); we use it to denote the respective term constructor, type constructor and judgment that arise from the application of rule. For instance, in all rules except $\lambda$, we actually have $\Gamma_1=\cdots=\Gamma_h$ and $\odot(\Gamma_1,\dots,\Gamma_h)=\Gamma_1$.
iii) The content of the dotted parts is always uniquely determined by the given information (remembering Figure~\ref{fig:tITrules} and Definition~\ref{def:adequate}).
One last comment: in all the cases the structure of $\pi$ mimics the one of $\Pi$ in the expected way, except the ones for $\YY$ and for the cast. In the former, we unfold $\YY$ as a (finite) application following Definition~\ref{def:unfold_Pi} and Definition~\ref{def:unfold_pi}; in the latter, we ignore the case rule in $\Pi$ and jump to its previous last rule.}
\label{fig:refines}
\end{figure}

We now associate an ordinal size to each \pPCF-derivation, and define the notion of refinement by induction on such size:

\begin{definition}
We associate, by structural induction on the derivation tree, each derivation $\Pi$ of \pPCF\ with an ordinal size $o(\Pi)$, as follows\footnote{Notice that it follows from the definition that, if the derivation does not contain $Y$, then the size is simply the number of rules in the derivation tree.}:
\[\begin{array}{ccl}
    o(\textit{any axiom derivation}) & := & 1\in\N \\
    o(\odot(\Pi_1,\dots,\Pi_h)) & :=  & (\sum_{i=1}^h
    o(\Pi_i))+1 \quad \textit{for } \odot=\textit{successor, predecessor, $\oplus$, ifz, $@$, cast, $\lambda x$} \\
    o(\YY(\Pi)) & := & \bigvee\limits_{n\in\N} ((o(\Pi)\cdot n)+1+n)
\end{array}\]
%Finally, again by structural induction on $\Pi$, we define the size of $\Pi$ as the pair $(r(\Pi),o(\Pi))$, where $r(\Pi)$ is the name of the last rule of $\Pi$ (i.e.\ it is one of ax $0$ bool, ax $1$ bool, ax $1$ nat, succ, pred, $\oplus$, ifz, $@$, cast, $\lambda x$ (for any variable $x$), $\YY$).Such pairs are ordered by $(r,o)\leq (r',o')$ iff $o\leq o'$.
\end{definition}

The definition makes sense, because $o(\Pi)\cdot n\leq o(\Pi)\cdot (n+1)$ for all $n\in\N$.

\begin{remark}\label{rmk:sizeY}
One easily sees that 
$(o(\Pi)\cdot n)+1+n=o(\mathbf{unfold}^{(n)}.\Pi)$.

So, $o(\YY(\Pi))=\bigvee\limits_{n\in\N} o(\mathbf{unfold}^{(n)}.\Pi)$.
\end{remark}

We can finally give the following

\begin{definition}[Refinement]\label{def:refines}
For all \pPCF-derivations $\deriv{\Pi}{\Gamma\vdash M:A}$, we define a predicate $\appr{\pi}{\Pi}$ on $\TIT$-derivations $\pi$ adequate for $\Gamma\vdash M:A$, and we say that $\pi$ \emph{refines $\Pi$}.
The predicate is defined by transfinite induction on the size of $\deriv{\Pi}{\Gamma\vdash M:A}$ according to the rules in Figure~\ref{fig:refines}.
\end{definition}

Also, one can straightforwardly see that the definition makes sense: on the one hand, the condition of being adequate propagates correctly through the inductive steps, reading the rules top to bottom).
On the other hand, the induction is well-founded because thanks to Remark~\ref{rmk:sizeY} it is immediately seen that each case calls the definition for strictly smaller sized derivations.

\subsection{Soundness and completeness}

Remember that, given $\deriv \Pi {\Gamma\vdash M:A}$ in \pPCF, for $a\in\model{A}$ and $\gamma(x)\in !\model{\Gamma(x)}$ for all $x$ declared in $\Gamma$, we have $\model{\Pi}^{\mathbb X}_{\gamma \ | \ a}\in\fps{\NINF}{\mathbb X}$ ($\mathbb X$ being the set $\set{X_1,\OV X_1,\dots,X_n,\OV X_n}$).

The following result relates the n.i.i.-type system to the WRS. %As usual, it shows that they are but different presentations of the same entity (namely, the weights of all the reductions of a term to normal form).

\begin{theorem}[Soundness of $\TIT$ wrt WRS]\label{thm:soundwrtwrs}
    %Let $\Gamma\vdash M:A$, $\gamma\in !\model{\Gamma}$.
    %Let $a\in\model A$ and $s$ a minimal polynomial (i.e.\ $s=\VN(s)$) over $X,\OV X$.
    %Then
    %$\gamma\cdot\mu\in\mathrm{supp}(\model{\Gamma\vdash M:A}^{\mathbb X}_a)$ for all $\mu\in\mathrm{supp}(s)$, iff $M: \big \langle \gamma\vdash^{s} a  \big \rangle$.
%
    Let $\deriv \Pi \Gamma\vdash M:A$ and fix $(\gamma,a)$ adequate for $(\Gamma,A)$.
    Then\footnote{To be pedantic, in the statement below we mean $\mathrm{supp}(\model{\Pi}^{\mathbb X}_{{\gamma}_{|_{\mathrm{dom}(\Gamma)}}\,|\, a})$.}:
    \[
    \bigcup\limits_{\tiny{\appr{\deriv \pi {M: \big \langle \gamma\vdash^{s_\pi} a \big \rangle \ }}{ \ \Pi}}}\!\!\!\!\!\!\!\!\!\!\!\!\mathrm{supp}(s_\pi) 
    \subseteq
    \mathrm{supp}(\model{\Pi}^{\mathbb X}_{\gamma \,|\, a}).
    \]
    %Here for a derivation $\pi$ with polynomial $s_\pi$, we set $\mathrm{NP}(\pi):=\mathrm{supp}(s_\pi)$.
    \end{theorem}
\begin{proof}
The result follows by taking $K$ a singleton in the claim below, which states a more general statement so that an induction argument goes through smoothly.
We claim that the following holds:
if $\deriv \pi {M: \big \langle \gamma_k\vdash^{s_k} a_k \big \rangle_{k\in K}}$ under the hypotheses of the statement, then $\mathrm{supp}(s_k)
\subseteq
\mathrm{supp}(\model{\Pi}^{\mathbb X}_{\gamma_k\,|\, a_k})$ for all $k\in K$.
This claim is precisely stated and proven as the following Lemma~\ref{lm:sound}.
\end{proof}

Before giving the proof of Lemma~\ref{lm:sound} (which requires some non trivial argument), let us discuss the general picture that follows from Theorem~\ref{thm:soundwrtwrs}.

The trained eye would recognise the similarity with \cite[Equation 10, Lemma 10]{EhrhardTassonPagani14}.
In fact, our formula is a version of that formula where we keep the parameters (instead of instantiating them), and where we are only interested in the support of the interpretation (this comes from the fact that we only need the tropicalisation $\mathsf t$), in addition to having only an inclusion instead of an equality.

In fact, an equivalent formulation of the above is the following.

\begin{remark}
Remember that, given $\deriv \Pi {x_1:\Gamma(x_1),\dots,x_k:\Gamma(x_k)\vdash M:A}$ in \pPCF, for $a\in\model{A}$ we have $\model{\Pi}^{\mathbb X}_{a}\in\fps{(\fps{\NINF}{\mathbb X})}{x_{\Gamma(x_1)},\dots,x_{\Gamma(x_k)}}$ ($\mathbb X$ being the set $\set{X_1,\OV X_1,\dots,X_n,\OV X_n}$), where $x_{\Gamma(x_i)}$ is sugar for $\model{\Gamma(x_i)}$ (so there is actually a possibly infinite amount of variables).
More concretely, $\model{\Pi}^{\mathbb X}_{a}$ is the formal power series 
\[
\model{\Pi}_{a}=\sum\limits_{\gamma \textit{ adequate for }\Gamma} \model{\Pi}_{\gamma\,|\,a}\, x_{\Gamma(x_1)}^{\gamma(x_1)}\cdots x_{\Gamma(x_k)}^{\gamma(x_k)}
\]
where $\model{\Pi}_{\gamma\,|\,a}\in \fps{\NINF}{\mathbb X}$ can be looked at in \cite[Def.\ IV.10 and Figure 3]{Manzo2013}, by taking the semiring of coefficients ($\mathcal{R}$, with their notations) the semiring $\fps{\NINF}{\BB X}$.
\end{remark}

Writing the above as a fps in $\fps{\NINF}{\mathbb X,x_{\Gamma(x_1)},\dots,x_{\Gamma(x_k)}}$ instead, we can see that\footnote{To be pedantic, in the formula below we are secretly casting $a\in\model A$ to a $\TIT$-type $\tilde a$, in the obvious way: if $A=\NAT$ then $\tilde a:=a$, if $A=\Bool$ then $\tilde a:=a\in\set{0,1}\subseteq\N$ and if $A=E\to F$ then $a=:([e_1,\dots,e_h],f)$ and $\tilde a:=[\tilde e_1,\dots,\tilde e_h]\multimap \tilde f$.
As a side note, notice that not all $\TIT$-types are of shape $\tilde a$: only the ``coherent'' ones are.}, for all $\appr{\deriv \pi {M: \big \langle \gamma\vdash^{s} a \big \rangle  }}{\Pi}$ and $X_1^{i_1}\OV X_1^{j_1}\cdots X_n^{i_n}\OV X_n^{j_n}\in\mathrm{supp}(s)$, we have:
\[
x_{\Gamma(x_1)}^{\gamma(x_1)}\cdots x_{\Gamma(x_k)}^{\gamma(x_k)}X_1^{i_1}\OV X_1^{j_1}\cdots X_n^{i_n}\OV X_n^{j_n}
\in
\mathrm{supp}(\model{\Pi}^{\mathbb X}_a)
\]

This formulation, in turn, corresponds to \cite[Equation 11, Definition 11]{EhrhardTassonPagani14} and shows how $\TIT$ characterises \emph{a part} of the tropicalisation of the parametric interpretation.

From either formulation, in the case $\Gamma=\emptyset$ and $A=\NAT$, we immediately obtain:
\[
\trop\model{\deriv \Pi {\vdash M:\NAT}}^{\mathbb X}_{\mathsf n}(q)
\leq
\inf\set{\mu\cdot q \mid \mu\in\mathrm{supp}(s) \textit{ for some } M: \big \langle \vdash^{s} n  \big \rangle}
\]
for all $n\in\N$ and $q\in\mathbb T^{2n}$ ($n$ being the cardinality of $\mathbb X$).

With this, we immediately have the main statement of Theorem \ref{thm:correct}:

\begin{proof}[Proof of the first statement of Theorem \ref{thm:correct}]
The statement is now that:

$\trop \model{\deriv \Pi {\vdash M:\NAT}}_{\mathsf n}(q)\leq s^!(q)$ for all $s$ generated by an adequate $\appr\pi\Pi$.

This is immediate from the inequality above.
\end{proof}

In fact, we can say more: the part captured by $\TIT$ is at least the minimal one:

\begin{theorem}[Completness of $\TIT$ wrt \emph{minimal} WRS]\label{th:completeness}
Let $\deriv \Pi \Gamma\vdash M:A$ and fix $(\gamma,a)$ adequate for $(\Gamma,A)$.
    Then\footnote{To be pedantic, in the statement below we mean $\mathrm{supp}(\model{\Pi}^{\mathbb X}_{{\gamma}_{|_{\mathrm{dom}(\Gamma)}}\,|\, a})$.}:
    \[
    \mathrm{supp}(\model{\Pi}^{\mathbb X}_{\gamma \,|\, a})_{\mathrm{min}}
    \subseteq
    \bigcup\nolimits_{\tiny{\appr{\deriv \pi {M: \big \langle \gamma\vdash^{s_\pi} a \big \rangle \ }}{ \ \Pi}}}\mathrm{supp}(s_\pi).
    \]
\end{theorem}
\begin{proof}
The result follows by taking $K$ a singleton in the claim below, which states a more general statement so that an induction argument goes through smoothly.
We claim that the following holds:
for all $k\in K$, for all $\mu\in \mathrm{supp}(\model{\Pi}^{\mathbb X}_{\gamma_k \,|\, a_k})_{\mathrm{min}}$, there is $\deriv{\pi_{k,\mu}}{M: \big \langle \gamma_k\vdash^{s} a_k \big \rangle}$ such that $\mu\in\mathrm{supp}(s)$.
This claim is precisely stated and proven in the following Lemma~\ref{lm:min_compl}.
\end{proof}

Before giving the proof of Lemma~\ref{lm:min_compl}, let us discuss the general picture that follows from Theorem~\ref{thm:soundwrtwrs}.

As before, we can reformulate what we got as: 
all $\mu\in\mathrm{supp}(\model{\Pi}^{\mathbb X}_a)_{\mathrm{min}}$ are of shape
\[\mu=x_{\Gamma(x_1)}^{\gamma(x_1)}\cdots x_{\Gamma(x_k)}^{\gamma(x_k)}X_1^{i_1}\OV X_1^{j_1}\cdots X_n^{i_n}\OV X_n^{j_n}\]
for some $\appr{\deriv \pi {M: \big \langle \gamma\vdash^{s} a \big \rangle  }}{\Pi}$ with $X_1^{i_1}\OV X_1^{j_1}\cdots X_n^{i_n}\OV X_n^{j_n}\in\mathrm{supp}(s)$.

Now, remember the definition of tropicalisation and that in general $((\model{\_}^{\mathbb X}_a)_{\mathrm{min}})^!=\trop\model{\_}^{\mathbb X}_a$ as discussed at the end of Section 6.2.
Therefore, we obtain from the equality above:
%\[\trop\model{\vdash M:\NAT}^{\mathbb X}_{\mathsf n}(x) \leq \Big(\sum_{M: \big \langle \vdash^{s} n  \big \rangle} s\Big)^!(x) \leq (\trop\model{\vdash M:\NAT}^{\mathbb X}_{\mathsf n})_{\mathrm{min}}(x) = \trop\model{\vdash M:\NAT}^{\mathbb X}_{\mathsf n}(x)\]
\[
\trop\model{\deriv \Pi {\vdash M:\NAT}}^{\mathbb X}_{\mathsf n}(q)
=
\inf\set{\mu\cdot q \mid \mu\in\mathrm{supp}(s) \textit{ for some } M: \big \langle \vdash^{s} n  \big \rangle}
\]
for all $n\in\N$ and $q\in\mathbb T^{2n}$ ($n$ being the cardinality of $\mathbb X$).

Let us now prove the claimed Lemma~\ref{lm:sound} and Lemma~\ref{lm:min_compl}.

We need two auxiliary Lemmas.

\begin{lemma}\label{lm:supp:dot=+}
Let $Q$ be a continuous commutative semiring $Q$ and $s,t\in\fps{Q}{\Sigma}$.
    If $Q$ has no non-zero zero-divisors (e.g.\ $\NINF$), then $\mathrm{supp}(st)=\mathrm{supp}(s)+\mathrm{supp}(t)$.
\end{lemma}
\begin{proof}
We have:
\[\begin{array}{ccl}
    \mathrm{supp}(st) & = & \set{\rho\in !\Sigma \mid \sum_{\mu+\eta=\rho} s_\mu t_\eta \neq 0_Q} \\
    & = &
    \set{\rho\in !\Sigma \mid \exists \mu,\eta\in!\Sigma \textit{ s.t. } \mu+\eta=\rho \textit{ and } s_\mu t_\eta \neq 0_Q} \\
    & = &
    \set{\rho\in !\Sigma \mid \exists \mu,\eta\in!\Sigma \textit{ s.t. } \mu+\eta=\rho \textit{ and } s_\mu \neq 0_Q \textit{ and } t_\eta \neq 0_Q} \\
    & = & 
    \mathrm{supp}(s)+\mathrm{supp}(t).
\end{array}
\]
The first equality is by definition of Cauchy product, the second because $Q$ is positive as it is continuous, and the third is because it has no non-zero zero-divisors.
\end{proof}

\begin{lemma}\label{lm:supp:sup=cup}
Let $Q$ be a continuous commutative semiring $Q$ (e.g.\ $\NINF$) and $(t_n)_{n\in\N}\subseteq\fps{Q}{\Sigma}$.
Then $\mathrm{supp}(\bigvee_{n\in\N}t_n)=\bigcup_{n\in\N}\mathrm{supp}(t_n)$.
\end{lemma}
\begin{proof}
We have:
\[\begin{array}{ccl}
    \mathrm{supp}(\bigvee_{n\in\N}t_n) & = & \mathrm{supp}(\sum_\mu (\bigvee_n (t_n)_\mu)\mu) \\
    & = &
    \set{\mu\in !\Sigma \mid \bigvee_n (t_n)_\mu\neq 0_Q} \\
    & = &
    \set{\rho\in !\Sigma \mid \exists n\in\N \textit{ such that } (t_n)_\mu\neq 0_Q} \\
    & = & 
    \bigcup_{n\in\N}\mathrm{supp}(t_n).
\end{array}
\]
The first equality is because the definition of supremum of fps's is pointwise, the others are clear.
\end{proof}

Now we can finally state and prove, first of all, Lemma~\ref{lm:sound}.

\begin{lemma}\label{lm:sound}
Let $\deriv \Pi \Gamma\vdash M:A$ and, for finite $K\neq\emptyset$, fix a family $(a_k)_k$ of $\TIT$-types and $(\gamma_k)_k$ a family of $\TIT$-context such that $(\gamma_k,a_k)$ is adequate for $(\Gamma,A)$ for all $k\in K$.

For all $\deriv \pi {M: \big \langle \gamma_k\vdash^{s_k} a_k \big \rangle_k}$ such that $\appr{\pi}{\Pi}$, we have\footnote{To be pedantic, in the statement below we mean $\mathrm{supp}(\model{Pi}^{\mathbb X}_{{\gamma_k}_{|_{\mathrm{dom}(\Gamma)}}\,|\, a_k})$.} $\mathrm{supp}(s_k) 
    \subseteq
    \mathrm{supp}(\model{\Pi}^{\mathbb X}_{\gamma_k \,|\, a_k})$ for all $k\in K$.
\end{lemma}
\begin{proof}
We go by transfinite induction on the size of $\Pi$.

$\star$ Minimal size cases:

$\bullet$ Case $\Pi$ is the axiom $\Gamma\vdash \ZERO:\NAT$, or the axiom $\Gamma\vdash \ZERO:\Bool$ or the axiom $\Gamma\vdash \ONE:\Bool$.

In the first two cases $\pi$ is the axiom $\ZERO: \big \langle \vdash^{1_{\fps{\NINF}{\mathbb X}}} 0  \big \rangle$. %via the axiom rule $\ZERO$ and $a=0\in\N$.
Now, $\model{\Pi}_{\vec{[]} \,|\, 0}=1_{\fps{\NINF}{\mathbb X}}$, so we are done.
In the third case $\pi$ is the axiom $\ONE: \big \langle \vdash^{1_{\fps{\NINF}{\mathbb X}}} 1  \big \rangle$. %via the axiom rule $\ZERO$ and $a=0\in\N$.
Now, $\model{\Pi}_{\vec{[]} \,|\, 1}=1_{\fps{\NINF}{\mathbb X}}$, so we are done.
%=\set{\prod_{j}X_j^0\OV X_j^0}

$\bullet$ Case $\Pi$ is the axiom $\Gamma, y:A\vdash y:A$.

Then $\pi$ is an axiom $y: \big \langle y:[a_k] \vdash^{1_{\fps{\NINF}{\mathbb X}}} a_k  \big \rangle_k$. %via the axiom rule $\mathrm{id}$.
Now, $\model{\Pi}_{\gamma(x\neq y)=\vec{[]},\gamma(y)=[a_k] \,|\, a_k}=1_{\fps{\NINF}{\mathbb X}}$, so we are done.
%=\set{\prod_{j}X_j^0\OV X_j^0}

$\star$ Successor size cases:

$\bullet$ Case $\deriv{\Pi}{\Gamma\vdash\NAT}$ via a cast rule on $\deriv{\Pi'}{\Gamma\vdash\Bool}$.

Then $\deriv{\pi}{M:\big \langle \gamma_k \vdash^{s_k} n_k \big \rangle_k}$ for $n_k\in\N$ (because the output type has to be adequate for $\NAT$).
Moreover, $\appr{\pi}{\Pi'}$.
From this we get two facts:
first, by IH, $\mathrm{supp}(s_k)\subseteq \model{\Pi'}_{\gamma_k \,|\, n_k}$.
Second, $(\gamma_k,n_k)$ is adequate for $(\Gamma,\Bool)$, and in particular $n_k\in\set{0,1}$.
But from the latter it follows, by definition of the interpretation, that $\model{\Pi}_{\gamma_k \,|\, n_k}=\model{\Pi'}_{\gamma_k \,|\, n_k}$, so we are done.

$\bullet$ Case $\deriv{\Pi}{\Gamma\vdash \SUCC M:\NAT}$ via a successor rule on $\deriv{\Pi_1}{\Gamma\vdash M:\NAT}$.

Then $\deriv{\pi}{\SUCC M:\big \langle \gamma_k \vdash^{s_k} n_k+1 \big \rangle_k}$ via a successor rule on a $\deriv{\pi_1}{M:\big \langle \gamma \vdash^{s_k} n_k \big \rangle_k}$, where $n_k\in\N$ for all $k$.
Moreover, $\appr{\pi_1}{\Pi_1}$. 
By IH, we have $\mathrm{supp}(s_k)\subseteq\mathrm{supp}(\model{\Pi_1}_{\gamma_k \,|\, n_k})$.
Let us now analyse $\model{\Pi}_{\gamma_k \,|\, n_k+1}$.
It is equal to $\model{\Pi_1}_{\gamma_k \,|\, n_k}$, so we are done.

$\bullet$ Case $\deriv{\Pi}{\Gamma\vdash \PRED M:\NAT}$ via a predecessor rule on $\deriv{\Pi_1}{\Gamma\vdash M:\NAT}$.

Then $\deriv{\pi}{\PRED M:\big \langle \gamma_k \vdash^{s} n_k\dotdiv 1 \big \rangle_k}$ via a predecessor rule on a $\deriv{\pi_1}{M:\big \langle \gamma_k \vdash^{s_k} n_k \big \rangle_k}$ with $n_k\in\N$ for all $k$.
Moreover, $\appr{\pi_1}{\Pi_1}$. 
By IH, we have $\mathrm{supp}(s_k)\subseteq\mathrm{supp}(\model{\Pi_1}_{\gamma_k \,|\, n_k})$.
Let us now analyse $\model{\Pi}_{\gamma_k \,|\, n_k\dotdiv1}$.
We have two cases:
Either $n_k\dotdiv1=0$, and then either $n_k=0$ or $n_k=1$; or $n_k\dotdiv1\geq 1$ and then $n_k\dotdiv1=n_k-1$.
In the case $n_k\dotdiv1=0$, the interpretation equals $\model{\Pi_1}_{\gamma_k \,|\, 0}+\model{\Pi_1}_{\gamma_k \,|\, 1}$ and so its support is the union of the supports of the two addends. %, which in particular contains $\mathrm{supp}(\model{\Gamma\vdash M:\NAT}_{\gamma_k \,|\, 1})$.
In the case $n_k\dotdiv1\geq 1$, it is equal to $\model{\Pi_1}_{\gamma_k \,|\, (n_k\dotdiv1)+1}$. 
%We observe that in both cases $\mathrm{supp}(\model{\Gamma\vdash \PRED M:\NAT}_{\gamma_k \,|\, n_k})\supseteq \mathrm{supp}(\model{\Gamma\vdash M:\NAT}_{\gamma_k \,|\, n_k+1})$, and s
One immediately sees that so we are done.

$\bullet$ Case $\deriv{\Pi}{\Gamma\vdash \lambda y.M:E\to F}$ via a $\lambda y$ rule (for $y$ non-declared in $\Gamma$) on $\deriv{\Pi_1}{\Gamma,y:E\vdash M:F}$.

Then $\deriv{\pi}{\lambda y. M:\big \langle \gamma_k \vdash^{s_k} \eta_k\multimap f_k \big \rangle_k}$ via a $\lambda y$ rule on a $\deriv{\pi_1}{M:\big \langle \gamma_k,y:\eta_k \vdash^{s_k} f_k \big \rangle_k}$ with $\eta_k\in !\model E$ and $f_k\in\model F$. % and $a_k=(\eta_k,f_k)$.
Moreover, $\appr{\pi_1}{\Pi_1}$.
By IH, we have $\mathrm{supp}(s_k)\subseteq\mathrm{supp}(\model{\Pi_1}_{\gamma_k[y\mapsto\eta_k] \,|\, f_k})$.
Let us now analyse $\model{\Pi}_{\gamma_k \,|\, (\eta_k,f_k)}$.
It is equal to $\model{\Pi_1}_{x\neq y\mapsto\gamma_k(x),y\mapsto\eta_k \,|\, f_k}$, so we are done.

$\bullet$ Case $\deriv{\Pi}{\Gamma\vdash M\oplus_X N:A}$ via a $\oplus$ rule on $\deriv{\Pi_1}{\Gamma\vdash M:A}$ and $\deriv{\Pi_2}{\Gamma\vdash N:A}$.

Then $\deriv{\pi}{M\oplus_X N:\big \langle \gamma_k \vdash^{s_k} a_k \big \rangle_k}$ via a $\oplus$ rule on a $\deriv{\pi_1}{M:\big \langle \theta_i \vdash^{t_i} c_i \big \rangle_{i\in I}}$ and a $\deriv{\pi_2}{N:\big \langle \delta_j \vdash^{u_j} d_j \big \rangle_{j\in J}}$.
Moreover, $\appr{\pi_1}{\Pi_1}$ and $\appr{\pi_2}{\Pi_2}$.
By IH we have $\mathrm{supp}(t_i)\subseteq \mathrm{supp}(\model{\Pi_1}_{\theta_i\,|\, c_i})$ and $\mathrm{supp}(u_j)\subseteq \mathrm{supp}(\model{\Pi_2}_{\delta_j\,|\, d_j})$ for all $i,j$.
Now, by definition of $\SELECT$ and because the family $(\theta_i,c_i)_i$ is made of pairwise distinct elements by construction, and the same for the family $(\delta_j,d_j)_j$, for each $k$ we are in exactly one case among the three below:
%this means that there is a function $g:L \to I+J+(I\times J)$ such that for all $k$,
\[\begin{array}{ll}
    \textit{case }1: (\gamma_k,s_k,a_k) = (\theta_i,\VN(Xt_i),c_i) & \textit{for a unique }i\in I %g(l)=:i\in I 
    \\
    \textit{case }2: (\gamma_k,s_k,a_k) =  (\delta_j,\VN(\OV X u_j),d_j) & \textit{for a unique }j\in J %g(l)=:j\in J 
    \\
    \textit{case }3: (\gamma_k,s_k,a_k) = (\theta_i,\VN(Xt_i+\OV X u_j),c_i)=(\delta_j,\VN(Xt_i+\OV X u_j),d_j) & \textit{for a unique }(i,j)\in I\times J %g(l)=:(i,j)\in I\times J.
\end{array}\]
%So $K$ is partitioned in three sets $K_I:=g^{-1}I,K_J:=g^{-1}J,\hat K:=g^{-1}(I\times J)$.

Remember that $\VN(\_)=(\_)_{\mathrm{min}}$ and so, by definition of minimal polynomial, one has $\mathrm{supp}(\VN(\_))\subseteq\mathrm{supp}(\_)$.
Hence, for each $k$, we have:
\[\begin{array}{ll}
    \textit{case }1: \mathrm{supp}(s_k)\subseteq X\mathrm{supp}(t_i) \subseteq & \!\!\!\!X\mathrm{supp}(\model{\Pi_1}_{\theta_i\,|\, c_i}) \\
    \textit{case }2:\mathrm{supp}(s_k)\subseteq \OV X\mathrm{supp}(u_j) \subseteq & \phantom{X\mathrm{supp}(\model{\Pi_1}_{\theta_i\,|\, c_i}) \cup}\!\!\OV X\mathrm{supp}(\model{\Pi_2}_{\delta_j\,|\, d_j}) \\
    \textit{case }3: \mathrm{supp}(s_k)\subseteq X\mathrm{supp}(t_i)\cup \OV X\mathrm{supp}(u_j) \subseteq & \!\!\!\!X\mathrm{supp}(\model{\Pi_1}_{\theta_i\,|\, c_i}) \cup \OV X\mathrm{supp}(\model{\Pi_2}_{\delta_j\,|\, d_j}).
\end{array}\]

Finally, let us analyse $\model{\Pi}_{\gamma_k\,|\, a_k}$.
Following its definition, its support is precisely $X\mathrm{supp}(\model{\Pi_1}_{\gamma_k\,|\, a_k})\cup\OV X\mathrm{supp}(\model{\Pi_2}_{\gamma_k\,|\, a_k})$.
But for each $k$ we are in exactly one among the three cases above, and in the first $(\gamma_k,a_k)=(\theta_i,c_i)$ and similarly for the other two.
Therefore, we are done.

$\bullet$ Case $\deriv{\Pi}{\Gamma\vdash \ITE{M}{N}{P}:A}$ via a $\mathsf{ifz}$ rule on $\deriv{\Pi_1}{\Gamma\vdash M:\NAT}$ and $\deriv{\Pi_2}{\Gamma\vdash N:A}$ and $\deriv{\Pi_3}{\Gamma\vdash P:A}$.

Then $\deriv{\pi}{\ITE{M}{N}{P}:\big \langle \gamma_k \vdash^{s_k} a_k \big \rangle_k}$ is via a $\mathrm{ifz}$ rule an a $\deriv{\pi_1}{M:\big \langle \zeta_0\vdash^{v_0} 0 \ \big| \ \zeta_{i+1} \vdash^{v_{i+1}} i+1 \big \rangle_{i\in I}}$ and a $\deriv{\pi_2}{N:\big \langle \theta_j \vdash^{t_j} c_j \big \rangle_{j\in J}}$ and a $\deriv{\pi_3}{P:\big \langle \delta_l \vdash^{u_l} d_l \big \rangle_{l\in L}}$.
Moreover, $\appr{\pi_1}{\Pi_1}$, $\appr{\pi_2}{\Pi_2}$ and $\appr{\pi_3}{\Pi3}$.
Before applying the IH, let us observe that, by definition of $\SELECT$ and because the family $(\theta_j,c_j)_j$ is made of pairwise distinct elements by construction, and the same for the family $(\delta_l,d_l)_l$, for each $k$ we are in exactly one case among the two below:
\[\begin{array}{rll}
    \textit{case }1: & (\gamma_k,a_k) = (\zeta_{i_h+1}+\delta_{l_h},d_{l_h}), \ \forall h & \textit{for a unique } 1\leq m_k\leq \mathrm{Card}(I) \textit{ and} \\
    & s_k=\VN(\sum_{h=1}^{m_k}\VN(v_{i_h+1},u_{l_h})) & ((i_1,l_1),\dots,(i_{m_k},l_{m_k}))\in (I\times L)^{m_k} 
    \\[1ex]
    \textit{case }2: & (\gamma_k,a_k) = (\zeta_0+\theta_{j_0},c_{j_0})=(\zeta_{i_h+1}+\delta_{l_h},d_{l_h}), \ \forall h & \textit{for a unique } 0\leq m_k\leq \mathrm{Card}(I) \textit{ and} \\
    & s_k=\VN(\VN(v_0,t_{j_0})+\sum_{h=1}^{m_k}\VN(v_{i_h+1},u_{l_h})) & (j_0,(i_1,l_1),\dots,(i_{m_k},l_{m_k}))\in J\times (I\times L)^{m_k}
\end{array}\]
Now by the IH, what we just observed above, and remembering that $\mathrm{supp}(\VN(\_))\subseteq\mathrm{supp}(\_)$ and Lemma~\ref{lm:supp:dot=+}, we obtain that for all $k$:
\[\small\begin{array}{llll}
    \textit{case }1: & \mathrm{supp}(s_k) & \subseteq & \bigcup_h \,(\mathrm{supp}(v_{i_h+1})+\mathrm{supp}(u_{l_h})) \\
     & & \subseteq & \bigcup_h\, \left(\mathrm{supp}(\model{\Pi_1}_{\zeta_{i_h+1}\,|\, i_h+1}) + \mathrm{supp}(\model{\Pi_3}_{\delta_{l_h}\,|\, d_{l_h}})\right) \\
     & & \subseteq & \bigcup_h\, \left(\mathrm{supp}(\model{\Pi_1}_{\zeta_{i_h+1}\,|\, i_h+1}) + \mathrm{supp}(\model{\Pi_3}_{\delta_{l_h}\,|\, a_k})\right). \\ \\
    \textit{case }2: & \mathrm{supp}(s_k) & \subseteq & (\mathrm{supp}(v_0)+\mathrm{supp}(t_{j_0}))\cup\bigcup_h(\mathrm{supp}(v_{i_h+1})+\mathrm{supp}(u_{l_h})) \\
    & & \subseteq & (\mathrm{supp}(\model{\Pi_1}_{\zeta_{0}\,|\, 0}) + \mathrm{supp}(\model{\Pi_2}_{\theta_{j_0}\,|\, c_{j_0}})) \\
    & & & \cup\bigcup_h\,\left( \mathrm{supp}(\model{\Pi_1}_{\zeta_{i_h+1}\,|\, i_h+1}) + \mathrm{supp}(\model{\Pi_3}_{\delta_{l_h}\,|\, d_{l_h}})\right) \\
    & & = & (\mathrm{supp}(\model{\Pi_1}_{\zeta_{0}\,|\, 0}) + \mathrm{supp}(\model{\Pi_2}_{\theta_{j_0}\,|\, a_k})) \\
    & & & \cup\bigcup_h \,\left( \mathrm{supp}(\model{\Pi_1}_{\zeta_{i_h+1}\,|\, i_h+1}) + \mathrm{supp}(\model{\Pi_3}_{\delta_{l_h}\,|\, a_k})\right)
\end{array}\]
where in the respective last lines we simply used the fact that $d_{l_h}=a_k$ in the first case, and $c_{j_0}=d_{l_h}=a_k$ in the second case.
Finally, let us analyse $\model{\Pi}_{\gamma_k\,|\, a_k}$.
Following its definition, its support is the union of the following sets, for all $\xi^0,\xi^1$ $\TIT$-contexts such that $\xi^0+\xi^1=\gamma_k$:
\[
(
\mathrm{supp}(\model{\Pi_1}_{\xi^0\,|\, 0}) + \mathrm{supp}(\model{\Pi_2}_{\xi^1\,|\, a_k})
) \cup\bigcup\limits_{n\in\mathbb N}(
\mathrm{supp}(\model{\Pi_1}_{\xi_0\,|\, n+1}) + \mathrm{supp}(\model{\Pi_3}_{\xi^1\,|\, a_k})
)
\]
Now, fix $k$ and let $\mu\in\mathrm{supp}(s_k)$.
We show that $\mu$ is in the support of the parametric interpretation as above: 
We know that we are in exactly one among the two cases above; 

if we are in case $1$, then there is $h$ such that $\mu$ belongs to the sum of supports as above. We can then take $\xi^0:=\zeta_{i_h+1},\xi^1:=\delta_{l_h}$ (since $\zeta_{i_h+1}+\delta_{l_h}=\gamma_k$ in case $1$), and it is immediate to see that then $\mu$ belongs to the right union in the interpretation, for $n:=i_h$.

if we are in case $2$, then we have two cases: either $\mu$ belongs to the left sum above, and we can take $\xi^0:=\zeta_{0},\xi^1:=\theta_{j_0}$ (since $\zeta_{i_h+1}+\theta_{j_0}=\gamma_k$ in case $2$), immediately having that $\mu$ belongs to the left sum of supports in the interpretation. 
Or there is $h$ such that $\mu$ belongs to the right sum of supports, in which case we can then take $\xi^0:=\zeta_{i_h+1},\xi^1:=\delta_{l_h}$ (since in case $2$ also $\zeta_{i_h+1}+\delta_{l_h}=\gamma_k$), and it is immediate to see that then $\mu$ belongs to the right union for $n=i_h$.

Therefore, we are done.

$\bullet$ Case $\deriv{\Pi}{\Gamma\vdash MN:A}$ via $\deriv{\Pi_1}{\Gamma\vdash M:B\to A}$ and $\deriv{\Pi_2}{\Gamma\vdash N:B}$.

Then $\deriv{\pi}{MN:\big \langle \gamma_k \vdash^{s_k} a_k \big \rangle_k}$ via a $@$ rule on a $\deriv{\pi_1}{M:\big \langle \theta_k \vdash^{t_k} \zeta_k\multimap a_k \big \rangle_k}$ and a $\deriv{\pi_2}{N:\big \langle \delta_{1}^{k,b} \vdash^{u_1^{k,b}} b \,\big|\, \cdots \,\big|\, \delta_{\zeta_k(b)}^{k,b} \vdash^{u_{\zeta_k(b)}^{k,b}} b \big \rangle_{b\in\mathrm{supp}(\zeta_k),k\in K}}$ where $\zeta_k\in !\model B$ and $\theta_k,\delta_j^{k,b}$. 
Moreover, we have $\gamma_k = \theta_k+\sum_{b\in\mathrm{supp}(\zeta_k)}\sum_{j=1}^{\zeta_k(b)} \delta_j^{k,b}$ and $s_k=\VN(t_k,\prod_{b\in\mathrm{supp}(\zeta_k)}\prod_{j=1}^{\zeta_k(b)} u_j^{k,b})$.
Moreover, $\appr{\pi_1}{\Pi_1}$ and $\appr{\pi_2}{\Pi_2}$.

By IH, we have that for all $k$, $\mathrm{supp}(t_k)\subseteq \mathrm{supp}(\model{\Pi_1}_{\theta_k\,|\, (\zeta_k,a_k)})$ and $\mathrm{supp}(u_j^{k,b})\subseteq \mathrm{supp}(\model{\Pi_2}_{\delta_j^{k,b}\,|\, b})$ for all $b\in\mathrm{supp}(\zeta_k),\, j\in\set{1,\dots,\zeta_k(b)}$.
Remembering the shape of $s_k$, and using Lemma~\ref{lm:supp:dot=+} we have 
\[\begin{array}{ccl}
    \mathrm{supp}(s_k) & \subseteq & \mathrm{supp}(t_k\prod_{b\in\mathrm{supp}(\zeta_k)}\prod_{j=1}^{\zeta_k(b)} u_j^{k,b})   \\
    & = & \mathrm{supp}(t_k)+\sum_b\sum_{j} \mathrm{supp}(u_j^{k,b}) \\
    & \subseteq & \mathrm{supp}(\model{\Pi_1}_{\theta_k\,|\, (\zeta_k,a_k)})
+\sum_b\sum_{j}
\mathrm{supp}(\model{\Pi_2}_{\delta_j^{k,b}\,|\, b})
\end{array}\] 
where in the last inclusion we used what we found in the lines just above.
Now let us analyse $\model{\Pi}_{\gamma_k \,|\,a_k}$.
By looking at its definition, its support is the union of the following sets, for all $\zeta\in !\model{B}$ and $\theta,\delta_j^b$ $\TIT$-contexts such that $\gamma_k=\theta+\sum_{b\in\mathrm{supp}(\zeta)}\sum_{j=1}^{\zeta(b)}\delta_j^b$:
\[
\mathrm{supp}(\model{\Pi_1}_{\theta \,|\,(\zeta,a_k)})
+\sum_{b\in\mathrm{supp}(\zeta)}\sum_{j=1}^{\zeta(b)}
\mathrm{supp}(\model{\Pi_2}_{\delta_j^b \,|\,b}).\]
But we just found, for each $k$, such a multiset $\zeta$ (namely, $\zeta_k$) and decomposition $\theta,\delta_j^b$ (namely, $\theta_k,\delta_j^{k,b}$) of $\gamma_k$.
Therefore, we are done.

$\star$ Limit size case:

$\bullet$ Case $\deriv{\Pi}{\Gamma\vdash \YY M:A}$ via a $\YY$ rule on $\deriv{\Pi'}{\Gamma\vdash M:A\to A}$.

Then $\deriv{\pi}{\YY M:\big\langle \gamma_k\vdash^{s_k} a_k\big\rangle_k}$ and $\appr{\mathbf{unfold}.\pi \ }{ \ \mathbf{unfold}^{(\mathbf{cl}(\pi))}.\Pi'}$ and we remember, from Definition~\ref{def:unfold_Pi}, that in general $\deriv{\mathbf{unfold}^{(n)}.\Pi'}{\Gamma,y:A\vdash M^{(n)}y:A}$.
Now, it is easy to see, using the definition of the interpretation, that $\model{\Pi}_{\gamma_k \,|\, a_k}=\bigvee_{n\in\N} \model{\mathbf{unfold}^{(n)}.\Pi'}_{x\neq y\mapsto \gamma_k(x), y\mapsto[] \,|\, a_k}$.
Therefore, using Lemma~\ref{lm:supp:sup=cup}, we have 
\[
\mathrm{supp}(\model{\Pi}_{\gamma_k \,|\, a_k})=\bigcup_{n\in\N}\mathrm{supp}(\model{\mathbf{unfold}^{(n)}.\Pi'}_{x\neq y\mapsto \gamma_k(x),y\mapsto[] \,|\, a_k}).
\]
Remember that, by Definition~\ref{def:unfold_pi}, we have $\deriv{\mathbf{unfold}.\pi}{M^{(\mathbf{cl}(\pi))}y:\big\langle \gamma_k\vdash^{s_k} a_k\big\rangle_k}$.
But then, using the IH on $\mathbf{unfold}^{(\mathbf{cl}(\pi))}.\Pi'$, we obtain 
\[
\mathrm{supp}(s_k)\subseteq \mathrm{supp}(\model{\mathbf{unfold}^{(n)}.\Pi'}_{x\neq y\mapsto \gamma_k(x),y\mapsto[] \,|\, a_k}).
\]
Therefore, we are done.
\end{proof}

%%%%%%%%%%%%%%%%%%%%%%%%%%%%%%%%%%%%%

For Lemma~\ref{lm:min_compl} we need the following lemmas:
the first is needed in the subcase $\oplus_X$ of the ``merge lemma'' below, the second in the subcase $\YY$.

\begin{lemma}\label{lm:aux1}
Let $s$ be a polynomial. Then $(Xs)_\mathrm{min}=Xs_\mathrm{min}$
\end{lemma}

\begin{lemma}\label{lm:aux2}
Let $\pi,\Pi$ be $\TIT$ and \pPCF-derivations. Then
$\appr{\pi}{\mathbf{wkn}^{(y:A)}.\Pi}$ iff $\appr{\pi}{\Pi}$.
\end{lemma}
\begin{proof}
Intuitively, it holds because $y:A$ plays no role in $\mathbf{wkn}^{(y:A)}.\Pi$, so $\pi$ can just ignore it. More precisely, one can go by induction on the size of $\Pi$.
If $\pi=\emptyset$ then it trivially holds. Otherwise, since the weakening commutes, by definition, with all the constructors of the \pPCF\ derivations, all the cases are trivial, except for the $Y$ one, which however immediately follows after remarking that weakening i) preserves clusters, and ii) commutes with any $n$-unfolding of \pPCF-derivations.
The i) is immediate, the ii) is done in the next Lemma~\ref{lm:aux3}.
\end{proof}

\begin{lemma}\label{lm:aux3}
Let $\Pi$ be a \pPCF-derivation. Then $\mathbf{wkn}^{(y:A)}.\mathbf{unfold}^{(n)}.\Pi=\mathbf{unfold}^{(n)}.\mathbf{wkn}^{(y:A)}.\Pi$.
\end{lemma}
\begin{proof}
Immediate induction on $n\in\N$, using the fact that weakenings commutes with each other, i.e.\ $\mathbf{wkn}^{(y:A)}.\mathbf{wkn}^{(z:B)}.\Pi=\mathbf{wkn}^{(z:B)}.\mathbf{wkn}^{(y:A)}.\Pi$. This latter fact is a trivial structural induction on~$\Pi$.
\end{proof}

\begin{lemma}[Derivation Merge Lemma]\label{lm:merge_deriv}
There is an algorithm that, given inputs a $\deriv{\Pi}{\Gamma~\vdash~M~:~A}$, a $\deriv{\pi}{M:\langle \gamma_i\vdash^{s_i} a_i\rangle_i}$ and a $\deriv{\rho}{M:\langle \delta_j\vdash^{t_j} a'_j \rangle_j}$, if  $\appr{\pi}{\Pi}$ and $\appr{\rho}{\Pi}$ returns a derivation $\deriv{\SELECT(\pi, \rho)}{M:\SELECT\big\langle \gamma_i\vdash^{s_i} a_i \ \big| \ \delta_j\vdash^{t_j} a'_j  \big\rangle_{i,j}}$ such that $\appr{\SELECT(\pi, \rho)}{\Pi}$.
\end{lemma}
\begin{proof}
Structural Induction on $\Pi$.
The axiom cases are easy.

Some cases are easy by the IH. The cases $\oplus_X$, $\mathsf{ifz}$, $@$ and $Y$ require much care though.
For example, for $\oplus_X$ there are 15 sub-cases. Luckily, they are all straightforward and, by symmetry, we only need to do a part of them. In some of them we use Lemma~\ref{lm:aux1}.

Let us detail the Case $\deriv{\Pi}{\Gamma\vdash \YY M:A}$ via a $\YY$-rule on $\deriv{\Pi'}{\Gamma\vdash M:A\to A}$:

By definition, $\appr{\mathbf{unfold}.\pi}{\mathbf{unfold}^{\mathbf{cl(\pi)}}.\Pi'}$ and $\appr{\mathbf{unfold}.\rho}{\mathbf{unfold}^{\mathbf{cl(\rho)}}.\Pi'}$.
Moreover, by the definition of unfolding of $\TIT$-derivations, $\mathbf{unfold}.\pi$ must be a cluster of $\mathbf{cl(\pi)}$ $@$-rules, with top-right premise %$\deriv{\pi^0}{y:\emptyset}$
$\frac{}{y:\emptyset}$, and left premises $\deriv{\pi_i}{M:\langle\cdots\rangle}$ for $i=1,\dots,\mathbf{cl(\pi)}$.
Similarly for $\mathbf{unfold}.\rho$ we have the respective $\frac{}{y:\emptyset}$ as top-right premise and $\rho_j$ for $i=j,\dots,\mathbf{cl(\rho)}$ as left premises.
We have% $\appr{\pi^0}{\frac{}{\Gamma,y:A\vdash y:A}}$ and
, by definition of refinement, $\appr{\pi_i}{\mathbf{wkn}^{(y:A)}.\Pi'}$, i.e.\ by Lemma~\ref{lm:aux2}, $\appr{\pi_i}{\Pi'}$.
Same for %$\rho^0$ and 
the $\rho_j$.
Therefore we can call the algorithm on $\Pi',\pi_1,\dots,\pi_{\mathbf{cl(\pi)}},\rho_1,\dots,\rho_{\mathbf{cl(\rho)}}$.
Let $\theta$ be the output.
By IH% (since the size of $\Pi'$ is strictly smaller than the one of $\Pi$)
, we have $\appr{\theta}{\Pi'}$, i.e.\ $\appr{\theta}{\mathbf{wkn}^{(y:A)}.\Pi'}$ by Lemma~\ref{lm:aux2}.
%We can also call the algorithm on $\pi^0,\rho^0$, obtaining output $\theta^0$ which is, by construction, $\appr{\theta^0}{\frac{}{\Gamma,y:A\vdash y:A}}$.
Now let us consider the $\TIT$-derivation $\deriv{\zeta}{My:\langle\cdots\rangle}$ obtained via a single $@$-rule on $\theta$ as left premise and $\frac{}{y:\emptyset}$ as right premise.
By construction, $\zeta=\mathbf{unfold}.\xi$, where $\deriv{\xi}{\YY M:\langle\cdots\rangle}$ is the $\TIT$-derivation obtained via a $Y$-rule on $\theta$ as left premise and $\frac{}{y:\emptyset}$ as right premise.
Now we remember that we already obtained that $\appr{\theta}{\mathbf{wkn}^{(y:A)}.\Pi'}$ and,
moreover, $\appr{\frac{}{y:\emptyset}}{\frac{}{\Gamma,y:A\vdash y:A}}$. 
Therefore $\appr\xi{\mathbf{unfold}^{(1)}.\Pi'}$.
But by construction, $\mathbf{cl}(\xi)=1$, so by definition of refinement, $\appr\xi\Pi$.
Finally, one can tediously check that $\deriv{\xi}{M:\SELECT\big\langle \gamma_i\vdash^{s_i} a_i \ \big| \ \delta_j\vdash^{t_j} a'_j  \big\rangle_{i,j}}$.
Therefore, we can return $\xi$ as output.
\end{proof}

We can now state and prove Lemma~\ref{lm:min_compl}.

\begin{lemma}\label{lm:min_compl}
We define in the proof an algorithm\footnote{To be precise, we give a non-deterministic algorithm, in the sense that, in some cases, we need to recursively call it in parallel on different inputs, without knowing which is the correct input that will return an answer (but knowing that one eventually will).} that, given $\deriv \Pi \Gamma\vdash M:A$, given a $\TIT$-type $a$ and a $\TIT$-context $\gamma$ such that $(\gamma,a)$ is adequate for $(\Gamma,A)$, and given\footnote{To be pedantic, in the following we mean $\mathrm{supp}(\model{\Pi}^{\mathbb X}_{{\gamma}_{|_{\mathrm{dom}(\Gamma)}}\,|\, a})$.} $\mu\in \mathrm{supp}(\model{\Pi}^{\mathbb X}_{\gamma \,|\, a})_{\mathrm{min}}$, returns a $\deriv{\pi}{M: \big \langle \gamma%_k
\vdash^{s%_k
} a%_k 
\big \rangle%_k
}$ %and a $k_0\in K$ 
such that %i) 
$\appr{\pi}{\Pi}$ and %ii) $(\gamma,a)=(\gamma_{k_0},a_{k_0})$ and 
$\mu\in\mathrm{supp}(s%_{k_0}
)$.

Its correctness and termination are guaranteed by the well-foundedness of the size ordinal.
\end{lemma}
\begin{proof}
Given inputs $\Pi,\gamma,a,\mu$, we go by pattern matching on $\Pi$, and the termination is obtained by transfinite induction on the size of $\Pi$.

$\star$ Minimal size cases:

$\bullet$ Case $\Pi$ is the axiom $\Gamma\vdash \ZERO:\NAT$, or the axiom $\Gamma\vdash \ZERO:\Bool$ or the axiom $\Gamma\vdash \ONE:\Bool$.

In the first two [resp.\ third] cases, IF $\mu$ in the support of $\model{\Pi}_{\gamma \,|\, a}$ then $\gamma(x)=[]$ for all $x$, $a=0$ [resp.\ $a=1$] and $\mu=\prod_j X_j^0\OV X_j^0$. 
Then it is immediate to check that we can return $\pi$ the axiom $\ZERO: \big \langle \vdash^{1_{\fps{\NINF}{\mathbb X}}} 0  \big \rangle$ and $k_0$ the unique element of the singleton.

$\bullet$ Case $\Pi$ is the axiom $\Gamma, y:A\vdash y:A$.

If $\mu$ is in the support of $\model{\Pi}_{\gamma \,|\, a}$ then $\gamma(x\neq y)=\vec{[]}$, $\gamma(y)=[a]$ and $\mu=\prod_j X_j^0\OV X_j^0$.
Then it is immediate to check that we can return $\pi$ the axiom $y: \big \langle y:[a] \vdash^{1_{\fps{\NINF}{\mathbb X}}} a  \big \rangle$ and $k_0$ the unique element of the singleton.

$\star$ Successor size cases:

$\bullet$ Case $\deriv{\Pi}{\Gamma\vdash\NAT}$ via a cast rule on $\deriv{\Pi'}{\Gamma\vdash\Bool}$.

By definition of the interpretation, if $\mu$ is in the support of $\model{\Pi}_{\gamma \,|\, a}$, then $a=:n\in\set{0,1}$ and $\mu\in\mathrm{supp}(\model{\Pi'}_{\gamma_k \,|\, n})$.
We can then call the algorithm on $\Pi',\gamma,n,\mu$.
Let $(\pi,k_0)$ the output.
It is immediate to check that we can then return $\pi$ and $k_0$.

$\bullet$ Case $\deriv{\Pi}{\Gamma\vdash \SUCC M:\NAT}$ via a successor rule on $\deriv{\Pi_1}{\Gamma\vdash M:\NAT}$.

By definition of the interpretation, if $\mu$ is in the support of $\model{\Pi}_{\gamma \,|\, a}$, then $a=:n+1$ with $n\in\N$, and $\mu\in\mathrm{supp}(\model{\Pi_1}_{\gamma_k \,|\, n})$.
We can then call the algorithm on $\Pi_1,\gamma,n,\mu$.
Let $(\pi_1,k_0)$ the output.
It is immediately seen that we can apply a successor rule to $\pi_1$. Let $\pi$ the obtained derivation.
It is immediate to check that we can then return then $\pi$ and $k_0$.

$\bullet$ Case $\deriv{\Pi}{\Gamma\vdash \PRED M:\NAT}$ via a predecessor rule on $\deriv{\Pi_1}{\Gamma\vdash M:\NAT}$.

By definition of the interpretation, if $\mu$ is in the support of $\model{\Pi}_{\gamma \,|\, a}$, then we have two cases:
either $a=0$ and $\mu\in\mathrm{supp}(\model{\Pi_1}_{\gamma \,|\, 0})\cup\mathrm{supp}(\model{\Pi_1}_{\gamma \,|\, 1})$.
Or $a=:n\geq 1$ and $\mu\in\mathrm{supp}(\model{\Pi_1}_{\gamma \,|\, n+1})$.
In the first case, depending on which part of the union $\mu$ is, we can call the algorithm on $\Pi_1,\gamma,i,\mu$, for $i=0$ or $1$.
Let $(\pi_1^i,k_0^i)$ be the output the respective case. 
In the other case, we can then call the algorithm on $\Pi_1,\gamma,n+1,\mu$.
Let $(\pi_1',k_0')$ be the output in this case.
Now we have:
$\deriv{\pi_1^i}{M:\big \langle \gamma_k \vdash^{s_k} m_k \big \rangle_k}$, where the $m_k$ are suitable natural numbers, because by IH $(\gamma_k,m_k)$ is adequate for $(\Gamma,\NAT)$.
Similarly for $\pi_1'$.
We can hence apply a predecessor rule to $\pi_1^i$ and, letting $\pi^i$ the obtained derivation, we return $(\pi^i,k_0^i)$.
Similarly, we can apply a predecessor rule to $\pi_1'$ and, letting $\pi'$ the obtained derivation, we return $(\pi',k_0')$.
Let us check that these choices work:
it is clear by definition of $\appr{}{}$ that the obtained $\pi^i$ or $\pi'$ refine $\Pi$, because by IH the respective $\pi_1^i$ or $\pi_1'$ refines $\Pi_1$.
Moreover, by IH we know that, in the case $\pi_1^i$, we have $(\gamma,i)=(\gamma_{k_0^i},m_{k_0^i})$, and in the case $\pi_1'$, we have $(\gamma,n+1)=(\gamma_{k_0'},m_{k_0'})$.
So, in $\pi'$, we have $(\gamma,a)=(\gamma,i\dotdiv 1)=(\gamma_{k_0^i},m_{k_0^i}\dotdiv 1)$, and in $\pi'$, we have $(\gamma,a)=(\gamma,n)=(\gamma_{k_0'},m_{k_0'}\dotdiv 1)$.

$\bullet$ Case $\deriv{\Pi}{\Gamma\vdash \lambda y.M:E\to F}$ via a $\lambda y$ rule (for $y$ non-declared in $\Gamma$) on $\deriv{\Pi_1}{\Gamma,y:E\vdash M:F}$.

By definition of the interpretation, if $\mu$ is in the support of $\model{\Pi}_{\gamma \,|\, a}$, then $\mu\in\mathrm{supp}(\model{\Pi_1}_{x\neq y\mapsto\gamma(x),y\mapsto\eta \,|\, f})$ and $a=(\eta,f)$, for some $(\eta,f)$ adequate for $(E,F)$.
We can then call the algorithm on $\Pi_1,(x\neq y\mapsto\gamma(x),y\mapsto\eta),f,\mu$.
Let $(\pi_1,k_{0})$ be the output.
It is immediately seen that we can apply a $\lambda y$ rule to $\pi_1$.
Let $\pi$ be the obtained derivation.
It is immediate to check that we can then return $(\pi,k_{0})$.

$\bullet$ Case $\deriv{\Pi}{\Gamma\vdash M\oplus_X N:A}$ via a $\oplus$ rule on $\deriv{\Pi_1}{\Gamma\vdash M:A}$ and $\deriv{\Pi_2}{\Gamma\vdash N:A}$.

By definition of the interpretation, if $\mu$ is in the support of $\model{\Pi}_{\gamma\,|\, a}$, then it is either of shape $\mu=X\mu_1$ with $\mu_1\in\mathrm{supp}(\model{\Pi_1}_{\gamma\,|\, a})$, or it is of shape $\mu=\OV X\mu_2$ with $\mu_2\in\mathrm{supp}(\model{\Pi_2}_{\gamma\,|\, a})$.
We can then call the algorithm on $\Pi_1,\gamma,a,\mu_1$ and on $\Pi_2,\gamma,a,\mu_2$.
If the first returns, let $(\pi_1,k_0^1)$ its output.
If the second returns, we proceed similarly to the first case, which we detail below.
In the first case, we can apply a $\oplus$ rule on $\deriv{\pi_1}{M:\big\langle \gamma_k\vdash^{s_k} a_k \big\rangle_k}$ and $N:\emptyset$ in order to obtain a derivation $\deriv{\pi}{M\oplus_X N:\SELECT\big\langle \gamma\vdash^{Xs_k} a_k \big\rangle}$.
But since the $(\gamma_k,a_k)$ are all pairwise distinct by construction, the merge does not change the judgment (we are in ``case 1'', with the terminology of the proof of the previous lemma), which is thus $\deriv{\pi}{M:\oplus_X N:\big\langle \gamma_k\vdash^{Xs_k} a_k \big\rangle}$.
Now, by IH we have $(\gamma_{k_0^1},a_{k_0^1})=(\gamma,a)$ and $\mu_1\in\mathrm{supp}(s_{k_0^1})$.
Therefore we can return $(\pi,k_0^1)$.

$\bullet$ Case $\deriv{\Pi}{\Gamma\vdash \ITE{M}{N}{P}:A}$ via a $\mathsf{ifz}$ rule on $\deriv{\Pi_1}{\Gamma\vdash M:\NAT}$ and $\deriv{\Pi_2}{\Gamma\vdash N:A}$ and $\deriv{\Pi_3}{\Gamma\vdash P:A}$.

Similar to the above.

$\bullet$ Case $\deriv{\Pi}{\Gamma\vdash MN:A}$ via $\deriv{\Pi_1}{\Gamma\vdash M:B\to A}$ and $\deriv{\Pi_2}{\Gamma\vdash N:B}$.

By definition of the interpretation, if $\mu$ is in the support of $\model{\Pi}_{\gamma \,|\, a}$, then we have:
\[\mu=\mu_0+\sum_{b\in\mathrm{supp}(\zeta)}\sum_{j=1}^{\zeta(b)} \mu_j^b\]
for some $\zeta\in !\model{B}$ and $\theta,\delta_j^b$ $\TIT$-contexts such that $\gamma=\theta+\sum_{b\in\mathrm{supp}(\zeta)}\sum_{j=1}^{\zeta(b)}\delta_j^b$, with
\[\begin{array}{rcl}
    \mu_0 & \in &\mathrm{supp}(\model{\Pi_1}_{\theta \,|\,(\zeta,a)}) \\
    \mu_j^b & \in & \mathrm{supp}(\model{\Pi_2}_{\delta_j^b \,|\,b}).
\end{array}\]
It is easily seen that we can call the algorithm on $\Pi_1,\theta,(\zeta,a),\mu_0$ and on $\Pi_2,\delta_j^b,b,\mu_j^b$.
Let $(\pi_1,k_0^1)$ and $(\pi_2^{j,b},k_0^{j,b})$ be respective the outputs.
One can now merge all the (finitely many) $\pi_2^{j,b}$ using Lemma~\ref{lm:merge_deriv}, in order to obtain a $\pi_2$.
With this, one can conclude.

$\star$ Limit size case:

$\bullet$ Case $\deriv{\Pi}{\Gamma\vdash \YY M:A}$ via a $\YY$ rule on $\deriv{\Pi'}{\Gamma\vdash M:A\to A}$.

Similar argument, first decomposing the interpretation and then putting together the derivations obtained by IH.
\end{proof}

 %la vera sezione 7

\end{document}